\documentclass[iop, apj, numberedappendix, appendixfloats]{emulateapj}

\usepackage{amsmath}
\usepackage{amssymb}
\usepackage{graphicx}

\graphicspath{{./figs/}}
\defcitealias{ivison16}{Paper~I}

\newcommand{\beam}{\textrm{beam}}
\newcommand{\fir}{far-IR}
\newcommand{\hermes}{\textit{Her}MES}
\newcommand{\lfir}{L_{\textrm{\fir{}}}}
\newcommand{\lsol}{L_{\odot}}
\newcommand{\megayear}{\textrm{Myr}}
\newcommand{\mdm}{M_{\textrm{DM}}}
\newcommand{\msol}{M_{\odot}}
\newcommand{\millijanksy}{\textrm{mJy}}
\newcommand{\millimeter}{\textrm{mm}}
\newcommand{\pur}{P_{\textrm{UR}}}
\newcommand{\scuba}{\mbox{SCUBA-2}}
\newcommand{\zphot}{z_{\textrm{phot}}}
\newcommand{\zspec}{z_{\textrm{spec}}}

\begin{document}

\title{Ultra-Red Galaxies Signpost Candidate Proto-Clusters at High Redshift}

\shorttitle{Ultra-Red Galaxies Signpost High-Redshift Proto-Clusters}
\shortauthors{Lewis et al.}

\author{A.\,J.\,R.~Lewis\altaffilmark{1}}
\author{R.\,J.~Ivison\altaffilmark{2,1}}
\author{P.\,N.~Best\altaffilmark{1}}
\author{J.\,M.~Simpson\altaffilmark{1}}
\author{A.~Weiss\altaffilmark{3}}
\author{I.~Oteo\altaffilmark{1,2}}
\author{Z-Y.~Zhang\altaffilmark{1,2}}

\author{V.~Arumugam\altaffilmark{2,1}}
\author{M.\,N.~Bremer\altaffilmark{13}}
\author{S.~C.~Chapman\altaffilmark{19}}
\author{D.~L.~Clements\altaffilmark{14}}
\author{H.~Dannerbauer\altaffilmark{9,10,11}}
\author{L.~Dunne\altaffilmark{1,5}}
\author{S.~Eales\altaffilmark{5}}
\author{S.~Maddox\altaffilmark{1,5}}
\author{S.\,J.~Oliver\altaffilmark{17}}
\author{A.~Omont\altaffilmark{7,8}}
\author{D.\,A.~Riechers\altaffilmark{4}}
\author{S.~Serjeant\altaffilmark{18}}
\author{E.~Valiante\altaffilmark{5}}
\author{J.~Wardlow\altaffilmark{12}}
\author{P.~van~der~Werf\altaffilmark{6}}
\author{G.\,De~Zotti\altaffilmark{16}}

\altaffiltext{1}{Institute for Astronomy, University of Edinburgh,
  Royal Observatory, Blackford Hill, Edinburgh EH9 3HJ, UK}
\altaffiltext{2}{European Southern Observatory, Karl Schwarzchild
  Stra{\ss}e 2, D-85748 Garching, Germany}
\altaffiltext{3}{Max-Planck-Institut f\"{u}r Radioastronomie,
  Auf dem H\"{u}gel 69, D-53121 Bonn, Germany}
\altaffiltext{4}{Cornell University, Space Sciences Building, Ithaca, NY 14853, USA}
\altaffiltext{5}{School of Physics \& Astronomy, Cardiff University,
  Queen's Buildings, The Parade, Cardiff CF24 3AA, UK}
\altaffiltext{6}{Leiden Observatory, Leiden University, P.O.\ Box
  9513, NL-2300 RA Leiden, The Netherlands}
\altaffiltext{7}{UPMC Univ Paris 06, UMR 7095, IAP, 75014, Paris, France}
\altaffiltext{8}{CNRS, UMR7095, IAP, F-75014, Paris, France}
\altaffiltext{9}{IAC, E-38200 La  Laguna, Tenerife, Spain}
\altaffiltext{10}{Departamento de Astrofisica, Universidad de La
  Laguna, E-38205 La Laguna, Tenerife, Spain}
\altaffiltext{11}{Universit\"{a}t Wien, Institut f\"{u}r Astrophysik,
  T\"{u}rkenschanzstr.\ 18, 1180 Wien, Austria}
\altaffiltext{12}{Centre for Extragalactic Astronomy, Department of
  Physics, Durham University, South Road, Durham DH1 3LE, UK}
\altaffiltext{13}{H.\,H.\ Wills Physics Laboratory, University of
  Bristol, Tyndall Avenue, Bristol BS8 1TL, UK}
\altaffiltext{14}{Astrophysics Group, Imperial College London,
  Blackett Laboratory, Prince Consort Road, London SW7 2AZ, UK}
\altaffiltext{15}{SISSA, Via Bonomea 265, I-34136, Trieste, Italy}
\altaffiltext{16}{INAF-Osservatorio Astronomico di Padova, Vicolo
  dell’Osservatorio 5, I-35122 Padova, Italy}
\altaffiltext{17}{Astronomy Centre, Department of Physics and Astronomy, University of Sussex, Brighton BN1 9QH}
\altaffiltext{18}{Department of Physical Sciences, The Open University, Milton Keynes, MK7 6AA, UK}

\slugcomment{To be submitted to The Astrophysical Journal}

\begin{abstract}
  We present images obtained with the Atacama Pathfinder Experiment
  (APEX) telescope's Large APEX BOlometer CAmera (LABOCA) of a sample
  of 22 galaxies selected via their red \textit{Herschel} SPIRE
  (Spectral and Photometric Imaging Receiver) 250-, 350- and
  \mbox{$500$-$\mu\textrm{m}$} colors. We aim to see if these
  luminous, rare and distant galaxies are signposting dense
  regions in the early Universe. Our $870\textrm{-}\micron{}$
  survey covers an area of $\approx0.8\,\textrm{deg}^2$ down to an
  average r.m.s.\ of $3.9\,\millijanksy{}\,\beam{}^{-1}$, with our five deepest maps going $\approx2\times$ deeper still.
  We catalog 86 dusty star-forming galaxies (DSFGs) around our `signposts', detected above a significance of $3.5\sigma$.
  This implies a $100^{+30}_{-30}\%$ over-density of $S_{870}>8.5\,\millijanksy{}$ DSFGs, excluding our signposts, when comparing our number counts to those in `blank fields'.
  Thus, we are 99.93\% confident that our signposts are pinpointing
  over-dense regions in the Universe, and $\approx95$\% [50\%] confident
  that these regions are over-dense by a factor of at least $\ge1.5\times$
  [$2\times$].  Using template spectral energy distributions and
  SPIRE/LABOCA photometry we derive a median photometric redshift of
  $z=3.2\pm0.2$ for our signposts, with an interquartile range of
  $z=2.8\textrm {--}3.6$, somewhat higher than expected for
  $\sim850\,\mu\textrm{m}$-selected galaxies.  We constrain the DSFGs
  likely responsible for this over-density to within
  $|\Delta z|\le0.65$ of their respective signposts; over half of our
  ultra-red targets ($\approx55\%$) have an average of two DSFGs within
  $|\Delta z|\le0.5$. These `associated' DSFGs are radially
  distributed within (physical) distances of $1.6\pm0.5\,\textrm{Mpc}$
  from their signposts, have median star-formation rates (SFRs) of
  $\approx(1.0\pm0.2)\times10^3\,M_ {\odot}\,\textrm{yr}^ {-1}$ (for a
  Salpeter stellar initial mass function) and median gas reservoirs of
  $\sim1.7\times10^{11}\,\msol{}$. These candidate proto-clusters
  have average total SFRs of at least
  $\approx (2.3\pm0.5)\times10^3\,\msol{}\,\textrm{yr}^{-1}$ and
  space densities of $\sim9\times10^{-7}\,\textrm{Mpc}^{-3}$,
  consistent with the idea that their constituents may evolve to
  become massive early-type galaxies in the centers of the rich
  galaxy clusters we see today.
\end{abstract}


\section{Introduction}
\label{sec:introduction}

Galaxy clusters whose cores are rich with early-type galaxies (ETGs, i.e.\
relatively passive ellipticals and lenticulars) mark the densest regions in
the distribution of dark matter (DM), regions which have grown
hierarchically from initial, Gaussian fluctuations, supposedly etched into
the Universe at some arbitrarily early epoch \citep[e.g.][]{peebles70,
  white78, spergel03}. In the local Universe, these galaxy clusters harbor
the majority of ETGs, which in turn harbor over half of the present-day
stellar mass ($M_{ \textrm{stars}}$). Thus studying their cosmic evolution
can place valuable constraints on models of galaxy formation
\citep[e.g.][]{springel05, robertson07, overzier09a, lacey16}.

ETGs obey a tight scaling relation between their color and magnitude, where
magnitude equates roughly to $M_{\textrm{stars}}$. This is known as the
`red sequence', in which more massive galaxies are typically redder with
older stellar populations and less ongoing star formation
\citep[e.g.][]{bower98, baldry04, bower06, bell04}. Furthermore, ETGs in
local galaxy clusters appear redder (and thus more massive, since they
follow the scaling relation) as their distance to the cluster center
decreases \citep{bernardi06}.  These properties are consistent with the
concept of `cosmic downsizing' (\citeauthor{cowie96}, \citeyear{cowie96};
and see Fig.~9 in \citeauthor{thomas10}, \citeyear{thomas10}), whereby
the most massive ETGs formed their stars early ($z\gtrsim3$) and over
relatively short timescales \citep[$\lesssim0.5\,\textrm{Gyr}$
---][]{nelan05, thomas05, thomas10, snyder12, tanaka13a, tanaka13b}.

ETGs have commonly been viewed as transformed late-type galaxies (LTGs,
i.e.\ star-forming spirals) which have had their star formation quenched
via some mechanism, leaving behind an ETG on the red sequence
\citep{dressler97, gerke07}.  In local galaxy clusters this quenching is
brought about rapidly via ram pressure stripping \citep{gunn72} or by
so-called `starvation' and/or `strangulation' processes\footnote{Galaxy
  clusters reside in deep gravitational potentials which heat the
  intracluster medium (ICM). As a consequence, the ICM strips the cold gas
  from infalling LTGs and subsequently starves/strangles them of cold gas,
  the fuel for further star formation.}  \citep{larson80, balogh00,
  elbaz07, cooper08, tanaka13a, casado15}. However, at higher redshifts,
could the most-massive ETGs, in the centers of galaxy clusters, be the
remnants of colossal merger events instead?

An extreme event like this would require wildly different behavior for the precursors of ETGs at $z>3$, with such systems exhibiting immensely high star-formation rates (SFRs, $\psi\sim10^3\,\msol{}\,\textrm{yr}^{-1}$).
In a hierarchical context this large burst of star formation is driven by mergers in dense environments \citep{lacey93}.
Although the existence of such large systems at such high redshifts places stress on the hierarchical paradigm \citep{granato04}, it is conceivable that dusty star-forming galaxies \citep[DSFGs --- e.g.][]{blain02, casey14} are associated with these distant events at an epoch when the merger rates are comparatively high \citep{hine16, delahaye17}.

Conventional wisdom places this dusty population at $z\sim2.5$ \citep{chapman05, simpson14}, but recent work by \citet{riechers13}, \citet{dowell14}, \citet{asboth16} and \citet[][hereafter \citetalias{ivison16}]{ivison16}, to name but a few, suggests that a rare, $z\gtrsim3$ subset can be identified via their red, far-infrared (far-IR) colors as measured by the Spectral and Photometric Imaging Receiver \citep[SPIRE ---][]{griffin10} on board the \textit{Herschel} Space Observatory \citep{pilbratt10}.
Lensed DSFGs at similarly high redshifts have also been found by surveys at $\lambda_{\rm obs}>1\,\millimeter{}$ with the South Pole Telescope -- relying on flux-density ratios at even longer wavelengths to generate a sample of distant, dust-dominated sources \citep{vieira10, weiss13, strandet16}.

With remarkably high median rest-frame,
$8\textrm{--}1000\textrm{-}\mu\textrm{m}$ luminosities,
$\lfir{}=1.3\times10^{13}\,\lsol{}$, these so-called `ultra-red galaxies'
can provide the SFRs necessary to give birth to the most-massive ETGs in
the centers of galaxy clusters and, thus, the red sequence. In this work,
we go one step further than \citetalias{ivison16}
exploiting a representative sample of ultra-red galaxies to decipher
whether these $z\gtrsim3$ DSFGs exhibit evidence of clustering
consistent with their eventual membership of massive galaxy clusters
at $z\sim0$.

If ultra-red galaxies do indeed trace the precursors of the most massive
ETGs in the centers of present-day galaxy clusters, we would expect to
witness \emph{comparatively}\footnote{%
 This important semantic reflects the fact that on some scale any system could be virialized,
 for e.g., a single $10^{13}\textrm{--}10^{14}\,\msol{}$ proto-cluster may be unvirialized
 (for some foreseeable dynamical time) but it is comprised of many tens of virialized
 $10^{12}\textrm{--}10^{13}\,\msol{}$ components.
 }
unvirialized systems characterized by over-densities of (physically
associated) DSFGs \citep[i.e.\ a `proto-cluster' ---][]{muldrew15, casey16}.
Such systems have already been discovered in the $z>3$ Universe via their
submillimeter (submm) emission, with previous work typically relying either on
high-redshift radio galaxies \citep[HzRGs --- e.g.][]{ivison00, stevens03,
stevens04, rigby14}, \emph{pairs} of quasi-stellar objects
\citep[QSOs ---][]{uchiyama17}
or even strong over-densities of
Lyman-$\alpha$ emitters as signposts \citep{capak11, tamura09, tozzi15}.
Predictions by \citet{negrello05} suggested that bright-intensity peaks
within low-resolution data taken with the \textit{Planck} High Frequency Instrument, could
represent of clumps DSFGs. Indeed, over-densities of DSFGs at $z\sim3$ have been found
using this technique \citep[i.e.\ `HATLAS12-00' ---][]{clements16}.

Although DSFGs appear to be poor tracers of large structure below
$z\lesssim2.5$ \citep{miller15}, the situation appears to be quite different by
$z\sim5$ \citep{miller16, oteo17} -- albeit care must be taken when
discovering over-densities within a rare (thus low-numbered) population of galaxy.
%
At odds with this concept is the most-distant ($z\sim6$), ultra-red galaxy
discovered to date, `HFLS~3' \citep{riechers13}.
Confusion-limited observations of the environments surrounding this DSFG
showed little evidence that it signposted an over-density of DSFGs \citep{robson14}.
However, in light of new and improved comparison data, it appears that
HFLS~3 perhaps signposts regions that are over-dense by a factor of at least $\sim2\times$.

Thus, if our sample of ultra-red galaxies shows an excess of DSFGs compared to the
field then we will have confirmed this novel technique for pinpointing
primordial over-densities in the distant Universe. Combined with follow-up
optical imaging/spectroscopy of their environments
\citep[to detect so-called `Lyman-break' galaxies, LBGs ---][]{steidel96, madau96}
, we will be able to
place strong constraints on their $M_{\textrm{stars}}$ and DM components. A
joint approach -- combining models \citep[e.g.][]{springel05} and
observations -- is necessary to fully predict the eventual fate of these
proto-clusters at $z\sim0$ \citep{casey16, overzier16}  .

The structure of this paper is as follows. In the next section we outline
our target sample, as well as our data acquisition and reduction
methods. We analyze our data in \S\ref{sec:analysis} and discuss their
implications in \S\ref{sec:discussion}. Finally, our conclusions are
presented in \S\ref{sec:conclusion}. Throughout our analysis and
discussion, we adopt a `concordance cosmology' with
$H_0 = 71\,\textrm{km}\,\textrm{s}^ {-1}\,\textrm{Mpc}^{-1}$,
$\Omega_m = 0.27$ and $\Omega_{\Lambda} = 0.73$ \citep {hinshaw09}, in
which $1'$ corresponds to a (proper) distance of
$\approx0.5\,\textrm{Mpc}$ at $z=3.0$. For a quantity, $x$, we denote its
mean and median values as $\overline{x}$ and $x_{1/2}$, respectively.

\section{Target Sample and Data reduction}
\label{sec:data reduction}

\subsection{Target sample}
\label{sec:sample}

We selected 12 targets\footnote{%
 These targets were initially chosen for follow-up observations as their preliminary,
 albeit shallow, data at $\sim850\textrm{-}\mu\textrm{m}$ suggested that they were robust
 detections.}
 from the \textit{H}-ATLAS \citep[\textit{Herschel}
Astrophysical Terahertz Large Area Survey ---][]{eales10} imaging
survey. These targets are contained in the Data Release 1
\citep[\textsc{dr1} ---][] {valiante16, bourne16} \textit{H}-ATLAS images of the two
equatorial Galaxy And Mass Assembly (GAMA~09 and GAMA~15) fields and the
South Galactic Pole (SGP) field.  Our selection criteria are discussed
fully in \citetalias{ivison16}, which we briefly outline here.

We imposed color cuts of $S_{500}/S_{250}\ge1.5$ and
$S_{500}/S_ {350}\ge0.85$ in order to select rare, distant galaxies. We
increased the reliability of our ultra-red galaxy sample by imposing a
$500\textrm{-}\micron{}$ significance of $\ge3.5\sigma_{500}$, and by
requiring flux densities consistent with a high redshift in ground-based
snapshot images obtained at 850 or $870\,\micron{}$.

Additionally, we required that $S_{500}\lesssim100\,\millijanksy{}$ in order
to reduce the fraction of gravitationally lensed galaxies in favour of
intrinsically luminous galaxies \citep {negrello10, conley11}, though we
draw attention to SGP-28124, with a flux density
$S_{500}\approx120\,\millijanksy{}$, which is significantly higher than its
cataloged flux density at the time of our observations.

To this {\it H}-ATLAS sample, we added an additional ten targets from five
fields in the {\it Her}MES \citep[\textit {Herschel} Multi-tiered
Extragalactic Survey ---][]{oliver12} imaging survey -- ultra-red galaxies
selected in the \textit{Akari Deep Field}-South (ADF-S), the {\it
  Chandra} Deep Field-South Survey (CDFS), the European Large-Area Infrared
Survey-South~1 (ELAIS-S1) and the {\it XMM/Newton}-Large-Scale Survery
fields are contained in the \textsc{dr4.0} xID250 catalogs by
\citet{roseboom10, roseboom12}, whilst those selected from the {\it Her}MES
Large Mode Survey (HeLMS) are amongst the 477 red galaxies presented by
\citet{asboth16}. All {\it Her}MES images and catalogs were accessed
through the \textit{Herschel} Database in Marseille \citep[HeDaM
---][]{roehlly11}\footnote {\url{http://hedam.oamp.fr/hermes/}.}.

\subsection{Observing strategy}
\label{sec:observing strategy}

\begin{table*}
 \begin{center}
    \caption{Targets and their properties.}
    \label{tab: properties}
    \begin{scriptsize}
        \begin{tabular}{l l r c c c c c c c}
            \hline\hline
            \multicolumn{1}{l}{Nickname} &
            \multicolumn{2}{c}{$\alpha$ \quad (J2000) \quad $\delta$} &
            \multicolumn{1}{c}{$t_{\textrm{int}}$} &
            \multicolumn{1}{c}{$\overline{\tau}$} &
            \multicolumn{1}{c}{$\overline{\sigma}^{\dagger}$} &
            \multicolumn{1}{c}{$\Omega^{\ddagger}$} &
            \multicolumn{1}{c}{Date observed} &
            \multicolumn{1}{c}{Program}\\
            \multicolumn{1}{l}{} &
            \multicolumn{1}{c}{$^{\textrm{h}}$ \, $^{\textrm{m}}$ \, $^{\textrm{s}}$} &
            \multicolumn{1}{c}{$\, ^{\circ} \quad ' \quad ''$} &
            \multicolumn{1}{c}{hr} &
            \multicolumn{1}{c}{} &
            \multicolumn{1}{c}{$\millijanksy{}\,\beam{}^{-1}$} &
            \multicolumn{1}{c}{$\textrm{arcmin}^2$} &
            \multicolumn{1}{c}{yyyy--mm} &
            \multicolumn{1}{c}{} \\
            \hline
            SGP-28124 & 00:01:24.73 & $-$35:42:13.7 & 13.4 & 0.3 & 1.9 & 133 & 2013--04 & E-191.A-0748 \\
            HeLMS-42 & 00:03:04.39 & $+$02:40:49.8 & 0.8 & 0.3 & 6.3 & 121 & 2013--10 & M-092.F \\
            SGP-93302 & 00:06:24.26 & $-$32:30:21.4 & 16.6 & 0.3 & 1.7 & 129 & 2013--04 & E-191.A-0748 \\
            ELAIS-S1-18 & 00:28:51.23 & $-$43:13:51.5 & 0.9 & 0.2 & 5.3 & 117 & 2013--04 & M-091.F \\
            ELAIS-S1-26 & 00:33:52.52 & $-$45:20:11.9 & 4.4 & 0.4 & 4.0 & 118 & 2014--04 & M-093.F \\
            SGP-208073 & 00:35:33.82 & $-$28:03:03.2 & 4.9 & 0.3 & 3.2 & 130 & 2013--04 & M-091.F, E-191.A-0748, M-092.F \\
            ELAIS-S1-29 & 00:37:56.76 & $-$42:15:20.5 & 2.9 & 0.3 & 4.2 & 137 & 2013--10 & M-092.F, M-093.F \\
            SGP-354388 & 00:42:23.23 & $-$33:43:41.8 & 11.4 & 0.3 & 1.8 & 124 & 2013--10 & M-092.F, E-191.A-0748 \\
            SGP-380990 & 00:46:14.80 & $-$32:18:26.5 & 4.0 & 0.3 & 2.9 & 115 & 2012--11 & M-090.F \\
            HeLMS-10 & 00:52:58.61 & $+$06:13:19.7 & 0.5 & 0.3 & 8.0 & 114 & 2013--10 & M-092.F \\
            SGP-221606 & 01:19:18.98 & $-$29:45:14.4 & 1.3 & 0.4 & 6.0 & 112 & 2014--05 & M-093.F \\
            SGP-146631 & 01:32:04.35 & $-$31:12:34.6 & 2.4 & 0.3 & 5.0 & 119 & 2014--04 & M-093.F \\
            SGP-278539 & 01:42:09.08 & $-$32:34:23.0 & 3.2 & 0.4 & 4.4 & 121 & 2014--04 & M-093.F \\
            SGP-142679 & 01:44:56.46 & $-$28:41:38.3 & 3.0 & 0.4 & 4.3 & 116 & 2014--04 & M-093.F \\
            XMM-LSS-15 & 02:17:43.86 & $-$03:09:11.2 & 2.0 & 0.3 & 4.4 & 118 & 2013--10 & M-092.F \\
            XMM-LSS-30 & 02:26:56.52 & $-$03:27:05.0 & 4.1 & 0.3 & 3.4 & 132 & 2013--09 & E-191.A-0748, M-090.F, M-092.F \\
            CDFS-13 & 03:37:00.91 & $-$29:21:43.6 & 1.0 & 0.2 & 5.3 & 118 & 2013--10 & M-092.F \\
            ADF-S-27 & 04:36:56.47 & $-$54:38:14.6 & 3.4 & 0.3 & 3.7 & 135 & 2012--09 & M-090.F \\
            ADF-S-32 & 04:44:10.30 & $-$53:49:31.4 & 2.0 & 0.3 & 5.0 & 129 & 2013--04 & M-091.F, M-092.F \\
            G09-83808 & 09:00:45.41 & $+$00:41:26.0 & 9.2 & 0.3 & 1.8 & 125 & 2013--10 & E-191.A-0748 \\
            G15-82684 & 14:50:12.91 & $+$01:48:15.0 & 6.7 & 0.3 & 2.3 & 116 & 2014--03 & M-093.F \\
            SGP-433089 & 22:27:36.98 & $-$33:38:33.9 & 13.2 & 0.3 & 1.8 & 117 & 2012--09 & M-090.F, M-091.F, M-093.F \\
            \hline
        \end{tabular}
    \end{scriptsize}
    \end{center}
    \vspace{-\baselineskip}
    \noindent
    $^{\dagger}$ Average depth computed across each beam-smoothed LABOCA map,
    where the resulting FWHM of a beam is $27''$.

    \noindent
    $^{\ddagger}$ Extent of LABOCA map.

    \noindent
    \textbf{Note.} Targets are listed in order of increasing right ascension.
\end{table*}

Our sample of 22 ultra-red galaxies were imaged with the Atacama Pathfinder
Experiment (APEX) telescope's Large APEX BOlometer CAmera \citep[LABOCA
---][]{kreysa03, siringo09} instrument over six observing runs from 2012
September to 2014 March\footnote{ESO program E-191.A-0748 and MPI programs
  M-090.F-0025-2012, M-091.F-0021-2013 and M-092.F-0015-2013.}.
The passband response for this instrument is centered on $870\,\mu\textrm{m}$
($345\,\textrm{GHz}$) and has a half-transmission width of
$\sim150\,\mu\textrm{m}$ ($\sim60\,\textrm{GHz}$).

Targets were observed in a compact-raster scanning mode, whereby the
telescope scans in an Archimedean spiral for
$35\,\textrm{sec}$ at four equally spaced raster positions
in a $27''\times27''$ grid. Each scan was approximately
$\approx7\,\textrm{min}$ long such that each raster
position was visited three times, leading to a fully sampled map over the
full $11'$-diameter field of view of LABOCA. An average time of
$t_{\textrm {int}}\approx4.6\,\textrm{hr}$ was spent integrating on each
target. Maps with longer integration times
($t_{\textrm{int}}\gtrsim10\,\textrm{hr}$) provide deeper data sensitive to
less luminous DSFGs in the vicinity of our signposts. Our shallower maps
($t_{\textrm {int}}\lesssim1\,\textrm{hr}$) help constrain the abundances
of the brightest DSFGs, thus reducing the Poisson noise associated with these
rare galaxies.
These deep/shallow $870\textrm{-}\micron{}$ data are necessary to constrain
the photometric redshifts of the brighter/fainter DSFGs within the vicinities
of our signposts, therefore allowing us to identify members of any
candidate proto-cluster found.

During our observations, we recorded typical precipitable water vapor (PWV)
values between $0.4\text{--}1.3\,\textrm{mm}$, corresponding to a zenith
atmospheric opacity of $\tau=0.2\text{--}0.4$. Finally, the flux density
scale was determined to an r.m.s.\ accuracy of
$\sigma_{\textrm{calib}}\approx7\%$ using observations of primary
calibrators, Uranus and Neptune, whilst pointing was checked every hour
using nearby quasars and found to be stable to
$\sigma_{ \textrm{point}}\approx3''$ (r.m.s.).

\subsection{From raw timestreams to maps}
\label{sec:data reduction}

The data were reduced using the Python-based BOlometer data Analysis
Software package \citep[BOA \textsc{v4.1} ---][]{schuller12}, following the
prescription outlined in \S10.2 and \S3.1 of \citet{siringo09} and
\citet{schuller09}, respectively. We briefly outline the reduction steps
below.

\begin{itemize}
    \renewcommand\labelitemi{--}
  \item Timestreams for each scan were calibrated onto the
    $\textrm {Jy}\,\beam{}^{-1}$ scale using primary or secondary
    flux density calibrators.

  \item Channels exhibiting strong cross talk with their neighbors, showing no signal or high noise were flagged, whilst the remaining channels were flatfielded.

  \item Timestreams were flagged in regions where the speed and acceleration
    of the telescope are too severe to guarantee reliable positional
    information at every timestamp.

  \item In an iterative manner, the following sequence was performed:

    \begin{enumerate}
        \item Noisy channels were $n\sigma$-clipped relative to all channels,
        where $n=5\text{--}3$ with each iteration.
        \item Sky noise determined across all channels was removed from each
        channel.
        \item Each channel's timestreams were `despiked' about their mean value.
        \item An $n^{\textrm{th}}$-order polynomial baseline was subtracted
        from the timestreams to remove any low frequency drifts, where
        $n=1\text{--}4$ with each iteration.
    \end{enumerate}

  \item Large discontinuities (jumps) in the timestreams, seen by all
    channels, and correlated noise between groups of channels (e.g.\
    channels sharing the same part of the electronics or being connected to
    the same cable) were removed.

  \item The Fourier spectrum of the timestreams were high-pass filtered
    below $0.5\,\textrm{Hz}$ using a noise-whitening algorithm to remove
    the $1/f$ noise. At this stage, the mean noise-weighted point-source
    sensitivity of all channels was calculated to remove scans corrupted by
    electronic interference. Uncorrupted scans were opacity-corrected using
    skydips and radiometer opacity values before being pixelated onto a
    map.  We over-sampled the pixelization process by a factor of four to
    preserve the spatial information in the map. This results in a final
    map for a given scan with a pixel scale,
    $p\approx4.8''\,\textrm{pix}^{-1}$.
\end{itemize}

\begin{figure}
    \includegraphics[width=0.475\textwidth]{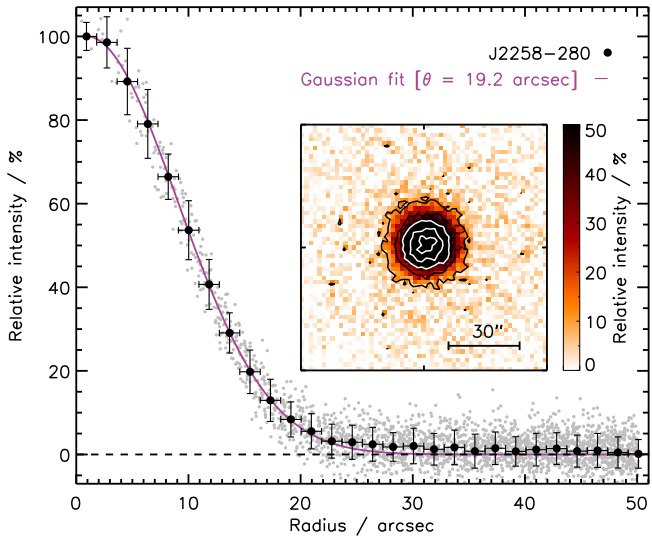}
    \caption{\textit{Main}: radially-averaged beam profile of J2258$-$280,
      the most frequently visited pointing source for this work, reduced in
      the same manner as our maps. Black points indicate radial bin averages and their
      respective r.m.s. values, after sky subtraction.
      The beam is well described by a Gaussian with
      full-width-at-half-maximum $\theta=19.2''$ (purple line), which
      we use to beam-smooth our final maps. \textit{Inset}: normalized flux
      map of J2258$-$280 ($S_{\nu}=765.4\pm 26.2\,\millijanksy{}$) with contours
      indicating the 10, 30 (black), 50, 70 and $99\,\textrm{(white)}\%$
      peak flux levels.}
    \label{fig: beam profile}
\end{figure}

We coadded, with inverse weighting, all of the reduced maps for each scan
before beam smoothing the final map to remove any high-frequency noise on
scales smaller than the beam. The effect of convolving with a Gaussian with
full-width-at-half-maximum $\theta=19.2''$ (i.e.\ the beam width, see
Fig.~\ref {fig: beam profile}) degrades the spatial resolution to
$\theta\approx27''$. Thus we appropriately scale the final map in order to preserve
the peak intensity.


We repeated these reduction steps, this time using the final reduced map as
a model to mask significant sources before flagging the timestreams.  Using
a model in this fashion helps to increase the final signal-to-noise ratio
($S/N$) of detections \citep{schuller09, nord09, belloche11}. We find that
one repetition is sufficient to achieve convergence in the $S/N$ of a point
source, in agreement with the findings of \citet{weiss09b} and
\citet{gomez10}. We present the final $S/N$ maps for all of our ultra-red
galaxies in Appendix~\ref {sec:maps}.

To model the instrumental noise of our maps, we generated so-called
`jackknife' maps by randomly inverting (i.e.\ multiplying by $-1$) half of
our reduced scans before coadding them. The result is a map free of
astronomical sources and confusion, which we estimate to be $\approx0.9\,\millijanksy{}$
in our deepest maps, and thus these realizations will underestimate the true noise.
For each map, we created 100 jackknife realizations of the instrumental
noise.

\begin{figure*}
    \centering
    \includegraphics[width=0.4725\textwidth]{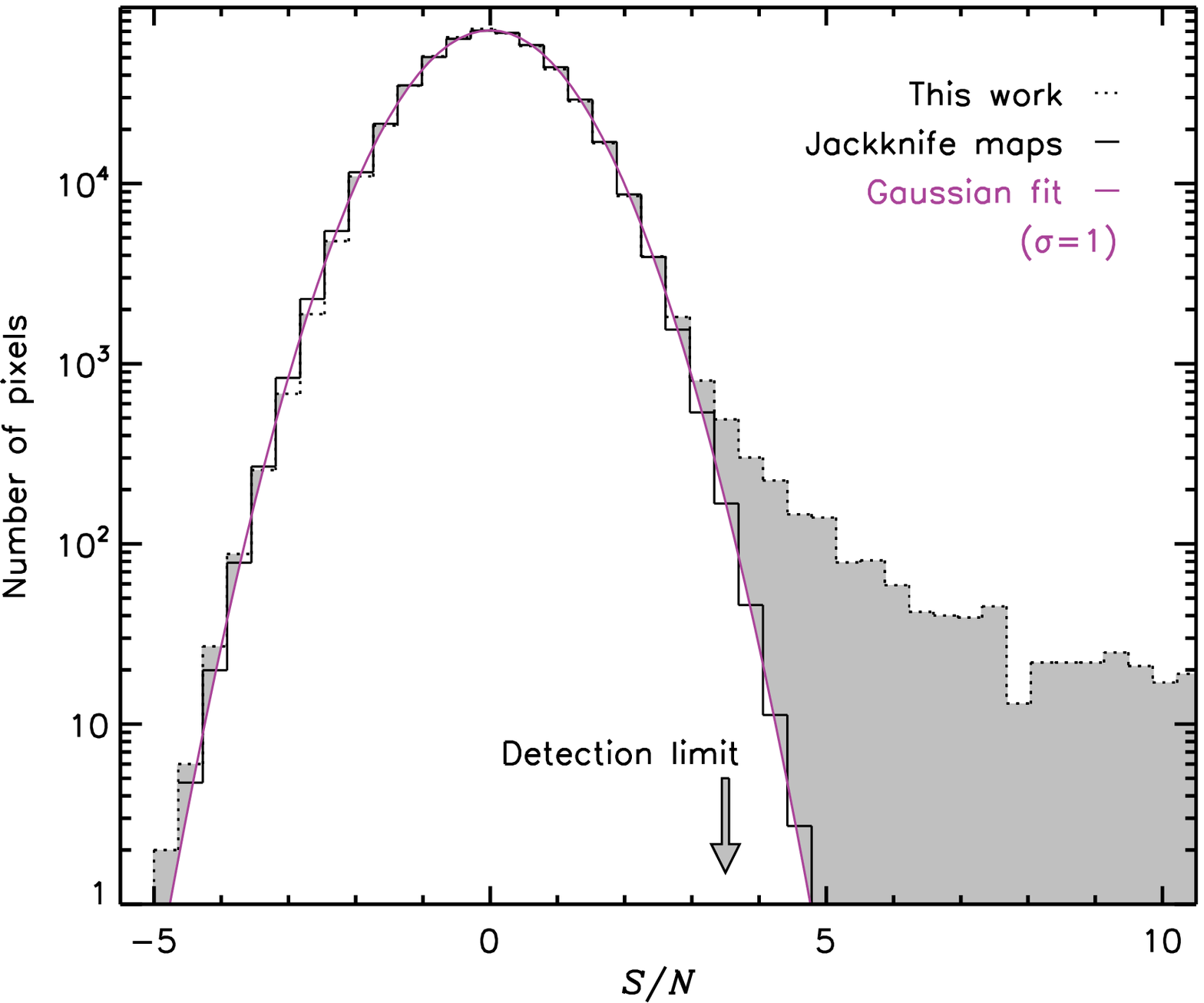}
    \hspace{0.0166667\textwidth}
    \includegraphics[width=0.4775\textwidth]{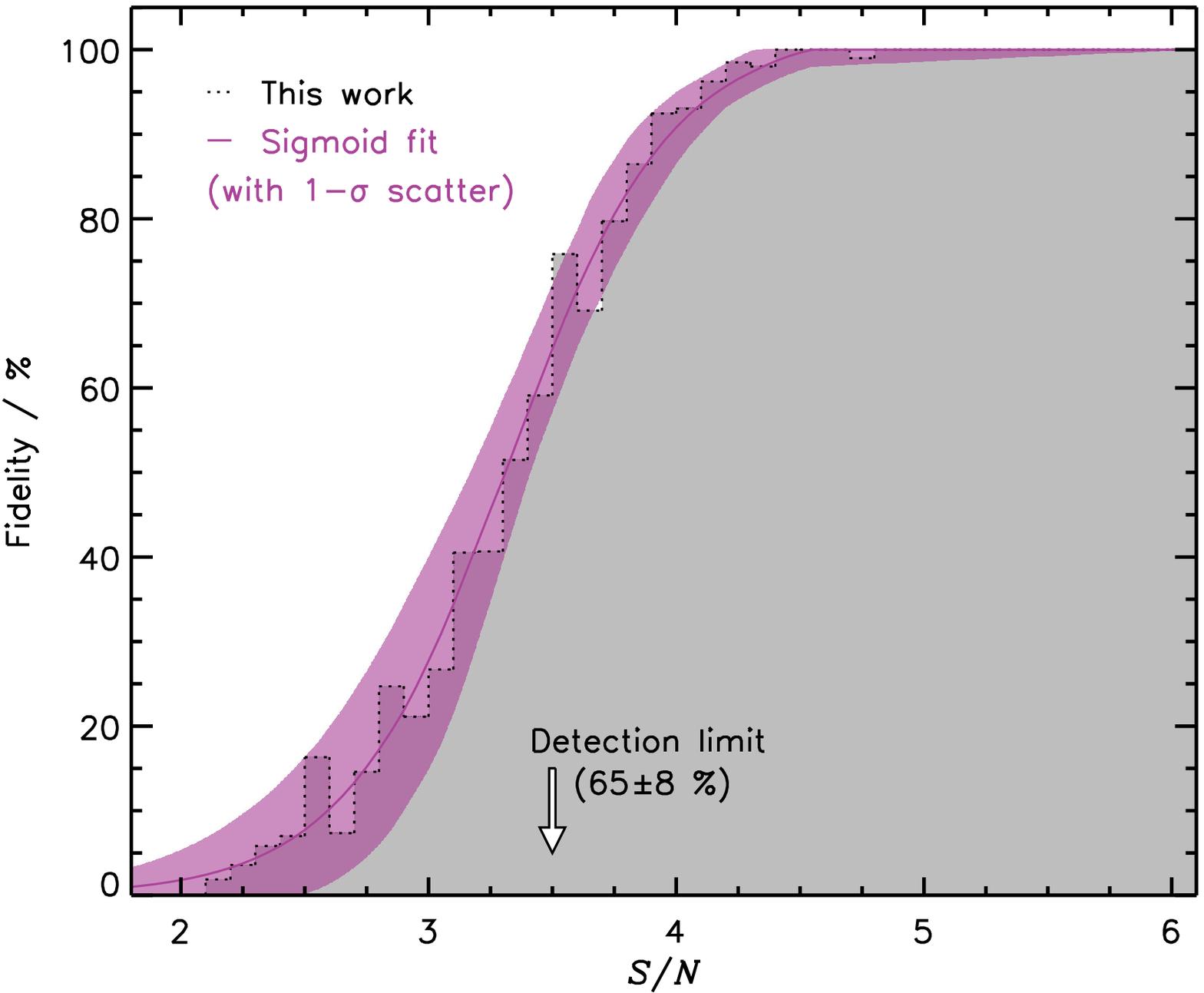}
    \caption{%
      \textit{Left}: beam-smoothed $S/N$ pixel distribution for our maps
      (dotted, black histogram) which shows an excess above our detection
      threshold due to the presence of astrophysical sources (gray
      region). We also plot the beam-smoothed $S/N$ pixel distribution of
      our jackknife maps (black solid histogram, see \S\ref{sec:data
        reduction}), whose mean is well modeled by a Gaussian (solid,
      purple line) centered on $\mu=0$ with a standard deviation
      $\sigma=1$, as expected.  \textit{Right}: mean fidelity (black, solid
      histogram --- $\mathcal{F}$) as a function of detection $S/N$ for our
      maps using our extraction algorithm (see \S\ref{sec:source
        extraction}). We parameterize the histogram by a sigmoid function
      (purple, solid line), which we use to deduce the fidelity of each
      source detected. We draw attention to the fact that this is a
      statistical measurement and that on average $65\pm8\%$ of sources
      detected at $3.5\sigma$ will be trustworthy, i.e.\ a third of these
      sources may be spurious.}
    \label{fig: snr pixel distribution}
\end{figure*}

In Fig.~\ref{fig: snr pixel distribution}, we show the pixel distributions
of the final $S/N$ maps and their respective jackknife realizations. There is
clearly a positive excess above $S/N\gtrsim3$ in the final reduced maps compared
to the jackknife maps. This excess is caused by the presence of astronomical sources.

\section{Analysis}
\label{sec:analysis}

We chose a detection threshold ($\Sigma_{\textrm{thresh}}$) based on the
values of a `fidelity' or `trustworthiness' parameter, $\mathcal{F}$,
similar to that outlined in \citet{aravena16}. For all of our maps, we ran
our extraction algorithm (\S\ref{sec:source extraction}) and compared the
number of sources detected in our maps, $\mathcal{N}$, to the mean number
of sources detected in our 100 jackknife realizations for each map,
$\overline {\mathcal{N}}_ {\textrm{jack}}$, as a function of detection
$S/N$:

\begin{equation}
    \label{eq: ensemble fidelity}
    \mathcal{F} = 1 -
    \frac{\overline{\mathcal{N}}_{\textrm{jack}}}{\mathcal{N}} .
\end{equation}

We show the average fidelity in the right-hand panel of Fig.~\ref{fig: snr
  pixel distribution} which illustrates that by increasing the detection
$S/N$ we increase our confidence in the recovered sources. We reach a
fidelity of $\mathcal{F}\approx100\%$ at $\gtrsim5\sigma$ and a fidelity of
$\mathcal {F}=50\%$ at $\approx3\sigma$, the latter indicating that we
would expect about half of our sources to be spurious at $S/N\approx 3$. We
chose -- as a compromise between reliability and the number of cataloged
sources -- a detection threshold of $\Sigma_{\textrm{thresh}}>3.5$, where
we have a fidelity, $\mathcal{F}\approx65\pm8\%$.

The intrinsic map-to-map scatter in the fidelity is caused by the varying
abundance of sources in each map, due to the effects of cosmic variance and
the differing r.m.s.\ noise levels. This scatter decreases with increasing
detection threshold and is $\sigma_{\mathcal{F}}\approx\pm3\%$ at
$5\sigma$.

\subsection{Source extraction}
\label{sec:source extraction}

We used a custom-written Interactive Data Language \citep[\textsc{idl} ---][]
{landsman93} source extraction algorithm to identify and extract sources in the
beam-smoothed $S/N$ maps, noting that the beam-smoothing step described above
optimizes the detection of point sources.

In a top-down fashion, we searched for pixels above\footnote{To accommodate
  sources whose true peak falls between pixels we temporarily lowered
  $\Sigma_{\textrm{thresh}}$ by $\approx95\%$, keeping sources with
  bicubically interpolated sub-pixel values that meet our original $S/N$
  detection threshold.}  our floor $S/N$ detection threshold
$\Sigma_{\textrm{thresh}}>3.5\sigma$. In Table~\ref{tab: photometry}, we
catalog the peak flux density, noise and position determined from a
three-parameter Gaussian fit made inside a box of width $\theta$ (i.e.\ $\approx27''$)
centered
on a source. After removing the fit from the map, we searched for and cataloged
subsequent peaks until no more could be found.

During the extraction process we performed some additional steps: sources
deemed too close to each other ($\Delta r<\theta/2$) have their parameters
re-evaluated, fitting multiple three-parameter Gaussians simultaneously;
sources deemed too close ($\Delta r<\theta/2$) to the map edges were rejected.

\subsubsection{Completeness, flux boosting and positional offsets}
\label{sec:simulations}

We inserted simulated sources into our jackknife maps to quantify the
statistical properties of our cataloged sources. To ensure that we did not
encode any clustering, we randomize the injection sites of our simulated
sources. We drew model fluxes densities down to
$S_{\textrm{mod}}=1\,\millijanksy{}$ from a Schechter function
parameterization of the number counts

\begin{equation}
    \label{eq: schechter function}
    \frac{dN}{dS_{\textrm{mod}}} \propto \left(\frac{S_{\textrm{mod}}}{S_{0}}\right)^
    {-\alpha}e^{-S_{\textrm{mod}}/S_{0}},
\end{equation}

\noindent
where $S_{0}=3.7\,\millijanksy{}$ and $\alpha=1.4$ \citep{casey13}, which we
scale to $870\,\mu\textrm{m}$ using a spectral index of $\nu^2$, i.e.\ we divide
the model fluxes by $(\nu_{870}/\nu_{850})^2\approx1.05$.

For each simulated source, we ran our source-extraction algorithm and if we
detected a peak within a threshold radius,
$r_{\textrm{thresh}}\le1.5\times\theta$, of the injection site then we recorded
the best-fitting Gaussian parameters. If we recovered multiple peaks within
our threshold radius\footnote{We note that due to the Gaussian nature of
  our jackknife maps, we expect $5\pm 2$ peaks at $>3.5\sigma$ in each
  $130\,\textrm{arcmin}^2$ map.}  we took the most significant.  Finally,
if we failed to recover a simulated source, we recorded the model flux
density and the instrumental noise at the injection site.

This procedure was repeated $10,000$ times for each target so that
we generated a large, realistic catalog of simulated sources. We used this
to determine the noise-dependent completeness, $\mathcal{C}$, i.e.\ the
fraction of recovered sources to input sources, as well as the flux
boosting, $\mathcal{B}$, i.e.\ the ratio of recovered to input flux
densities, and the radial offsets, $\mathcal{R}$, i.e.\ the distance between
recovered and input positions for each cataloged source.

\begin{figure}
    \centering
    \includegraphics[width=0.475\textwidth]{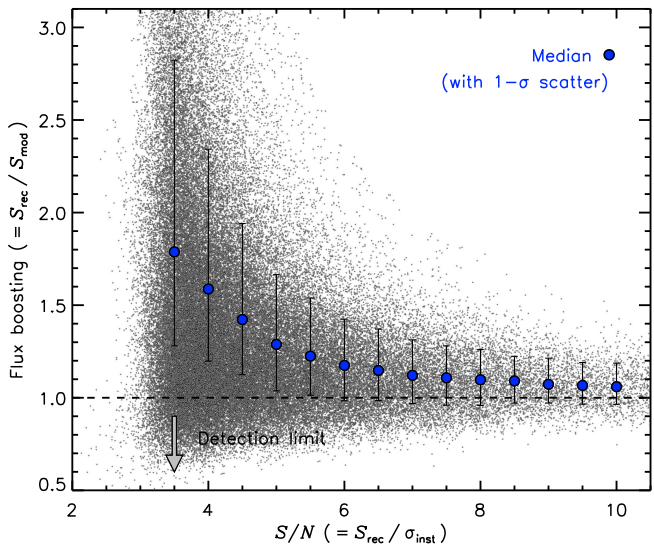}
    \caption{Flux boosting (i.e.\ recovered versus modeled flux density) as
      a function of recovered $S/N$ for SGP-93302. We generate a model flux
      density distribution using the Schechter parameterization of the
      number counts given in \citet {casey13} when determining these
      corrections. We record a negligible flux boosting factor,
      $\mathcal{B}<1.1$, at $S/N\gtrsim6.0$ and witness corrections of
      $\mathcal{B}\approx1.7$ at our detection threshold, comparable to
      that of S2CLS \citep[$\mathcal{B}\approx1.5$ ---][]{geach17}, despite
      the different noise levels.}
    \label{fig: flux boosting}
\end{figure}

\begin{figure}
    \centering
    \includegraphics[width=0.475\textwidth]{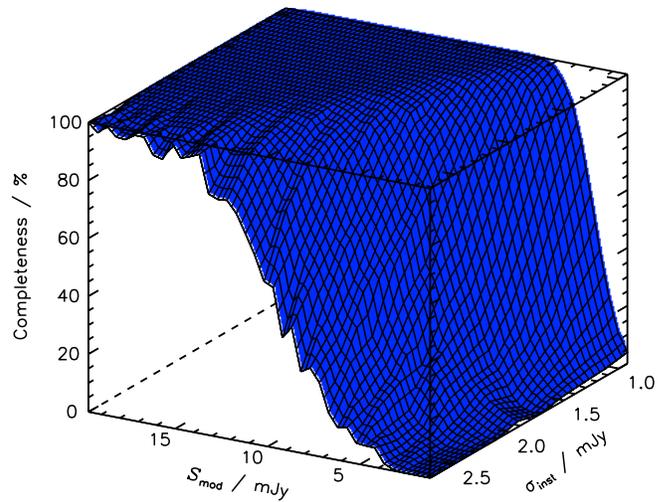}
    \caption{Completeness for SGP-93302 as a function of instrumental noise
      and model flux density. The two-dimensional treatment of our
      completeness is vital due to the radially varying sensitivity across
      our maps. We see that as the instrumental noise decreases and our
      model flux density increases, our completeness increases too. For
      this map, at an instrumental noise and model flux density of
      $\sigma_{\textrm{inst}}\approx1.2\,\millijanksy{}$ and
      $S_{\textrm{mod}}\approx1\,\millijanksy{}$, respectively, we recover
      hardly any sources, i.e.\ $\mathcal{C}\approx0\%$. However,
      increasing the model flux density to $\gtrsim5\,\millijanksy{}$ whilst
      keeping the noise constant results in most sources being recovered successfully,
      i.e.\ $\mathcal{C}\approx100\%$.}
    \label{fig: completeness}
\end{figure}

We calculated the median flux boosting in bins of recovered $S/N$, which we
use to translate the recovered flux densities of our detections into model
flux densities (see Fig.~\ref{fig: flux boosting}). After this stage, we
used our deboosted flux densities with their associated instrumental noise
levels to determine their completenesses and radial offsets. The former, we
compute from a spline interpolation of a two-dimensional surface of modeled
flux density and instrumental noise \citep[see Fig.~\ref{fig: completeness}
and, e.g.,][]{geach13}, whilst the latter we compute from a spline
interpolation of modeled $S/N$ (see left-hand panel of Fig.~\ref{fig:
  radial offset and spire flux correction}).

At our detection threshold, the flux density of a source in our deepest
map, SGP-93302, is typically boosted by $\mathcal{B}=1.7$, which is in
agreement with the literature at similar depths \citep[e.g.\
$\mathcal{B}\approx1.5$ ---][] {geach17} whilst at $S/N\gtrsim6$ the flux
boosting becomes negligible.  However, we draw attention to the relatively
severe deboosting factors recorded for our noisiest maps (e.g.\ central
r.m.s., $\sigma\gtrsim5\,\millijanksy{}$ for SGP-221606) due to the steep
bright end ($S_{\nu}>13\,\millijanksy{}$) slope of the number counts.

For SGP-93302, our two-dimensional completeness function indicates that we
are $\mathcal{C}\approx100\%$ complete at a deboosted flux density and
instrumental noise of $S_{\textrm{mod}}\approx5\,\millijanksy{}$ and
$\sigma_{\textrm{inst}}\approx1.2\,\millijanksy{}$, respectively. In this same flux
density plane, our completeness falls close to $\mathcal{C}\approx0\%$ as
the instrumental noise reaches $\sigma_{\textrm{inst}}\approx2.5\,\millijanksy{}$.

In the left-hand panel of Fig.~\ref{fig: radial offset and spire flux
  correction} we see that the mean radial offset is in good agreement with
that expected from Equation~B22 in \citet{ivison07}. There exists a large
scatter in the low $S/N\lesssim5$ bins, which indicates that our radial
offsets at a given $S/N$ value can vary by as much as
$\sigma_{\mathcal{R}}=\pm2.5''$. We also note that our brightest detections
with $S/N\approx30$ have radial offsets as little as $\mathcal{R}=0.5''$,
allowing us to accurately constrain the positions of such sources.

\subsection{\textit{Herschel} SPIRE photometry}
\label{sec:herschel photometry}

\begin{figure*}
    \centering
    \includegraphics[width=0.475\textwidth]{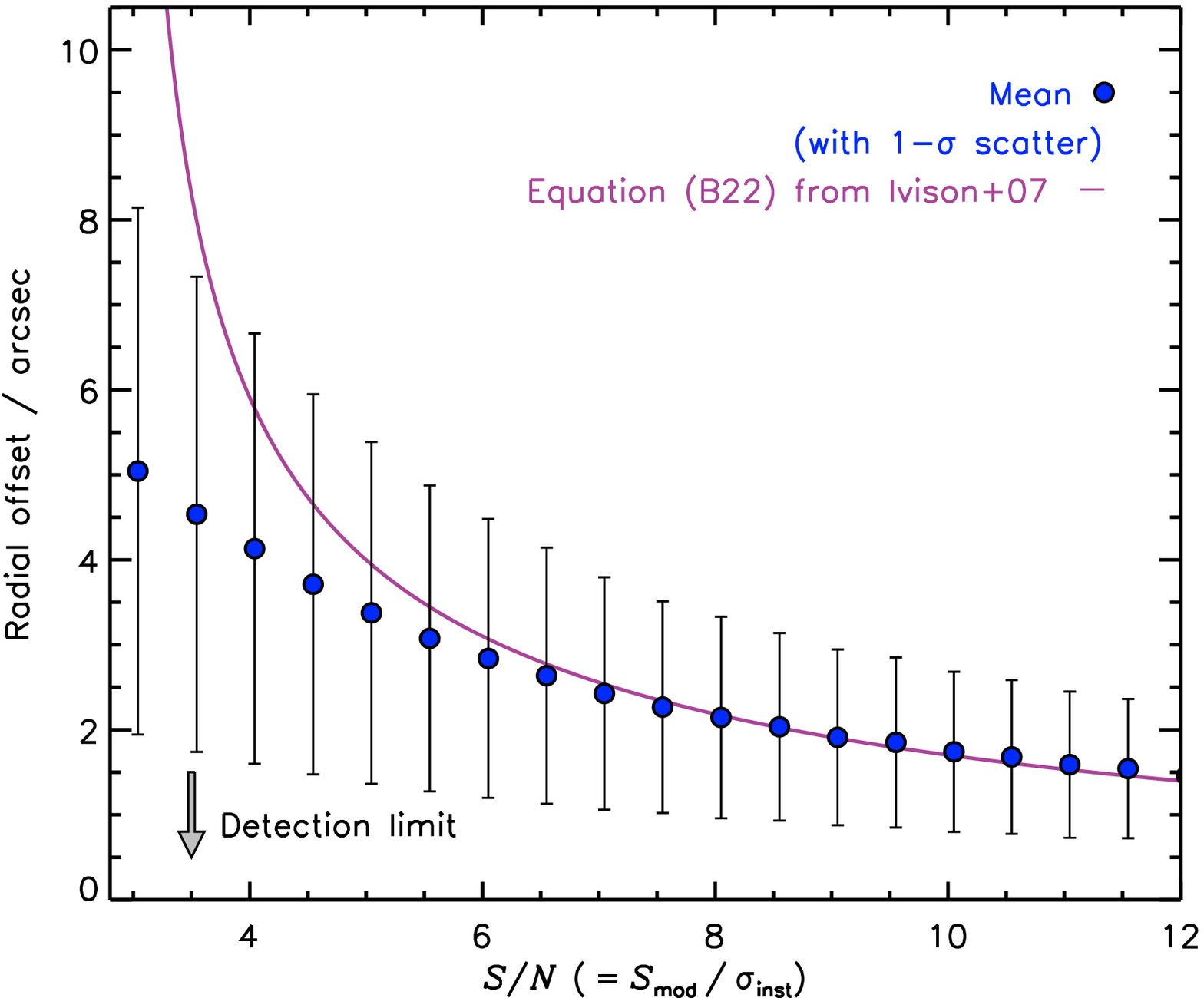}
    \hspace{0.0166667\textwidth}
    \includegraphics[width=0.475\textwidth]{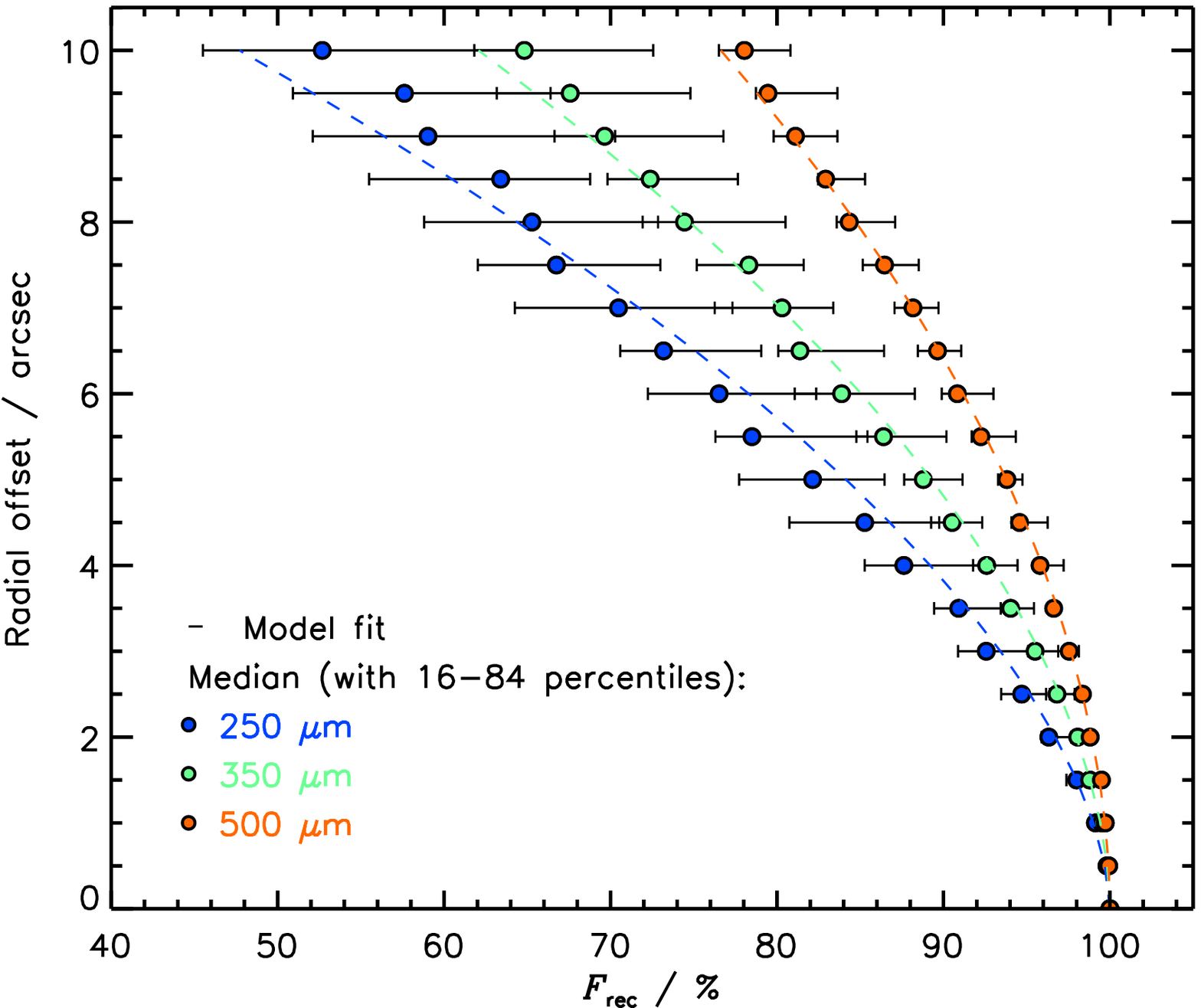}
    \caption{\textit{Left}: radial offset ($\mathcal{R}$, difference
      between the model and recovered source position) as a function of
      modeled $S/N$ for SGP-93302. The $1\sigma$ errors for each bin are
      taken from the r.m.s.\ of the radial offsets in that $S/N$ bin. We
      also plot the predicted form given by Equation~B22 in
      \citet{ivison07} using the number counts of LESS, which is in good
      agreement. \textit{Right}: SPIRE flux boosting to accommodate the
      drop in measured flux density due to the LABOCA radial offset
      (left-hand panel) as deduced from our Monte Carlo simulations. The
      shaded region represents the median with errors from the $15.865^{\textrm{th}}$ and $84.135^{\textrm{th}}$ percentiles from the median across all
      of the SPIRE survey fields. We see that at $\mathcal{R}=4''$,
      roughly equating to a modeled $S/N\approx5$, we recover 85, 92 and
      $97\%$ of the flux density across the 250, 350 and
      $500\,\mu\textrm{m}$ passbands, respectively. This decreasing loss
      of flux represents the increasing optimal pixel sizes due to the
      differing SPIRE beam sizes.}
    \label{fig: radial offset and spire flux correction}
\end{figure*}

In order to derive photometric redshifts for the LABOCA-detected DSFGs,
we bicubically interpolated the SPIRE flux-density maps at the LABOCA
source positions. We determined the errors
and local sky values from a box of width $\approx12\times\theta_{\textrm {SPIRE}}$
centered on each detection, where
$\theta_{\textrm{SPIRE}}\approx18$, $24$ and $35''$ for the 250-, 350- and
500-$\mu\textrm{m}$ passbands, respectively \citep{valiante16}.

To quantify the effect that the LABOCA radial offset has on determining our
SPIRE measurements, we analyzed how the `true flux density' of a source
varied as we tweaked the position at which we measured it. For each survey
field and passband, we selected a bright
($S_{250} \approx S_{350} \approx S_{500} \approx 1\, \textrm{Jy}$) point
source and measured the true flux density at its cataloged position. We
then performed 500 Monte Carlo simulations, drawing radial offset values from a
Gaussian distribution centered on the cataloged position with a standard
deviation\footnote{As $\mathcal{R}$ is defined as the radial distance from
  the injected to the recovered position of a simulated source, we vary
  each the coordinate of each spatial dimension ($\alpha$ and $\delta$) by
  $\mathcal{R}_{\alpha}= \mathcal{R}_ {\delta}=\mathcal{R}/\sqrt{2}$.}
$\sigma=\mathcal{R}$, which we allowed to range across
$\mathcal{R}=0\textrm{--}10''$. For each simulation, we measured the flux
density at the adjusted positions and compared them to the true flux
density. We used this ratio ($F_{\textrm{rec}}$) to flux-boost a SPIRE
photometric measurement, depending on the LABOCA radial offset it exhibited.
We parameterize this value using $F_{\textrm{rec}}\,/\,\% = 100 -
(\mathcal{R}\,/\,'' / \alpha)^{\beta}$, where $\alpha=1.0,1.4,1.9$ and
$\beta=1.7,1.8,1.9$ at 250-, 350- and 500-$\mu\textrm{m}$, respectively.

The right-hand panel of Fig.~\ref{fig: radial offset and spire flux
  correction} shows that the average flux boosting is passband related,
reflecting the different pixel scalings of 6, 8.3 and
$12''\,\textrm{pix}^{-1}$ for the 250-, 350- and 500-$\mu\textrm{m}$
passbands in \textit{H}-ATLAS, respectively (similar values are recorded in \hermes{}).
We see that for detections with low radial
offsets, $\mathcal{R}<2''$, and thus high $S/N\gtrsim8$ values, we recover
$\approx95\%$ of the true flux density. Due to the large
SPIRE $500\textrm{-}\mu\textrm{m}$ pixel size, even at the highest radial
offsets considered in this paper ($\mathcal{R}\approx10''$) for sources near to
or at our detection threshold, we still recover $\gtrsim80\%$ of the true flux density.
Conversely, however, we only recover $\gtrsim55\,\%$ and $\gtrsim65\,\%$ of the true
flux densities for these highest offsets at
$250\textrm{-}$ and $350\textrm{-}\mu\textrm{m}$, respectively.

We draw attention to 16 (i.e.\ $\approx15\,\%$) of our LABOCA sources that
are undetected at the $1\textrm{-}\sigma$ level in all SPIRE maps. The majority (12)
of these possibly spurious sources have detection $S/N\lesssim4.5$, in
agreement with our fidelity analysis. The number of sources with higher
$S/N$ values is also expected, once the intrinsic scatter in the fidelity
parameter is taken into account. These sources do not affect our number
counts as, on average, we correct for this effect. Thus our fidelity
$\mathcal{F}=65\pm8\%$ and high flux-boosting factors at these low $S/N$
thresholds weights these possibly-spurious sources accordingly. However, we
choose not to include any of these sources in our photometric redshift
analysis -- we are unable to meaningfully constrain them.


Finally, we note that the SPIRE fluxes derived in this manner, i.e.\ using a LABOCA
prior and a radial offset flux-boosting value, are consistent with those from which
they were originally selected - varying by $\pm1\sigma$.

\subsection{Photometric redshifts}
\label{sec:photometric redshifts}

We use a custom-written $\chi^2$-minimization routine in \textsc{idl} to
determine far-IR-based photometric redshifts for our catalog of sources,
which have at least one SPIRE detection above
$>1\sigma_{\textrm{SPIRE}}$. We fit to three well-sampled spectral energy
distributions (SEDs) used in \citetalias{ivison16}:
that of the Cosmic Eyelash \citep{swinbank10,
  ivison10d} and synthesized templates from \citet{pope08} and
\citet[][]{swinbank14}, ALESS\footnote{Fig.~4 in \citetalias{ivison16}
  highlights the diversity of these SEDs in the rest frame, each normalized
  in flux density at $\lambda_{\textrm{rest}}=100\,\mu\textrm{m}$.}.

We use the deboosted $870\textrm{-}\mu\textrm{m}$ and boosted SPIRE flux
densities during our template fitting. The fitting is done in linear space
(accommodating for negative fluxes) over a photometric redshift range
$0<\zphot{}<10$ down to a resolution of $\Delta z=0.01$.
We adopt the photometric redshift associated
with the template that produces the overall minimum $\chi^2$ value
($\chi^2_{\textrm{min}}$) and report
$1\textrm{-}\sigma$ errors based on the $\chi^2_{\textrm{min}}+1$ values. We find
that the errors determined in this way are consistent with the Monte
Carlo method used by \citetalias{ivison16}. However, they are inconsistent
with the intrinsic scatter deduced from a training sample of
spectroscopically-confirmed DSFGs that meet our ultra-red criteria.
In \citetalias{ivison16}, we find that
the accuracy and scatter in
$\Delta z/(1+z_
{\textrm{spec}})=(\zphot{}-\zspec{})/(1+\zspec{})$
are $\mu_{\Delta z}=-0.03(1+\zspec{})$ and
$\sigma_{\Delta z}=0.14(1+z_{ \textrm{spec}})$, respectively. This scatter
is representative of the minimum systematic uncertainty when determining
photometric redshifts using these three templates --- significantly larger
than that determined from both the $\chi^2_{\textrm{min}}+1$ values at
high redshift. 

The results of these fits, as well as the rest-frame,
$8\textrm{--}1000\,\micron{}$ luminosities
are presented in Table~\ref{tab: photometric redshifts}.

\section{Discussion}
\label{sec:discussion}

We catalog 108 DSFGs from our 22 maps above $\Sigma_{\textrm {thresh}}>3.5$
and list their SPIRE and LABOCA flux densities and their mean flux boosting,
$\overline {\mathcal{B}}$, and mean fidelity, $\overline{\mathcal{F}}$,
parameters in Table~\ref{tab: photometry}. Our signpost ultra-red galaxies
span a deboosted flux density range of
$S_{870}=2.9\text{--}42.8\,\millijanksy{}$, with a mean,
$\overline{S}_ {870}=17.0\,\millijanksy{}$. The surrounding field galaxies
span a deboosted flux density range of
$S_{870}=1.9\text{--}31.3\,\millijanksy{}$ with a mean,
$ \overline{S}_{{870}}=6.8\,\millijanksy{}$. There are two exceptionally
bright, new DSFGs, with $S_{870}\gtrsim25\,\millijanksy{}$, but on average
the new field galaxies are less bright than our target ultra-red galaxies.


We are unable to detect four of our target ultra-red galaxies above our
$S/N>3.5$ threshold; all of these are located in our shallower maps. In
such cases, we report the peak flux and r.m.s.\ pixel value within a $45''$
aperture centered on the telescope pointing position. We do not provide
completeness, flux boosting, fidelity or radial offset values for these
sources.

\subsection{Number counts}
\label{sec:number counts}

\begin{figure*}
    \centering
    \includegraphics[width=0.47\textwidth]{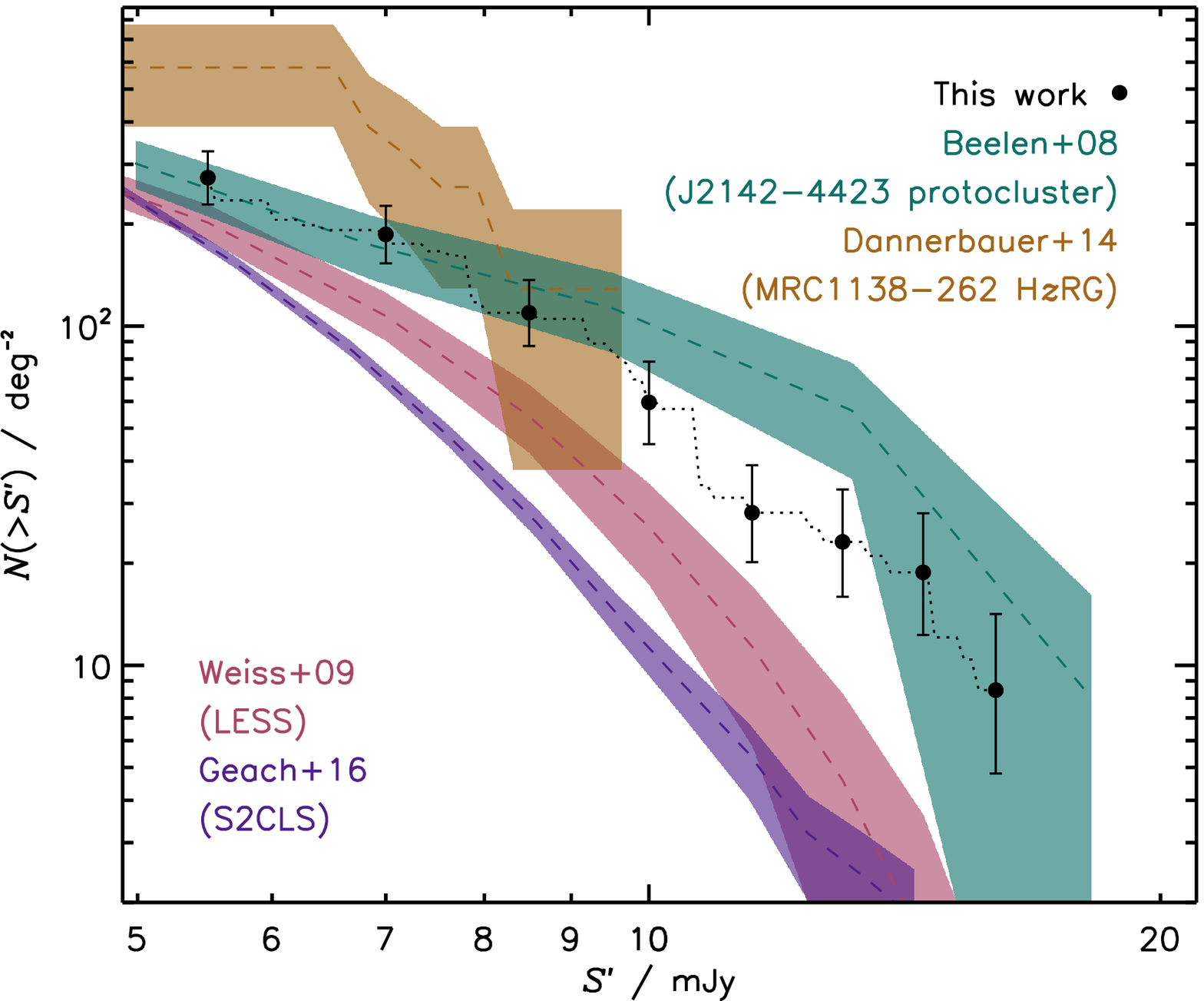}
    \hspace{0.0166667\textwidth}
    \includegraphics[width=0.48\textwidth]{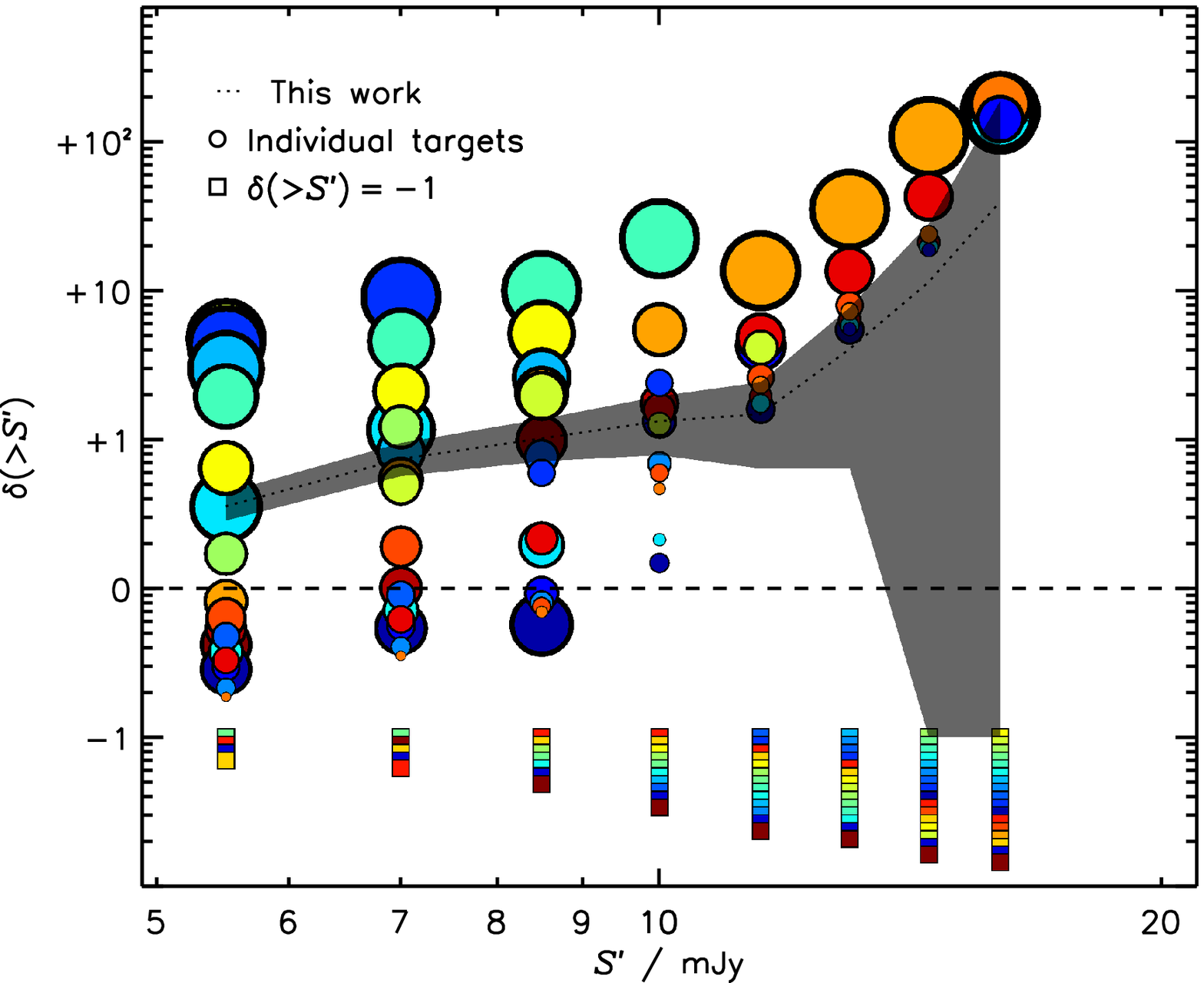}
    \caption{\textit{Left}: number counts (excluding our target ultra-red
      galaxies) as a function of $870\textrm{-}\mu\textrm{m}$ flux density
      (black squares) with $1\sigma$ double-sided Poisson errors
      \citep{gehrels86}. We show the blank-field number counts from LESS
      (pink region) and S2CLS (purple region --- scaled with a spectral
      index of $\nu^2$) surveys. We also show the number counts of two known proto-clusters,
      J2142$-$4423 \citep[green region ---][]{beelen08} and
      MRC\,1138$−$262 \citep[brown region ---][]{dannerbauer14}. It is
      clearly evident that our number counts are high at all flux density
      thresholds and exhibit a slight break at $S'>7\,\millijanksy{}$. We believe
      that the increasing excess at higher flux densities is the result of
      our ultra-red galaxies signposting similarly extreme DSFGs.  Our
      catalog contains five bright ($S_{870}>16\,\millijanksy{}$)
      sources. However, we concede that without high-resolution imaging we
      are unable to rule out gravitational lensing by chance alignment as a
      cause for the bright sources. \textit{Right}: number counts relative
      to LESS, i.e.\ the over-density parameter, $\delta(>S')$. In black we
      show the results for the entire sample (i.e.\ the circles from the left-hand panel),
      whilst in
      colored circles we show the over-density for each map. The size of each
      circle has
      been logarithmically scaled to show the influence that each target
      has in deducing the number counts for the whole sample.
      Maps where
      no sources are present above a given threshold flux are indicated by
      staggered squares starting from $\delta<-1$ for clarity.
      These squares highlight the
      deficit of sources due to intrinsic properties (i.e.\ cosmic
      variance) and varying map r.m.s.\ values. Hence, we see that some
      maps probe considerably more over-dense regions than others, with
      variations being sometimes as high $\approx\times5$. Finally, we color-code
      each target from blue to red in order of increasing right ascension, i.e.\
      in the order that our targets appear in Table~\ref{tab: properties} and the color
      that they have in Fig.~\ref{fig: reduced maps}.
      }
    \label{fig: number counts}
\end{figure*}

We deduce number counts, which we list in Table~\ref{tab: number counts}
and display in the left-hand panel of Fig.~\ref{fig: number counts}, using the
following equation:

\begin{equation}
    \label{eq: number counts}
    N(>S') = \sum_{\forall S_i>S'}
    \frac{\mathcal{F}}{\mathcal{C}\mathcal{A}}  ,
\end{equation}

\noindent
where the sum is over all deboosted flux densities, $S_{i}$, greater than
some threshold flux, $S'$. Fidelity corrections, $\mathcal{F}$, are made
using the detected $S/N$ values, whilst completeness corrections,
$\mathcal{C}$, are made using the deboosted flux densities and instrumental
noises. The area surveyed at a recovered flux density, $\mathcal{A}$, is
obtained by cumulating the area across all of our maps where a given flux
density is detected above our threshold. These three corrections account
for the varying map r.m.s.\ values in our sample.

We exclude the target ultra-red galaxies, to partially\footnote{%
 We note that this method does not fully remove all of the bias associated
 with imaging a region centered on a galaxy. This is due to the fact that
 galaxies themselves are known to cluster \citep{greve04,weiss09b}.
 Thus, these `galaxy-centric' regions will be,
 by definition, over-dense relative to arbitrarily selected regions.}
remove the bias associated with imaging a region where a galaxy is
already known to reside.

The errors on the number counts are deduced using

\begin{equation}
    \label{eq: number count errors}
    \sigma_{N(>S')} = N(>S')\frac{\sigma_{\textrm{G86}}}{\mathcal{N}(>S')} ,
\end{equation}

\noindent
where $\sigma_{\textrm{G86}}$ are the double-sided $1\textrm{-}\sigma$ Poisson errors
\citep{gehrels86} and $\mathcal{N}(>S')$ are the number of sources above
each threshold flux density.

Due to the large flux density uncertainties in some of the cataloged DSFGs, we
compare the method outlined above to drawing realizations of the flux densities
and computing Equation~\ref{eq: number counts} for each realization, adjusting
$\mathcal{B}$, $\mathcal{F}$ and $\mathcal{C}$ accordingly. We then take the
median and errors from the $15.865^{\textrm{th}}$ and $84.135^{\textrm{th}}$
percentiles from the median. We find no significant variation in the results
obtained from either method, which suggests that the large flux density
uncertainties are not severely affecting our number counts analysis.

Finally, we note that we recover the Schechter source counts
given in Equation~\ref{eq: schechter function} to within
$1\sigma$ using Equation~\ref{eq: number counts} on our simulated source maps
described in \S\ref{sec:simulations}.

\begin{table}
    \begin{center}
    \caption{Number counts and over-densities.}
    \label{tab: number counts}
    \begin{scriptsize}
        \begin{tabular}{l c c c c c c}
            \hline\hline
            \multicolumn{1}{l}{$S'^{\dagger}$} &
            \multicolumn{1}{c}{$N(>S')$} &
            \multicolumn{1}{c}{$\mathcal{N}(>S')^{\ddagger}$} &
            \multicolumn{1}{c}{$\delta(>S')$} &
            \multicolumn{1}{c}{$\overline{\mathcal{C}}$} &
            \multicolumn{1}{c}{$\overline{\mathcal{B}}$} &
            \multicolumn{1}{c}{$\overline{\mathcal{F}}$} \\
            \multicolumn{1}{l}{$\millijanksy{}$} &
            \multicolumn{1}{c}{$\textrm{deg}^{-2}$} &
            \multicolumn{1}{c}{} &
            \multicolumn{1}{c}{} &
            \multicolumn{1}{c}{} &
            \multicolumn{1}{c}{} &
            \multicolumn{1}{c}{} \\
            \hline
            5.5 & $273.9^{+53.7}_{-45.4}$ & $36^{+7}_{-5}$ & $+0.4^{+0.1}_{-0.1}$ & 0.68 & 1.54 & 0.98\\
            7.0 & $186.4^{+39.9}_{-33.3}$ & $31^{+6}_{-5}$ & $+0.7^{+0.2}_{-0.2}$ & 0.70 & 1.49 & 0.98\\
            8.5 & $109.5^{+27.2}_{-22.2}$ & $24^{+5}_{-4}$ & $+1.0^{+0.3}_{-0.3}$ & 0.74 & 1.45 & 0.99\\
            10.0 & $59.6^{+18.9}_{-14.8}$ & $16^{+5}_{-3}$ & $+1.3^{+0.6}_{-0.5}$ & 0.81 & 1.42 & 1.00\\
            11.5 & $28.2^{+10.7}_{-8.0}$ & $12^{+4}_{-3}$ & $+1.5^{+0.9}_{-0.8}$ & 0.88 & 1.25 & 1.00\\
            13.0 & $23.1^{+9.9}_{-7.2}$ & $10^{+4}_{-3}$ & $+4.0^{+3.6}_{-3.4}$ & 0.88 & 1.26 & 1.00\\
            14.5 & $18.8^{+9.3}_{-6.5}$ & $8^{+3}_{-2}$ & $+11.4^{+16.5}_{-16.0}$ & 0.87 & 1.26 & 1.00\\
            16.0 & $8.4^{+5.7}_{-3.6}$ & $5^{+3}_{-2}$ & $+39.2^{+146.3}_{-144.8}$ & 0.98 & 1.13 & 1.00\\
            \hline
        \end{tabular}
    \end{scriptsize}
    \end{center}
    \vspace{-\baselineskip}
    \noindent
    $^{\dagger}$ Flux density threshold levels are taken from \citet{weiss09b}
    to simplify the comparisons we made with LESS.

    \noindent
    $^{\ddagger}$ Represents the raw number of galaxy detected above a given flux density threshold.
\end{table}

Our number counts
are always $\gtrsim1\sigma$ above those from the LABOCA Extended Chandra
Deep Field South (ECDFS) Submillimetre Survey \citep[LESS ---][]{weiss09b}
and the \scuba{} Cosmology Legacy Survey \citep[S2CLS
---][]{geach17}. We see a slight break in the shape of counts at
$S'>7\,\millijanksy{}$, similar to that seen in the LESS number counts.

Fig.~\ref{fig: number counts} shows that there are similarities in the shape of our
number counts to those of
J2142$-$4423, a $\textrm{Ly-}\alpha$ proto-cluster \citep{beelen08}, at
$S'\leq7\, \millijanksy{}$ and $S'\geq14\,\millijanksy{}$. However, it is
unclear whether \citet{beelen08} removed the target source from their
number counts which, as mentioned earlier, will bias their results
higher. Futhermore, \citet{beelen08} claim that the environments around
J2142$-$4423 are only moderately over-dense compared to SHADES - but, as
discussed previously, we beleive this be evidence that J2142$-$4423 \emph{is}
over-dense compared to LESS and S2CLS.
Fig.~\ref{fig: number counts} also shows the number counts of
MRC\,1138$−$262 \citep[the so-called `Spiderweb galaxy' ---][]{miley06,
  dannerbauer14}, a HzRG with an over-density of sources compared to LESS at
$S'>7\,\millijanksy{}$ (i.e.\ $\approx385\,\textrm{deg}^{-2}$). This proto-cluster
is $\approx2\times$ more over-dense compared to our work, but it should be
noted that \citet{dannerbauer14} neither account for flux boosting nor
survey completeness, nor do the authors remove the target galaxy (DKB07).
We crudely correct for the first two of these differences
using the results obtained for SGP-93302, which was observed
under similar conditions for a similar integration time to
MRC\,1138$−$262. Adjusting for these corrections, we record less extreme
number counts of $N(>6\,\textrm {mJy})\approx394\pm176\,\textrm{deg}^{-2}$
($1\textrm{-}\sigma$ Poisson errors) that exhibit a sharp break at
$S'\approx6.5\,\millijanksy{}$. Thus, MRC\,1138$-$262 has number counts that
are only slightly higher than those presented in this work.

\begin{figure}
    \centering
    \includegraphics[width=0.475\textwidth]{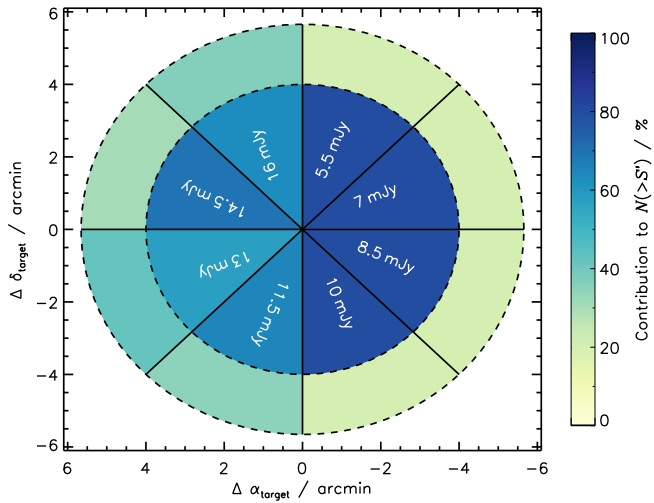}
    \caption{Contribution to the cumulative number counts from two signpost-centric
    annuli with equal area (i.e.\ $16\pi\,\textrm{arcmin}^2$).
    We separate each annuli by dashed, black lines and divide them into eight equally sized segments
    representing the $870\textrm{-}\micron{}$, flux-density thresholds listed
    in Table~\ref{tab: number counts}.
    We color-code the contribution to the total number counts from each annuli in a
    given segment (see scale).
    At $S'>8.5\,\millijanksy{}$, we see that the inner annuli contributes $\approx80\,\%$ of
    the sources responsible to the total number counts.
    However, by $S'>11.5\,\millijanksy{}$ the contribution is equally split between the
    two annuli, within the large Poisson errors ($\sigma\approx30\%$).
    This highlights the difficulty in claiming any radial dependence on the number
    counts due to variations in the instrumental noise (i.e.\ the noise increases as the distance from
    our signposts increases).
    }
    \label{fig: annuli number counts}
\end{figure}

In Fig.~\ref{fig: annuli number counts} we show how the contribution to our number counts
at the flux densities provided in Table~\ref{tab: number counts}
varies in two, signpost-centric annuli of equal area ($16\pi\,\textrm{arcmin}^2$).
We see that at $S'>8.5\,\millijanksy{}$, $\approx80\,\%$ of the contribution to the number
counts comes from DSFGs distributed within $r_{\textrm{target}}<4'$ of our signposts.
However, due to the low numbers of galaxies above these deboosted flux-density thresholds,
this excess contribution is not significant ($\approx1.5\sigma$).
However, the increasing instrumental noise with distance from our signposts makes
comparisons of the number counts at all but the highest flux densities heavily biased.
We see that at the higher flux-density thresholds this perceived
excess diminishes rapidly and above $S'>11.5\,\millijanksy{}$ the contribution appears to be
equally split between the two annuli. Thus, without uniformly wide imaging of these
environments, the number counts as a function of radial distance remains largely
unconstrained for this sample.

\begin{figure}
    \centering
    \includegraphics[width=0.47\textwidth]{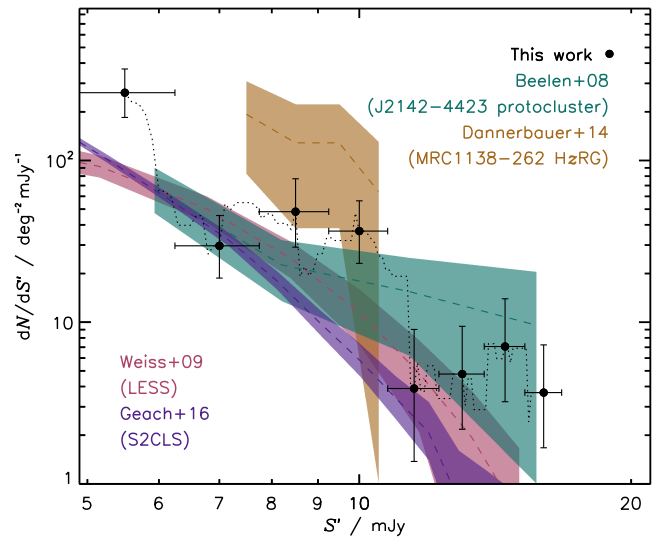}
    \caption{Differential number counts (excluding our target ultra-red
      galaxies) as a function of $870\textrm{-}\mu\textrm{m}$ flux density
      (black squares).
      As in Fig.~\ref{fig: number counts}, we also show the differential number counts
      for the LESS (pink) and S2CLS (purple) blank-fields as well as the two known
      proto-clusters J2142$-$4423 (green) and
      MRC\,1138$−$262 (brown).
      We see that above $S'>8.5\,\millijanksy{}$ our differential number counts are
      typically $1\sigma$ greater than those presented in LESS -- our comparison field
      of choice.}
    \label{fig: differential number counts}
\end{figure}

Finally, in Fig.~\ref{fig: differential number counts} we show the differential
number counts for this work alongside those of the LESS and S2CLS blank fields and
the two known proto-clusters J2142$-$4423 and MRC\,1138$−$262.

\subsection{Over-densities}
\label{sec:overdensities}

In order to make a statistical analysis of the significance of our number
counts, we employ an over-density parameter \citep{morselli14}:

\begin{equation}
    \label{eq: over-density parameter}
    \delta(>S') = \frac{N(>S')}{N(>S')_{\textrm{blank field}}} - 1 ,
\end{equation}

\noindent
where $N(>S')_{\textrm{blank field}}$ are the number counts expected in a
blank-field survey above some threshold flux density.

When choosing a blank-field survey suitable for comparison it is important
to compare `like-for-like' \citep[i.e.][]{condon07}.
For instance, broad-beam surveys can hide the
multiplicity of DSFGs predicted by models \citep[e.g.][]{cowley16} and proven by high-resolution observations \citep{wang11, simpson14, bussmann15, oteo17}. Furthermore, similar --- if not
identical --- data reduction techniques ensure consistency in the flux
densities and associated errors, which may otherwise lead to a lower or
higher estimate of the number counts (see \S~\ref{sec: under-density in less}).

Hence, we choose the LESS number counts (calculated directly from the
source catalog) to make comparisons. These data and ours were obtained from
the same instrument and are reduced in a similar manner using the same
software.  However, there are slight differences in the results when we run
our source extraction algorithm on the LESS \textsc{dr1.0}
$S/N$ map\footnote{%
 \url{http://archive.eso.org/cms/eso-data/data-packages/less-data-release-v1-0.html}.}.
Using a detection threshold of $\Sigma_{\textrm{thresh}}>3.7$, we
recover $95\%$ of their sources. Our $870\textrm{-}\micron{}$ flux
density measurements are comparable to those in \citet{weiss09b} as we
record a mean absolute offset of
$\overline{|\Delta S_{\nu}|}=0.4\,\millijanksy{}$.  These differences should
have a relatively minor effect on comparisons made with the number
counts. However, the computation of completeness and flux boosting
parameters do differ. We record $\lesssim15\%$ differences in the latter at
a detection $S/N\approx3.7$ for sources around SGP-433089, which has a
similar (albeit slightly higher) average depth to LESS.
We note that \citet{weiss09b} claim that LESS is under-dense
and also shows a deficit of bright sources relative to other blank fields.
However, Fig.~\ref{fig: number counts} shows that this is clearly not case when
adopting the much deeper and wider data from S2CLS as a reference.

We make over-density comparisons at a flux density threshold of $S'>8.5\,\millijanksy{}$,
which equates to a surveyed area of $\mathcal{A}\approx0.2\,\textrm
{deg}^{2}$ at our detection threshold. We choose this flux density threshold to
be directly comparable to LESS. Furthermore, this threshold is high enough to
minimize the correction effects needed for our low $S/N$ detections. At the same
time it is low enough such that our results should not drastically change if our
bright sources are magnified by $\mu\lesssim2$.

We add our number-count error bars in quadrature to those given
in \citet{weiss09b}. We determine an over-density of
$\delta=1.0^{+0.3}_{-0.3}$ at $S'>8.5\,\millijanksy{}$.
Or, put another way, we are $99.93\,\%$ confident that our signposts pinpoint
over-dense regions in the Universe, and are $\approx95(50)\,\%$ confident that these regions
are over-dense by a factor of at least $\ge1.5(2)\times$ compared to LESS.

However, we stress that by \emph{only} removing the target galaxy from our number
counts analysis we are left with a `residual bias' due to imaging a region where a galaxy is
known to reside.
We estimate this residual bias increases our over-density
parameter by $\delta_{\textrm{resid.~bias}}=0.23\pm0.02$ over the typical map areas
($\pi(6.2^{+0.3}_{-0.1})^2$) that we have surveyed in this work.

Furthermore, we crudely test what effect removing sources with $S/N\le4$
and $S/N\le4.5$ has on this over-density calculation.
This signal-to-noise regime is close enough to our detection threshold
such that the completeness corrections and surveyed area values that we apply
should be similar.
Thus, we derive over-density values of $\delta=1.0\pm0.3$ and $\delta=0.7\pm0.2$
for sources with $S/N>4$ and $S/N>4.5$, respectively.
This suggests that, despite a non-negligible fraction of sources near our detection
threshold potentially being spurious, our over-density above $8.5\,\millijanksy{}$ is
comprised of secure LABOCA detections.

There exists a strong correlation in flux density with our over-density
parameter, as seen in the right-hand panel of Fig.~\ref{fig: number counts}.
Here we plot the over-density parameter for each target, which
we have logarithmically scaled to reflect each target's contribution to our
overall number counts. We see a large scatter across our 22 maps indicative
of cosmic variance and varying levels of map noise. The evolution in
over-density increases $\sim50\times$ from $S'=7\text{--}16\,\millijanksy{}$,
although the Poisson error from the blank-field counts rises steeply at the
higher flux densities, exacerbated by the large relative error in the number
counts of bright sources in LESS.
We believe that this evolution is caused by our
ultra-red galaxies signposting regions that contain brighter DSFGs. However,
without high-resolution imaging of the environments around our ultra-red
galaxies, we cannot rule out gravitational lensing by chance alignment.

\subsubsection{Mundane, not cosmic, under-density in LESS}
\label{sec: under-density in less}

It is often claimed that LESS exhibits an under-density of DSFGs, which has resulted in the introduction, and use of \citep{swinbank14, dannerbauer14}, a multiplicative `fudge-factor' ($\sim2\times$) to the number counts presented in \citet{weiss09b}.
An `adjustment' of this magnitude would require us to significantly lower the value of our reported over-density parameter, if necessary.

This perceived under-density in LESS is often concluded against the results presented in SHADES \citep[SCUBA HAlf Degree Extragalactic Survey ---][]{coppin06} as it was the largest, `like-for-like' survey with which to compare against.
However, \scuba{} has uniformly reimaged
the entirety of the Subaru/XMM–Newton Deep Field (SXDF) -- one half of SHADES
-- improving upon its depth by a factor $\gtrsim2\times$ and thus allowing us to test the validity of this claim.

Using these new, deeper data\footnote{
  \url{https://zenodo.org/record/57792\#.WOtnkRiZNE5}.
}, we are only able to match 27/60 (i.e.\ $45\%$) of the detections cataloged in SHADES\footnote{
  \url{http://www.roe.ac.uk/ifa/shades/dataproducts.html}.
} to a counterpart cataloged in the S2CLS.
These `matched' sources -- with typical radial offsets of $4.7\pm3.0''$ --  have deboosted, $850\textrm{-}\micron{}$ flux densities that are on average $1.6(\pm0.1)\times$ greater than those reported in S2CLS.
The 33/60 ($55\%$) `unmatched' detections have a broad range of deboosted flux densities ($S=3.1\textrm{-}22.0\,\millijanksy{}$) that are typically $\approx4\times$ higher than Gaussian fits at their positions in the S2CLS UDS map suggest.

If these results were to be replicated for the Lockman Hole East, it would appear that the spurious fraction of sources and/or flux-boosting corrections within SHADES have been miscalculated.
Taken together, these findings suggest that the claimed under-density in LESS, and apparent deficit of bright DSFGs, is unlikely to be true and unlikely to be biasing our over-density parameter higher.
Furthermore, these findings are very reminiscent of those discussed by \citet{condon07}, who resolved the inconsistencies amid differing reports of the radio number counts at the time.
Thus, in homage, the variance in the number counts between SHADES and LESS appears to be `mundane' (likely due to instrumental and analysis effects) rather than `cosmic'.

\subsubsection{Probability of being ultra-red}
\label{sec: pur}

As can be seen Table~\ref{tab: photometry}, half of our signposts have SPIRE photometry
which is just consistent with them being ultra-red. This motivates us to derive, for
the first time, a probability that a galaxy is actually ultra-red ($\pur{}$) based on
its SPIRE photometry\footnote{
 These probabilities are calculated by assuming symmetric color uncertainties, and do not take account of the bias that more bluer galaxies will have had their colours scattered redward, into the ultra-red category, than vice-versa.
 However, these are only being used as a guide to the likelihood of being ultra-red.}.
To this end, we draw $10,000$ realizations of the SPIRE photometry from
a Gaussian distribution and
determine the number of times that these realizations meet our ultra-red criteria outlined in \citetalias{ivison16}.
By incorporating the photometric errors from all SPIRE bands, we are able to generate a subset of galaxies that are likely to be ultra-red.
Finally, we derive $1\textrm{-}\sigma$ errors assuming Poisson statistics
for these ultra-red galaxy probabilities, which we list in Table~\ref{tab: ultra-red probabilities}.

\begin{table}
 \begin{center}
    \caption{Targets and their probability of being ultra-red.}
    \label{tab: ultra-red probabilities}
    \begin{scriptsize}
        \begin{tabular}{l c}
            \hline\hline
            \multicolumn{1}{l}{Nickname} &
            \multicolumn{1}{c}{$\pur{}$} \\ 
            \multicolumn{1}{l}{} &
            \multicolumn{1}{c}{$\%$} \\ 
            \hline
            SGP-28124 & $94.6\pm0.4$  \\
            HeLMS-42 & $87.4\pm0.4$   \\
            SGP-93302 & $67.5\pm0.2$  \\
            ELAIS-S1-18 & $33.4\pm0.1$\\
            ELAIS-S1-26 & $61.4\pm0.2$\\
            SGP-208073 & $62.2\pm0.2$ \\
            ELAIS-S1-29 & $65.8\pm0.2$\\
            SGP-354388 & $93.2\pm0.4$ \\
            SGP-380990 & $71.1\pm0.3$ \\
            HeLMS-10 & $83.6\pm0.3$   \\
            SGP-221606 & $41.8\pm0.1$ \\
            SGP-146631 & $29.9\pm0.1$ \\
            SGP-278539 & $81.0\pm0.3$ \\
            SGP-142679 & $87.5\pm0.4$ \\
            XMM-LSS-15 & $29.5\pm0.1$ \\
            XMM-LSS-30 & $97.1\pm0.4$ \\
            CDFS-13 & $28.5\pm0.1$    \\
            ADF-S-27 & $43.1\pm0.1$   \\
            ADF-S-32 & $16.5\pm0.0$   \\
            G09-83808 & $89.0\pm0.4$  \\
            G15-82684 & $62.6\pm0.2$  \\
            SGP-433089 & $21.7\pm0.0$ \\
            \hline
        \end{tabular}
    \end{scriptsize}
    \end{center}
    \vspace{-\baselineskip}

    \noindent
    \textbf{Note.} Targets are listed in order of increasing right ascension, i.e.\ in the 
    same order that they appear in Table~\ref{tab: properties}.
\end{table}

\begin{figure}
    \centering
    \includegraphics[width=0.475\textwidth]{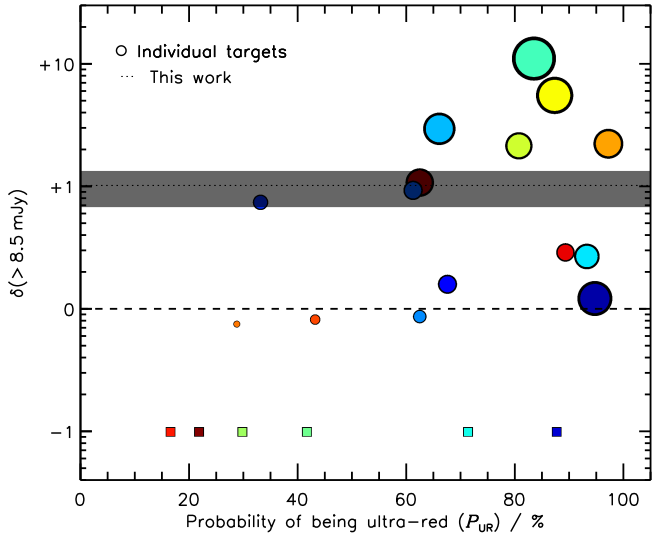}
    \caption{Over-density parameter above
    $S'>8.5\,\millijanksy{}$ versus the probability that our signposts are ultra-red
    using the method outlined in the text.
    Each target is color-coded and represented as a circle with a size reflecting its overall
    contribution to the number counts, i.e.\ as described in the caption of
    Fig.~\ref{fig: number counts}. The mean overdensity
    at this flux-density threshold is shown as a black dotted line, whilst the shaded area
    represents its $1\textrm{-}\sigma$ uncertainty.
    This shows that approximately half of our ultra-red galaxies have $\pur{}\lesssim68\,\%$
    once their SPIRE flux densities have been re-evaluated at their LABOCA position.
    Conversely, signposts that have a higher probability of being ultra-red contribute
    more to the mean over-density at this flux density threshold.}
    \label{fig: overdensity versus pur}
\end{figure}

In Fig.~\ref{fig: overdensity versus pur} we show how the over-density parameter above
$S'>8.5\,\millijanksy{}$ varies as a function of its probability of being ultra-red
for our signposts. Clearly evident is that galaxies that have a higher
probability of being ultra-red, typically have a much higher overdensity parameter.
Furthermore, over-dense signposts (i.e.\ signposts with $\delta>0$) all have a
probability of being ultra-red greater than $\pur{}\gtrsim30\,\%$. This lower limit
value is caused by galaxies lying at the boundaries of both of our SPIRE colour-cuts
outlined in \citetalias{ivison16}.
Above a probability of being ultra-red of $\pur{}\gtrsim60\,\%$, we see that only three
($\approx20\,\%$) of our signposts have environments that are consistent with being
under-dense (i.e.\ $\delta<0$). Such a low fraction of under-dense environments suggests
that using this novel ultra-red-probability technique in conjunction with
$870\textrm{-}\micron$ imaging provides a robust method for signposting over-densities
in the distant Universe.

\subsection{Colors}
\label{sec:colours}

\begin{figure}
    \centering
    \includegraphics[width=0.475\textwidth]{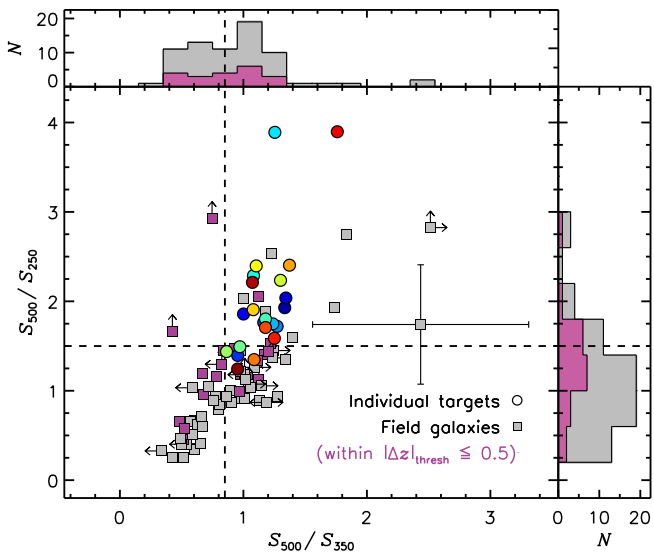}
    \caption{$S_{500}/S_{350}$ versus $S_{500}/S_{250}$ for our catalog of
      sources which have at least one SPIRE detection above $1\sigma$. We
      show our target (field) galaxies as circles (squares) and highlight in
      pink those field galaxies which lie within $|\Delta z| \leq 0.5$ of
      their signpost galaxy. We show our color-cut limits (dashed line),
      $S_{500}/S_{250}\ge1.5$ and $S_ {500}/S_{350}\ge0.85$, which a target
      is required to match in order to meet our ultra-red galaxy selection
      criteria (i.e.\ the top-right region of the plot). Five targets narrowly
      miss our $S_{500}/S_{250}$ color cut threshold, three by 0.1 and two
      by 0.2. This shift towards bluer colors is due to the larger
      250-$\micron{}$ boosting and the refined positions at which we
      make the SPIRE photometric measurements.
      A representative color uncertainty is shown and we
      use arrows to highlight $1\textrm{-}\sigma$ limits where applicable.}
    \label{fig: colour-colour}
\end{figure}

We analyze the $S_{500}/S_{250}$ and $S_{500}/S_{350}$ colors to see if our
field galaxies comprise similarly red galaxies as our signposts. Recall that
in all further analysis we exclude 16 LABOCA detections as we are unable to
constrain their photometric redshifts. This leaves us with $86-16=70$ DSFGs
around our 22 ultra-red signposts above $>3.5\sigma$.
Fig.~\ref
{fig: colour-colour} illustrates that only $7\%$ ($\approx5\,$DSFGs)
of our field galaxies meet our ultra-red galaxy criteria. Such a low fraction
might be expected as our ultra-red galaxy criteria selects the most luminous
and rare DSFGs. If we relax the $3.5\sigma_{500}$ threshold
(imposed in \citetalias{ivison16}) to $1\sigma_{500}$, our fraction of
field ultra-red galaxies increases to $17\%$ ($\approx12\,$DSFGs)
at the expense of being less reliable. 


Our field galaxies have median $S_{500}/S_{250}$ and $S_{500}/S_{350}$ colors of
$(S_{500}/S_{250})_{1/2}=1.1$ and $(S_{500}/S_{350})_{1/2}=0.9$, respectively,
with interquartile ranges of $S_{500}/S_{250}=0.7\text{--}1.4$ and
$S_{500}/S_{350}=0.7\text{--}1.2$. If we isolate the field galaxies that we
assume to be physically associated to their target galaxy (see
\S\ref{sec: physical associations}), we notice a redder change as the $S_{500}/S_{250}$
color increases to a median $(S_{500}/S_{250})_{1/2}=1.4$ with interquartile
range $S_{500}/S_{250}=1.2\text{--}1.5$. However, we see no appreciable change
in the $S_{500}/S_{350}$ color. As can be seen in
Fig.~\ref{fig: colour-colour}, this can be explained by five of signpost
galaxies narrowly missing our original ultra-red criteria once their SPIRE
photometry has been remeasured at their LABOCA position.

Thus, if we go one step further and isolate the associated field galaxies that
contribute to the overdensity at $S'>8.5\,\millijanksy{}$, we find that they have
redder median colors of $(S_{500}/S_{250})_{1/2}=1.0$
and $(S_{500}/S_{350})_{1/2}=1.4$. This is in part due to the exclusion of
SGP$-$433089 and its associated galaxies, which -- having had its SPIRE photometry
remeasured at the position of its LABOCA emission -- has a low probability of being ultra-red.
We remind the reader that this is shown in Table~\ref{tab: ultra-red probabilities} and
Fig.~\ref{fig: overdensity versus pur}, where galaxies with a higher probability of being
ultra-red, and are thus more distant, are primarily contributing to our
over-density parameter at $S'>8.5\,\millijanksy{}$.

\subsection{Physical associations}
\label{sec: physical associations}

To quantify whether the galaxies responsible for the over-density are
associated with their signpost ultra-red galaxy -- thus comprising a
proto-cluster -- we analyze their photometric redshifts.

The simplest analysis we could perform is to calculate the absolute difference
between the photometric redshifts of our field galaxies, $z_{\textrm{field}}$,
relative to their respective target ultra-red galaxy, $z_{\textrm{target}}$. We therefore
define a parameter

\begin{equation}
    \label{eq: absolute residual}
    |\Delta z| = |z_{\textrm{target}} - z_{\textrm{field}}|
\end{equation}

\noindent
in order to determine the fraction of galaxies which lie at or below some
association threshold, $|\Delta z|_{\textrm{thresh}}$. Choosing such a
threshold is complicated by the difficult task of determining photometric
redshifts using far-IR photometry alone.

\begin{figure}
    \centering
    \includegraphics[width=0.475\textwidth]{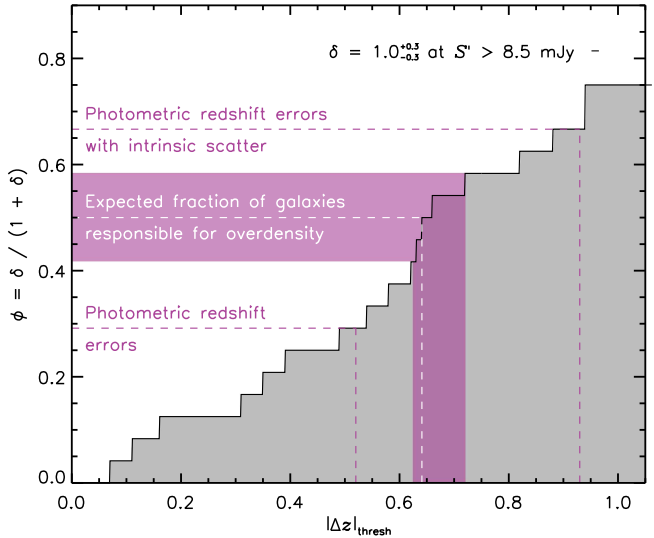}
    \caption{The fraction, $\phi=\delta/(1+\delta)$, of sources responsible for an
    over-density ($\delta$) as a function of association threshold, $|\Delta z|_
    {\textrm{thresh}}$. At $S'>8.5\,\millijanksy{}$ we expect $\phi=0.5^{+0.2}_
    {-0.2}$
    of our bright DSFGs to be associated, which we only achieve if our
    threshold is set to $|\Delta z|_{\textrm{thresh}}\le0.65$. We also show that
    we over-/under-account for DSFGs responsible this over-density if our
    threshold is based on the median photometric errors / added in quadrature
    with the intrinsic, template scatter. This motivates us to choose an association
    threshold of $|\Delta z|_{\textrm{thresh}}\approx0.5$.}
    \label{fig: zphot association threshold}
\end{figure}

For example, if we were to account for the fraction,
$\phi=\delta/(1+\delta)=0.5^ {+0.6}_{-0.4}$, of sources responsible for our
over-density, $\delta=1.0^{+0.6}_ {-0.5}$, at $S'>8.5\,\millijanksy{}$ we
would require an association threshold
$|\Delta z|_ {\textrm{thresh}}\le0.65$ (see Fig.~\ref{fig: zphot
  association threshold}). Put another way, we have an over-density of
$\delta=1.0$, comprised of 24 DSFGs with deboosted flux densities
$S>8.5\,\millijanksy{}$. We therefore expect $\phi=0.5$ (or 12) of these
DSFGs to be responsible for this over-density. We achieve this association
if we arbitrarily set our threshold to $|\Delta z|_ {\textrm{thresh}}\le0.65$
as shown in Fig.~\ref{fig: zphot association threshold} where we plot the
fraction of sources responsible for an overdensity against our association
threshold.

On the other hand, if we choose a threshold dependent on the median fitting
errors for our targets and field galaxies,
$|\Delta z|_{\textrm{thresh}} \le \left(\left(\sigma_
    {z_{\textrm{target}}}\right)_{1/2}^2 +
  \left(\sigma_{z_{\textrm{field}}}\right)_ {1/2}^2\right)^{1/2}=0.52$, we are unable
to account for $\approx20\%$ of the galaxies responsible for the
over-density.
Finally, if we were to include in quadrature the intrinsic scatter in
our three templates to the median fitting errors, our association threshold
would increase to $|\Delta z|_{\textrm {thresh}}\le0.93$.
As can be seen in Fig.~\ref{fig: zphot association threshold}, this threshold
includes all of the galaxies responsible for the over-density but is likely contaminated
by unassociated galaxies ($15\,\%$).

Both the former and latter association thresholds are too
large to make any reliable claim of association. We therefore compromise,
knowingly missing some of the galaxies responsible for the over-density by
choosing an association threshold, $|\Delta
z|_{\textrm{thresh}}\le0.52$. We do this in order to increase the
reliability of our further analysis of these potential proto-cluster
systems. Utilizing this approach for our entire catalog we find that half
of our target ultra-red galaxies have at least one associated DSFG.

We illustrate the results of this analysis in the top-panel of Fig.~\ref{fig:
redshift analysis}, where we have chosen to plot $\Delta z$ against the radial
distance between field galaxies and their targets ($\Delta r_{\textrm
{target}}$). Half of these associated DSFGs are within $\Delta r_{\textrm
{target}}\lesssim3'$ - suggesting that there is a slight dependence on
association with proximity, in agreement with the annuli analysis of our number
counts in \S\ref{sec:number counts}.
In terms of proper radial distances (derived at the redshift of the target),
we see that these galaxies are
distributed on scales of $\Delta r_{\textrm{target}}\sim2\,\textrm{Mpc}$, reporting
an average separation of $\overline{\Delta r}_{\textrm
{target}}=1.6\pm0.5\,\textrm{Mpc}$ with an interquartile range $\Delta r_
{\textrm{target}}=1.0\text{--}2.2\,\textrm{Mpc}$. We see no dependence on the
redshift of the target ultra-red galaxy and the average target separation from $z=2\text
{--}4$.

\begin{figure}
    \centering
    \includegraphics[width=0.475\textwidth]{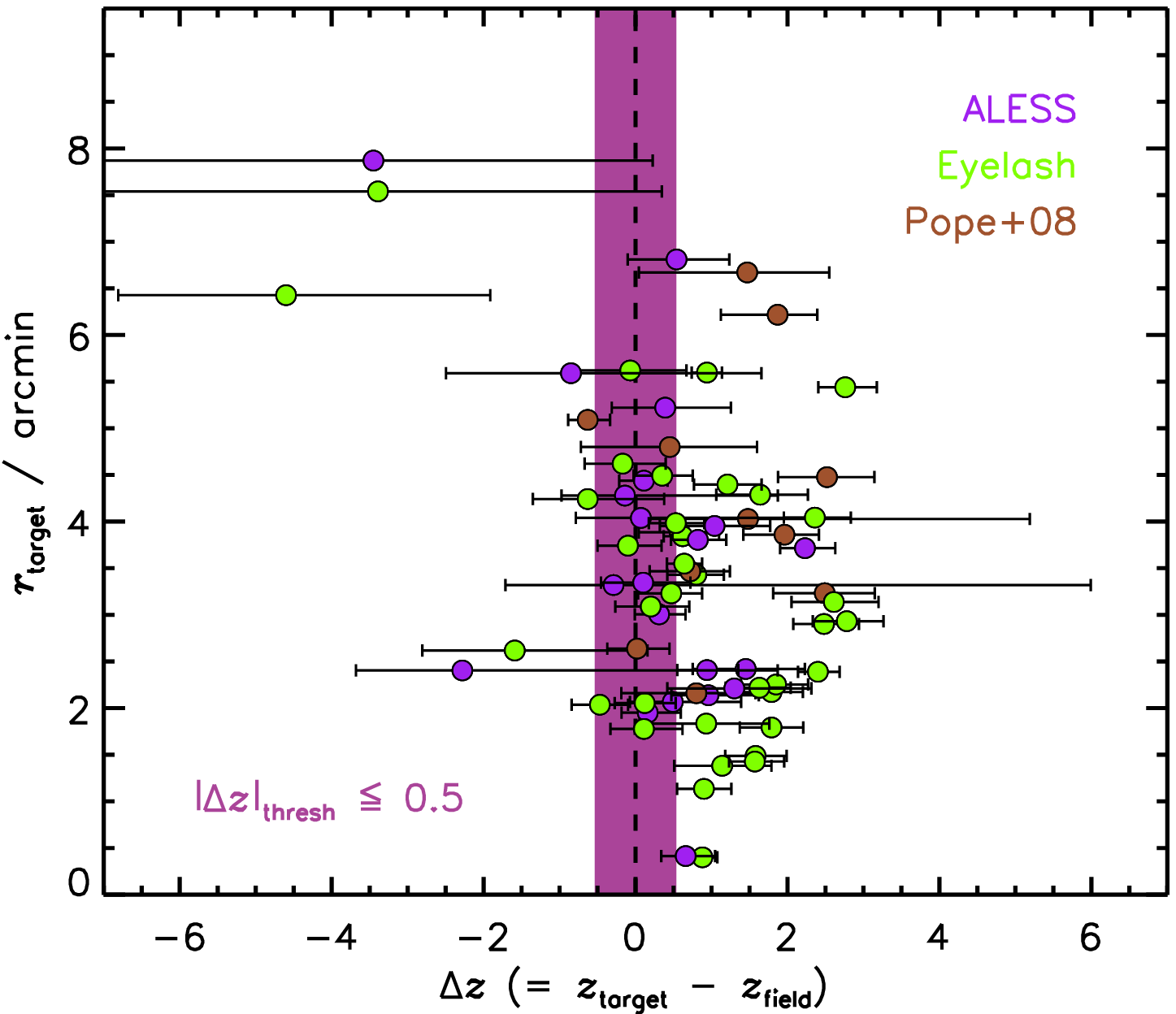}\\
    \vspace{0.0166667\textwidth}
    \includegraphics[width=0.475\textwidth]{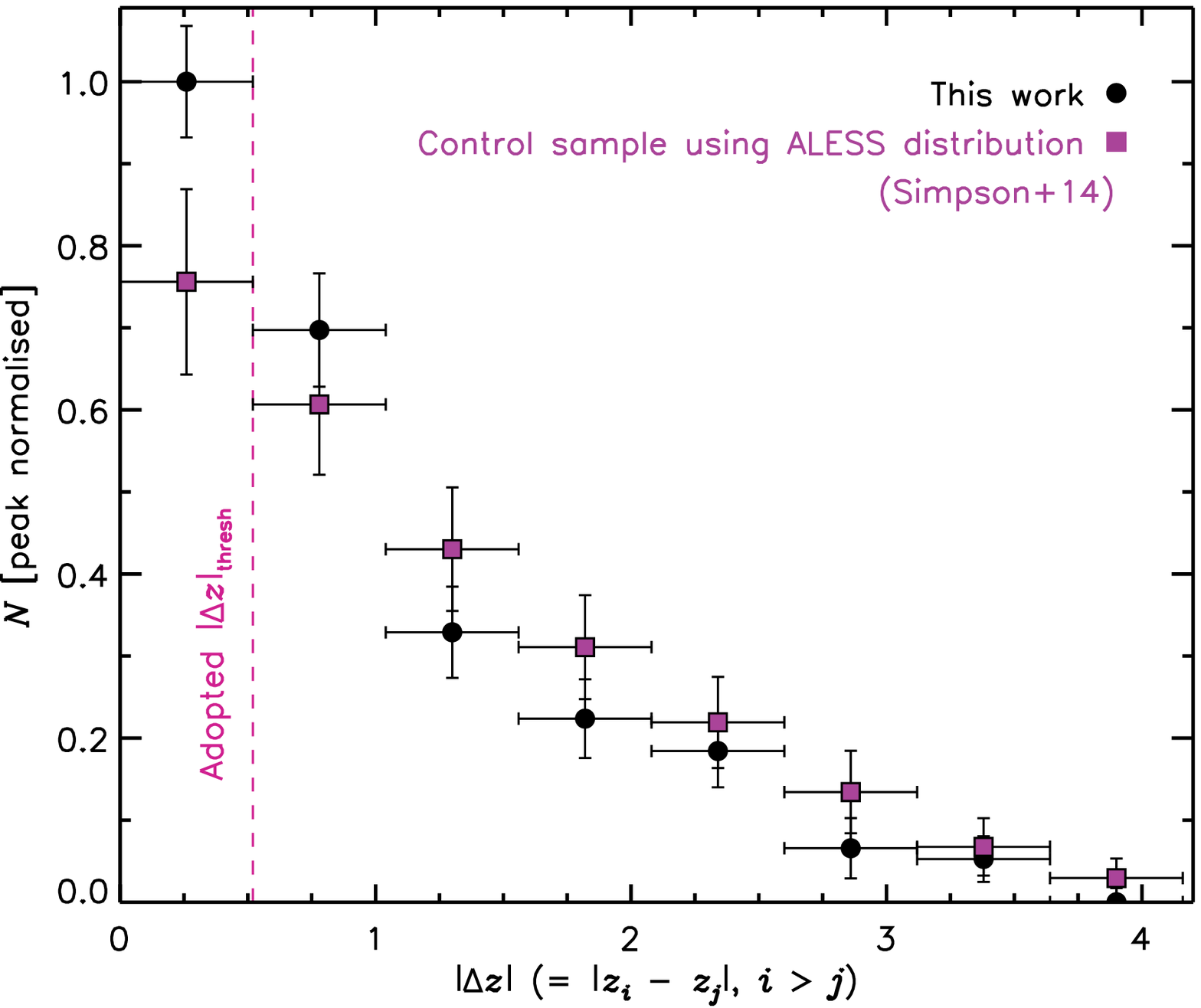}
    \caption{\textit{Top}:
      radial distance of our field galaxies to their signpost galaxies as a
      function of photometric redshift difference ($\Delta z$).
       Errors are deduced from the
      $\chi^2_{\textrm{min}}+1$ locations and are not added in quadrature
      with the intrinsic scatter. We note that the tail of sources with
      $\Delta z \ge 0$ reflects the fact that most galaxies are foreground
      to our targets, which sit at a median $z_{1/2}=3.2$. The pink region
      indicates our threshold boundaries for association, in which a
      fraction $\phi\approx0.3$ of our field galaxies lie.  The large
      errors in our photometric redshifts highlight the difficulty in
      accurately constraining the redshifts of our DSFGs. Finally, we color code
      each DSFG to indicate the best-fitting template adopted.
      \textit{Bottom}: alternative analysis of the absolute
      photometric redshift differences $|\Delta z_{i,j}|$ for all of our
      maps. We see a similar association excess to that of the top panel.}
    \label{fig: redshift analysis}
\end{figure}

The top-panel of Fig.~\ref{fig: redshift analysis} also shows that the
majority of our field galaxies are at a lower redshift compared to their
respective signpost galaxy, with the former lying at a median photometric
redshift, $z_{1/2}=2.6\pm0.2$, with interquartile range,
$z=1.9\text{--}3.1$, and the latter (our signposts) lying at a slightly
higher redshift, $z_{1/2}=3.2\pm0.2$, with an interquartile range,
$z=2.8\text{--}3.6$. If we remove the associated DSFGs, we refine the
median photometric redshift for the `interloper' galaxies to be
$z_ {1/2}=2.3\pm0.1$ with an interquartile range, $z=1.8\text{--}2.8$, in
good agreement with the general DSFG population \citep{chapman05,
  simpson14}.

Our associated DSFGs have a median rest-frame luminosity,
$(\lfir{})_{1/2}=10^{12.7}\,\lsol{}$, with an interquartile
range, $\lfir{}=10^{12.6}\text{--}10^{12.9}\,\lsol{}$. Between shells
of proper radial distance from the target of
$\Delta r_{\textrm{target}}=0.3\text{--}1.3\,\textrm{Mpc}$ and
$2.3\text{--}3.3\,\textrm{Mpc}$, we see an average difference in luminosity
of $\Delta \lfir{}=(3\pm2)\times10^{12}\,\lsol{}$.
This slight increase in luminosity perhaps hints at the existence of a mechanism
able to enhance the SF in denser environments \citep[e.g.][]{oteo17b}.

We translate rest-frame luminosities into SFRs using
$\psi / \,M_ {\odot}\,\textrm{yr}^{-1} \approx 1.7\times10^{10}
\lfir{} / \lsol{}$ \citep[see Equation~4 in][for starbursts
using a Salpeter initial mass function, IMF, noting that
a top-heavy IMF in distant dusty starbursts has been suggested multiple times ---
\citealt{romano17}]{kennicutt98}. Hence, these associated galaxies have
high median SFRs, $\psi_{1/2}=1000\pm200\,\msol{}\,\textrm{yr}^{-1}$,
with an average total star formation rate,
$\Psi =\Sigma \psi = 2200\pm500\,M_ {\odot}\,\textrm{yr}^{-1}$. This is
consistent with a scenario wherein these galaxies form the bulk of their
stellar mass quickly (in $<1\,\textrm {Gyr}$) at $z\sim3$ and evolve to
populate the centers of massive galaxy clusters seen today \citep{thomas05,
  thomas10, fassbender11, snyder12}.

To test the validity of this simplistic method for proto-cluster association, we
calculate the same residual parameter but this time for all galaxy pairs $i$
and $j$ in each map $k$, i.e.\ $|\Delta z_{i,j}|_k = |z_i - z_j|_k, \quad \forall
j > i$. We compare the average value of this parameter for all maps to that
of a control sample. We determine the latter by replacing all galaxies except
for our targets with a random galaxy drawn from the ALESS photometric redshift
distribution presented in \citet{simpson14}.

This alternative analysis is shown in the bottom-panel of Fig.~\ref{fig:
  redshift analysis} where we see a similar excess of $\phi\approx0.3$ to
that found in the previous analysis. Furthermore, this analysis shows that
there is a deficit of $|\Delta z|\ge1$ pairs, indicating that our field
galaxies are preferentially associated to their target galaxies below this
level. This alternative analysis, however, does not tell us which field galaxies are
associated with the signpost ultra-red galaxies.

The similarities between the findings of both methods suggests that we can
trust our analysis.

\subsection{Consequent fate at $z\sim0$}

Here, we briefly discuss the eventual fate of the ultra-red galaxy
environments that have at least one associated DSFG to their signpost.

To recap, just over half of our sample have at least one associated DSFG within $\Delta z \le 0.52$.
We have shown that these galaxies have high SFRs, with the candidate proto-clusters themselves having an average total SFR of $\Psi\sim2\times10^{3}\,\msol{}\, \textrm{yr}^{-1}$.
This supports a scenario wherein these galaxies evolve from $z\sim3$ to the present to populate the centers of the most massive galaxy clusters seen in the local Universe.

We now derive molecular gas masses, $M_{\textrm{H}_2}$, using the far-IR
continuum and an appropriate scaling constant \citep[$\alpha$
---][]{scoville14, scoville15}, determined from a sample of 28 SMGs with
CO(1--0) measurements at $z<3$

\begin{equation}
    \label{eq: gas mass}
    \alpha = \frac{L_{850\,\mu\textrm{m}}}{M_{\textrm{H}_2}} =
    1.0\pm0.5\times10^{20}\,\textrm{erg}\,\textrm{s}^{-1}\,\textrm{Hz}\,\msol{}^{-1} ,
\end{equation}

\noindent
where $L_{850\,\mu\textrm{m}}$ is the rest-frame luminosity at
$850\,\mu\textrm {m}$ determined from our best-fitting SEDs.
We derive median gas masses,
$(M_{\textrm{H}_2})_{1/2} = 1.7\times10^{11}\,\msol{}$, with an
interquartile range,
$M_{\textrm{H}_2} =
9.5\times10^{10}\textrm{--}2.1\times10^{11}\,\msol{}$, for our signpost
ultra-red galaxies and their associated DSFGs. Thus, if each DSFG converts its
reservoir of gas into stars, each would evolve into a present-day galaxy
with an average total stellar mass of at least
$\overline{M}_{\textrm {stars}}\gtrsim10^{11}\,\msol{}$. Furthermore, we
note that our signpost ultra-red galaxies have slightly elevated average gas
masses of $\overline{M}_{\textrm{H}_2}=2.5\pm1.2\times10^{11}\,\msol{}$
compared to their associated DSFGs. This is reminiscent of present-day
massive cD ETGs, which dominate the centers of present-day galaxy clusters
\citep{kelvin14}. However, we stress that without optical/near-IR imaging
of these ultra-red galaxy environments, we are potentially missing many
galaxies, each of which could contribute
$M_{\textrm{stars}}\approx10^{9}\textrm{--}10^{11}\,\msol{}$ worth of
stars to the final system \citep{overzier09b, casey15}; thus the eventual
stellar masses of these systems are largely unconstrained and all these
results should be regarded as firm lower limits.

Finally, we perform a crude space density calculation of our ultra-red-galaxy-selected
candidate proto-clusters.
We adjust the space-density redshift limits used for Equation~3 in
\citetalias{ivison16} to $2\lesssim z\lesssim6$ -- motivated by
the last epoch of virialized galaxy clusters \citep{casey16}
and the highest of our ultra-red galaxy redshifts \citep{fudamoto17, zavala17}, respectively.
We derive a space density of $\rho\sim3\times10^{-6}\,\textrm{Mpc}^{-3}$ for our ultra-red
galaxies within $2\lesssim z\lesssim6$ assuming a SF lifetime of
$t_{\textrm{burst}}=100\,\megayear{}$.
This roughly equates to the space density of $z<0.5$ galaxy clusters with DM masses of
$\mdm{}\sim4\times10^{14}\,\msol{}$, i.e.\ so-called `Virgo-type' galaxy clusters
\citep{chiang13, bahcall93}. Although, it should be noted that perhaps only $20\%\textrm{–}40\%$ of all
proto-clusters within $2\lesssim z \lesssim6$ are actually rich in DSFG \citep{casey16}.

However, as can be seen in the right-hand panel of
Fig.~\ref{fig: number counts} and Fig.~\ref{fig: overdensity versus pur}
not all of our ultra-red galaxies probe over-dense regions. In fact, we estimate that only
$33\pm8\%$ of our sample have over-density parameters above
$\delta(>8.5\,\millijanksy{})>1$.
Thus, we scale the space density of ultra-red galaxies accordingly to
derive a proto-cluster space density of
$\rho_{\textrm{proto-cluster}}\sim9\times10^{-7}\,\textrm{Mpc}^{-3}$.

\subsection{Remarks on selected ultra-red galaxies}
\label{sec:remarks on sources}

We discuss some of the most exciting and/or over-dense fields, each of which
clearly warrants further exploration. We remind the reader that the small areas
and varying r.m.s.\ levels of each map makes all further analysis heavily subject to
the effects of cosmic variance.

\begin{itemize}
\item \textit{SGP-93302}: this is our deepest map, reaching an average
  beam-smoothed r.m.s.\ of
  $\overline{\sigma}_{870}=1.7\,\millijanksy{}$. This 500-$\mu\textrm{m}$
  riser has a deboosted flux density of
  $S_{870}=30.9\pm1.3\,\millijanksy{}$. We estimate this ultra-red galaxy lies
  at $z=3.6^{+0.2}_ {-0.1}$ and note that one ($15\%$) of its field
  galaxies is an equally bright DSFG at $z=3.4^{+0.4}_{-0.3}$ with a
  deboosted flux density of $S_ {870}=31.0\pm1.9\,\millijanksy{}$. This
  associated DSFG also meets our strict criteria of being an ultra-red
  galaxy and is cataloged in \citetalias{ivison16} as SGP-261206 and reported by
  \citet{fudamoto17} to lie at $z=4.2$.
  Such an
  environment of robust ultra-red galaxies warrants spectroscopic follow-up and high-resolution
  imaging to explore the morphologies of its constituents. This map shows
  no particular over-/under-density compared to LESS in the low flux
  density regime, but it does show a $1\sigma$ excess at flux density
  thresholds of $S'>10\,\millijanksy{}$.
\item \textit{SGP-354388}: discussed by \citet{oteo17}, we revise the flux
  density of this extraordinary DSFG to
  $S_{870}=33.0\pm1.2\,\millijanksy{}$, assuming that it can be deblended
  into two, LABOCA point sources, separated by $\approx25''$ as our extraction
  algorithm suggests. The multiplicitous nature of this source is also seen
  at higher resolutions, where ALMA 3-mm continuum maps resolve the central
  fragments further, into three or more components \citep{oteo17}. Like
  SGP-93302, this ultra-red galaxy only shows an over-density of sources at
  flux density thresholds, $S'>10\,\textrm {mJy}$. We are only able to
  associate two of its nine field galaxies, although a further two DSFGs
  have unconstrained photometric redshifts. We refine its photometric
  redshift to $z=4.2\pm0.2$ using improved SPIRE measurements made at the
  $870-\mu\textrm{m}$ position, which is consistent with its spectroscopic
  redshift, $\zspec{}=4.002$ \citep{oteo17b}.
\item \textit{SGP-433089}: this galaxy marks the most over-dense
  field in our sample, which we place at a distance of $z=2.5\pm0.2$. We
  associate six of its ten field galaxies with the signpost, noting that
  one of its field galaxies has an unconstrained photometric redshift. This map shows a deficit of bright DSFGs, compared to the other maps explored here. Thus it does not contribute to our over-density parameter at
  $S>8.5\,\millijanksy{}$.
  Its brightest source (the signpost galaxy) has a deboosted
  flux density, $S_{870}=7.2\pm1.1\,\millijanksy{}$, while the mean deboosted
  flux density of the detected field galaxies is
  $\overline {S}_{870}=4.7\,\millijanksy{}$. The detection of these
  relatively faint DSFGs is due to the low average r.m.s.,
  $\overline{\sigma}_{870}=1.1\,\millijanksy{}$, which allows us to report an
  over-density factor of $\delta=0.7^{+0.9}_ {-0.6}$ at a flux density
  threshold of $S'>4\,\millijanksy{}$.
  \item \textit{ADFS-27}: $3\,\millimeter{}$ scans with ALMA suggest that this
  ultra-red galaxy lies at $z\approx5.7$ \citep{riechers17} - drastically different to the estimate that we provide in this paper.
  \citet{riechers17} derive a dust temperature of $T_{\textrm{dust}}\approx55\,\textrm{K}$
  for this source, which highlights the strong degeneracy between temperature and redshift
  when using far-IR photometry alone to derive photometric redshifts.
  For instance, when we use a hotter, but, on average, less accurate template for
  ultra-red galaxies \citepalias{ivison16}, such as HFLS~3,
  we revise the photometric redshift for this galaxy to $\zphot{}=5.9^{+0.5}_{-0.4}$,
  i.e.\ to within $1\sigma$ of its reported spectroscopic value.
  This source has 2 associated DSFGs that lie within $\Delta z\approx0.5$ - making it an ideal high-redshift, candidate proto-cluster to follow-up further.
  Finally, we note that our SPIRE flux densities are higher by
  $\approx2\textrm{--}5\,\millijanksy{}$ than those presented in \citet{riechers17}, i.e. from the \hermes{} xID250 catalog from which this source was originally selected.
  This is due to remeasuring these flux densities at the position of the LABOCA peak, resulting in photometry that makes ADFS-27 appear less red.
  \item \textit{G09-83808}: this is a gravitationally lensed ($\mu\approx9$) ultra-red galaxy,
  with a photometric redshift estimate that is also catastrophically lower that its spectroscopic value.
  Recent work by \citet{zavala17} shows
  that this galaxy actually resides at $z\sim6$, rather than $\zphot{}=4.45^{+0.4}_{-0.3}$
  as presented here.
  Again, this DSFG highlights the temperature-redshift degeneracy as adopting
  HFLS~3 as a template yields a photometric redshift that is more consistent with its
  spectroscopic one, $\zphot{}=6.2^{+0.5}_{-0.4}$.

\end{itemize}

\subsection{Caveats}
\label{sec:caveats}

\begin{itemize}
\item A larger sample of ultra-red galaxies would help to reduce the effects
  of cosmic variance. We could improve our fidelity by achieving a uniform
  depth, comparable to that of SGP-93302, for example, so
  $\overline {\sigma}=1.3\,\millijanksy{}$, for all existing ultra-red
  galaxies. This would reduce the number of potentially spurious LABOCA
  sources present in our catalog. A uniform, wide imaging survey would also
  allow the detection of less luminous DSFGs in the vicinity of our
  signposts, out to a radius of $\Delta R_{\textrm{target}}\approx6'$.
\item The intrinsic luminosity of our associated DSFGs will depend on the
  gravitational lensing that each may have suffered. Although we have made
  an effort to avoid lensing in our selection of the signpost galaxies, as
  outlined in \citetalias{ivison16}, a fraction of our ultra-red galaxies
  are gravitationally magnified by chance alignments \citep{oteo17}.
  Our SFRs, and average total SFRs, are thus upper limits, though the
  effect of invariant IMFs in these galaxies likely has a greater impact.
\item When we utilize the $850\textrm{-}\mu\textrm{m}$ number counts from
  S2CLS, our
  over-density parameter rises to
  $\delta_{\textrm{S2CLS}}=2.1^{+0.6}_{-0.5}$ at
    $S'>8.5\,\millijanksy{}$. Although the errors remain similar (as they are
    dominated by the Poisson noise) we find that
    $\delta_{\textrm{S2CLS}}$ is $\gtrsim2\sigma$ higher than that determined
    using LESS as a comparison.
  \item Our association analysis likely underestimates the number of true
    physical associations. Our template
    fitting algorithm is accurate to only $\sigma_z=0.14(1+z)$,
    typically much larger than the errors determined from the $\chi_{\textrm
    {min}}^2+1$ values at high redshift. Thus our
  fixed association threshold leads us to miss some
  associated DSFGs. Some galaxies not associated with a signpost galaxy
  will be falsely assigned until ALMA spectroscopy can
    improve upon the accuracy of our photometric redshifts.
  \item Optical identification of the surrounding LBGs is necessary if we are
    to accurately constrain the total stellar mass -- and thus DM
    component, and the eventual fate at $z\sim 0$ -- of these proto-clusters.
\end{itemize}

\section{Conclusion}
\label{sec:conclusion}

We have presented $870\textrm{-}\micron{}$ imaging
obtained with LABOCA on APEX for a sample of 22
ultra-red galaxies -- 12 and 10 from the \textit{H}-ATLAS and {\it Her}MES
imaging surveys, respectively -- selected originally via their red \textit
{Herschel} 250-, 350-, $500\textrm{-}\mu\textrm{m}$ flux-density ratios.

Our survey covers an area of $\mathcal{A}\approx0.8\,\textrm{deg}^2$ down
to an average r.m.s.\ depth of
$\overline{\sigma}=3.9\,\millijanksy{}\,\beam{}^{-1}$. Running our
extraction algorithm at a $S/N$ detection threshold of
$\Sigma_ {\textrm{thresh}}>3.5$, we detect 86 field galaxies around our 22
ultra-red galaxies. We compute number counts and compare them to those
reported in a comparable survey, LESS \citep{weiss09b}. We report an
over-density factor (excluding our target ultra-red galaxies) of
$\delta=1.0^{+0.3}_{-0.3}$ at $S'>8.5\,\millijanksy{}$. There exists a
positive correlation between over-density and $870\textrm{-}\mu\textrm{m}$ flux
density, such that our sample of ultra-red galaxies traces dense regions,
rich in brighter DSFGs.

We perform photometry on SPIRE maps at the positions of our LABOCA
detections to derive photometric redshifts using three template SEDs. We
find that our ultra-red galaxy sample has a median redshift
$z_{1/2}=3.2\pm0.2$, with interquartile range $z=2.8\textrm{--}3.6$. We
associate the field galaxies likely responsible for this over-density to
within $|\Delta z|\le0.65$ of their signpost ultra-red galaxy. Over half of
our ultra-red galaxies have an average of two associated DSFGs within
$|\Delta z|\lesssim0.5$. Once these associated DSFGs have been removed, the
median redshift of the field galaxies decreases to $z_ {1/2}=2.3\pm0.1$, in
line with the general DSFG population. The majority of the associated DSFGs
are distributed on scales of
$\overline{\Delta r}_{\textrm {target}}\sim2\,\textrm{Mpc}$ from their
signpost galaxy and have high median SFRs,
$\psi_ {1/2}\approx1000\pm200\,\msol{}\,\textrm{yr}^{-1}$. We determine
average total SFRs of
$\overline{\Psi}=2200\pm500\,\msol{}\,\textrm{yr}^{-1}$ for those
systems with at least one associated DSFG. We derive gas masses for our
ultra-red galaxies and their associated DSFGs, determining average total
stellar masses of $M_{\textrm{stars}}\sim10^{11}\,M_ {\odot}$ for these
systems if they convert all of their gas into stars by $z\sim0$. We
determine an ultra-red galaxy proto-cluster space density of
$\rho_{\textrm{proto-cluster}}\sim9\times10^{-7}\,\textrm{Mpc}^{-3}$
between $2\lesssim z\lesssim6$,
which is similar to that of the most-massive
($M_{\textrm{DM}}\sim10^{15}\,\msol{}$) galaxy clusters at $z<0.2$
\citep{casey16, overzier16, bahcall93}. It therefore seems plausible that these
systems of DSFGs may evolve into the massive ETGs which populate the
centers of rich galaxy clusters at $z=0$.

We have increased the number of potential distant, DSFG proto-clusters
using our novel signposting technique, based on ultra-red SPIRE flux-density
ratios. With deep optical imaging/spectroscopy of these environments, we
will be able to better determine their ultimate stellar masses -- and thus
DM properties, enabling us to predict the eventual fate of these systems.

Our catalogs and $870\textrm{-}\mu\textrm{m}$ images form part of a formal
data release.

\section*{Acknowledgments}

AJRL, RJI, JMS, IO, LD, VA and ZYZ acknowledge support from the European Research Council (ERC) in the form of Advanced Grant, 321302, \textsc{cosmicism}.

H.D. acknowledges financial support from the Spanish Ministry of Economy and Competitiveness (MINECO) under the 2014 Ram\'{o}n y Cajal program MINECO RYC-2014-15686.
JLW acknowledges support from an STFC Ernest Rutherford Fellowship.
D.R. acknowledges support from the National Science Foundation under grant number AST-1614213.
GDZ acknowledges support from ASI/INAF agreement n.~2014-024-R.1.
We pay special thanks to the useful feedback provided on the draft version of this work to D.~Farrah, J.~Greenslade, M.\,J.~Michalowski and I.~Valtchanov.
This research has made use of data from {\it Her}MES project
(\url{http://hermes.sussex.ac.uk/}). {\it Her}MES is a {\it Herschel} Key
Programme utilizing Guaranteed Time from the SPIRE instrument team, ESAC
scientists and a mission scientist.
The \textit{H}-ATLAS is a project with
\textit{Herschel}, which is an ESA space observatory with science
instruments provided by European-led Principal Investigator consortia and
with important participation from NASA. The \textit{H}-ATLAS website is
\url{www.h-atlas.org}. US participants in \textit{H}-ATLAS acknowledge
support from NASA through a contract from JPL.
This publication is based on
data acquired with the Atacama Pathfinder Experiment (APEX). APEX is a
collaboration between the Max-Planck-Institut fur Radioastronomie, the
European Southern Observatory, and the Onsala Space Observatory. Based on
observations made with APEX under European Southern Observatory program
E-191.A-0748 and Max Planck Institute (MPI) programs M-090.F-0025-2012,
M-091.F-0021-2013 and M-092.F-0015-2013.
\smallskip

\textit{Facilities}: \facility{APEX}, \facility{\textit{Herschel}}.

\bibliographystyle{apj}
\bibliography{ajrl}

\appendix

\section{LABOCA and SPIRE maps}
\label{sec:maps}

Here we present our LABOCA and \textit{Herschel} imaging.

\begin{figure*}
    \centering
    \begin{center}
        \includegraphics[height=0.22\textwidth]{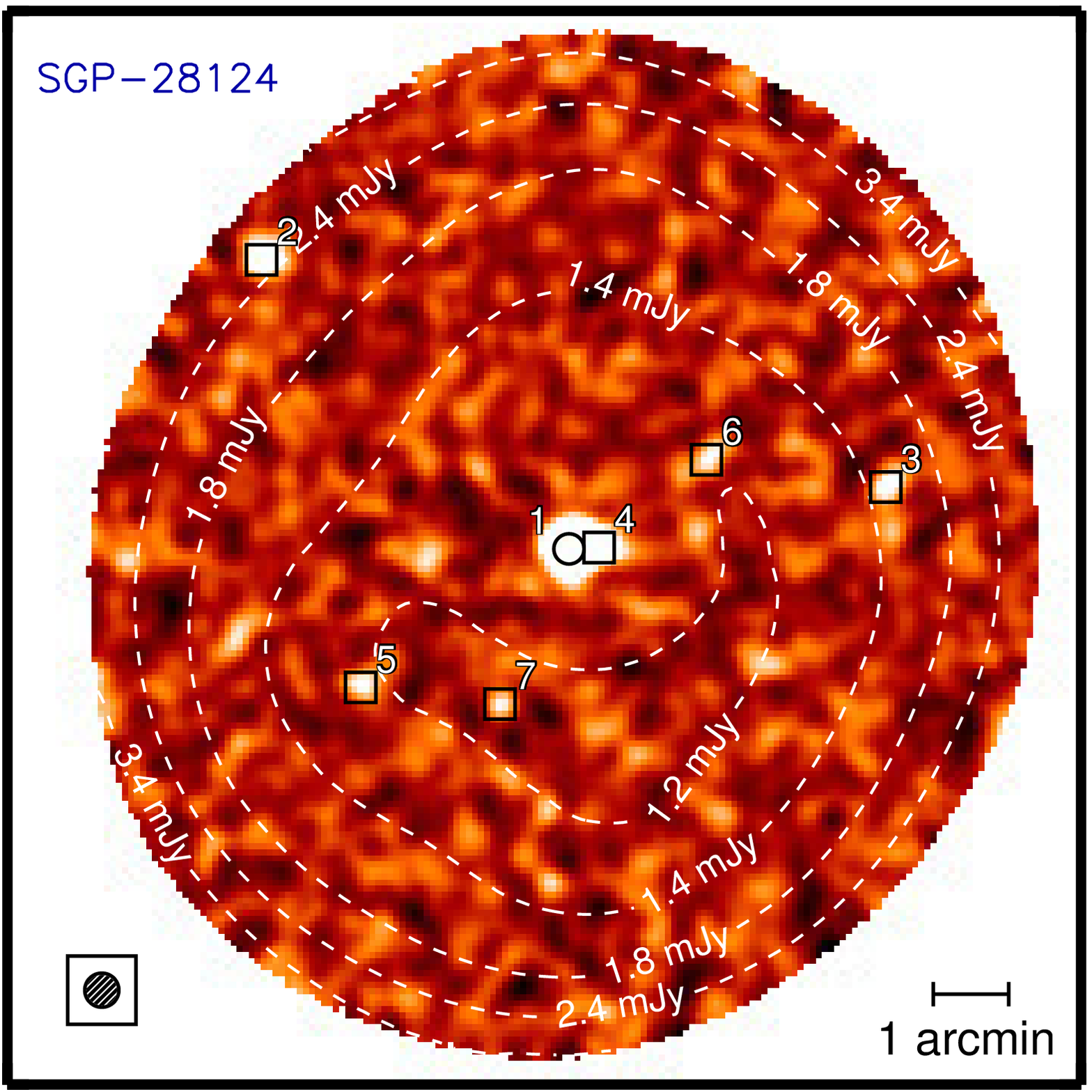}\hspace{-0.3em}
        \includegraphics[height=0.22\textwidth]{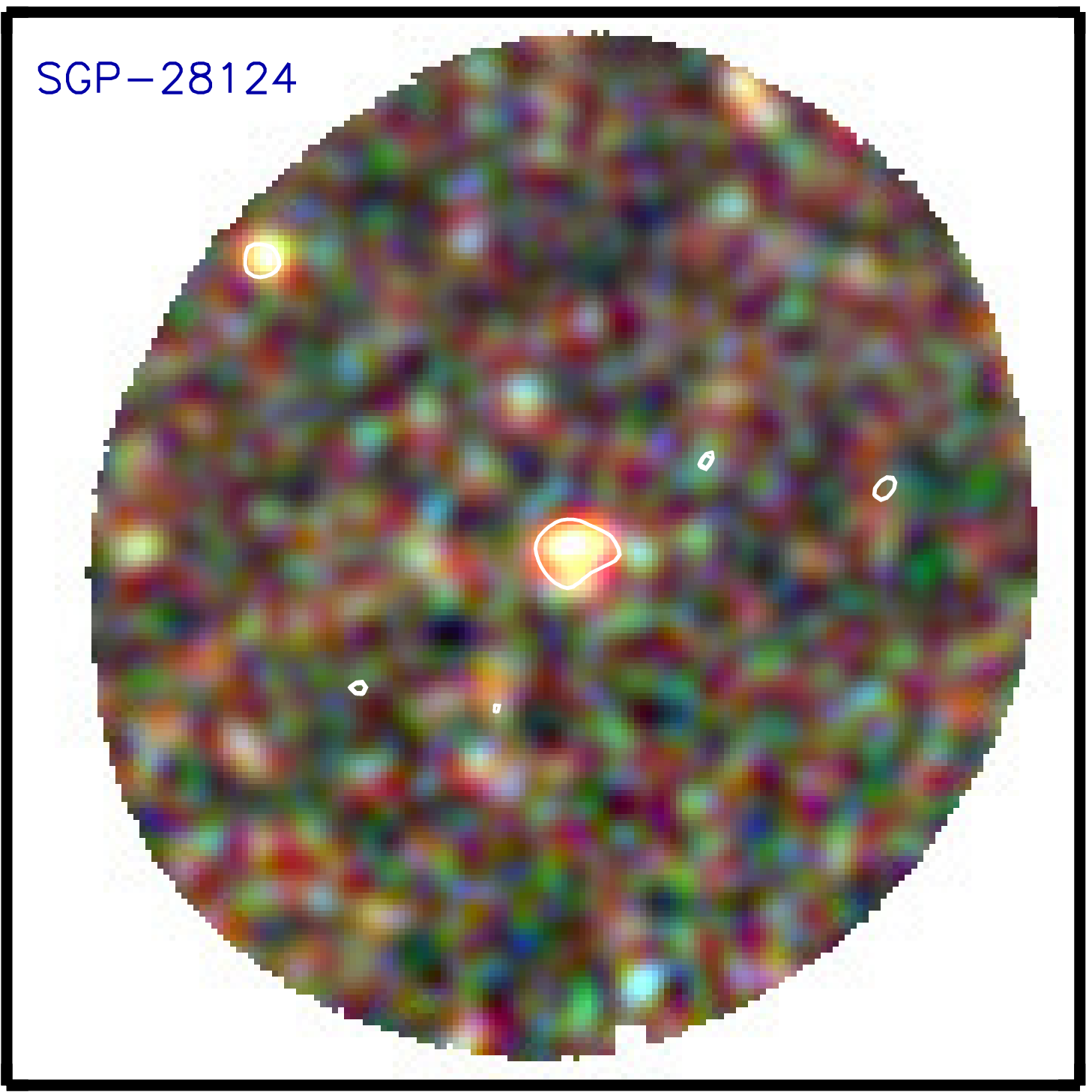}\hspace{1em}
        \includegraphics[height=0.22\textwidth]{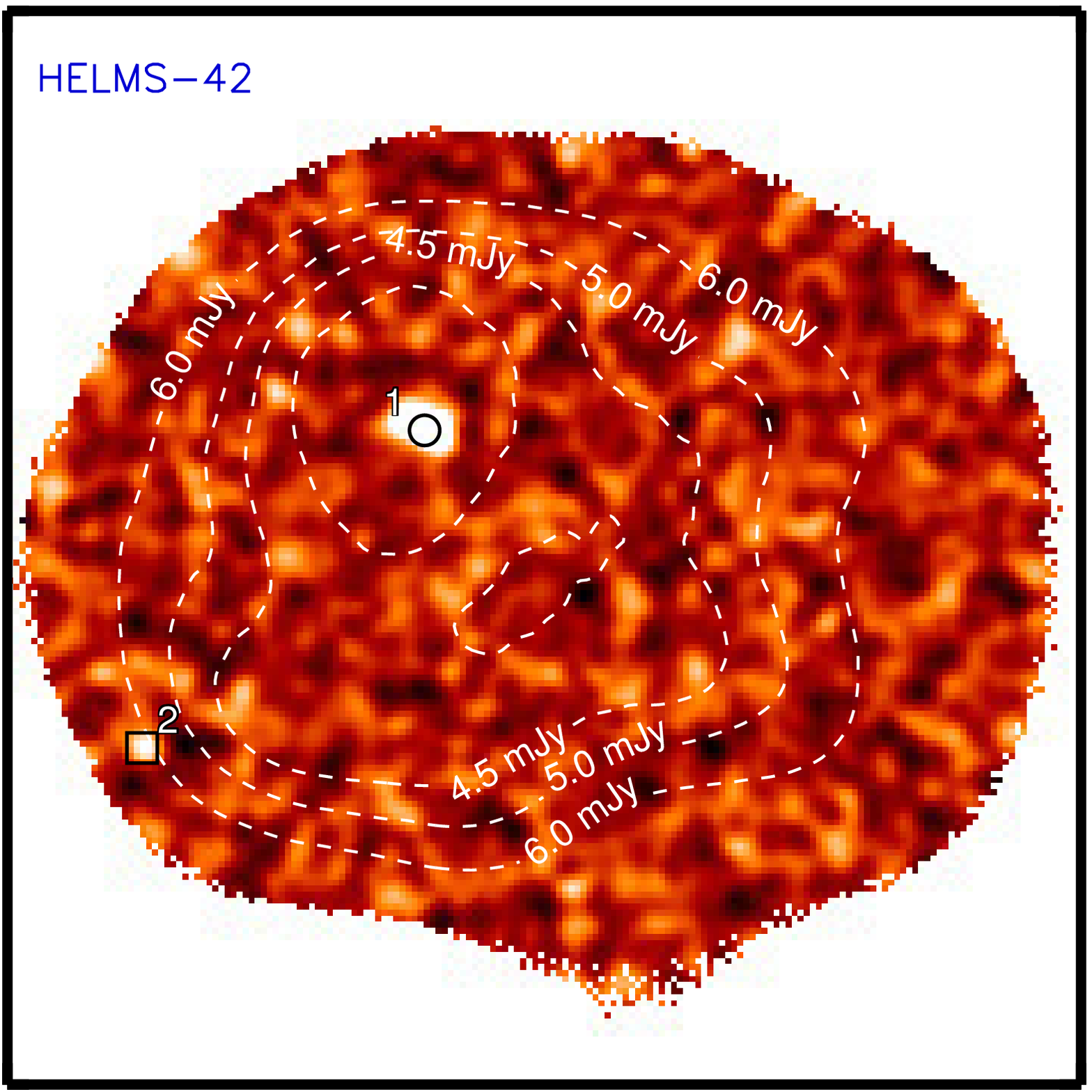}\hspace{-0.3em}
        \includegraphics[height=0.22\textwidth]{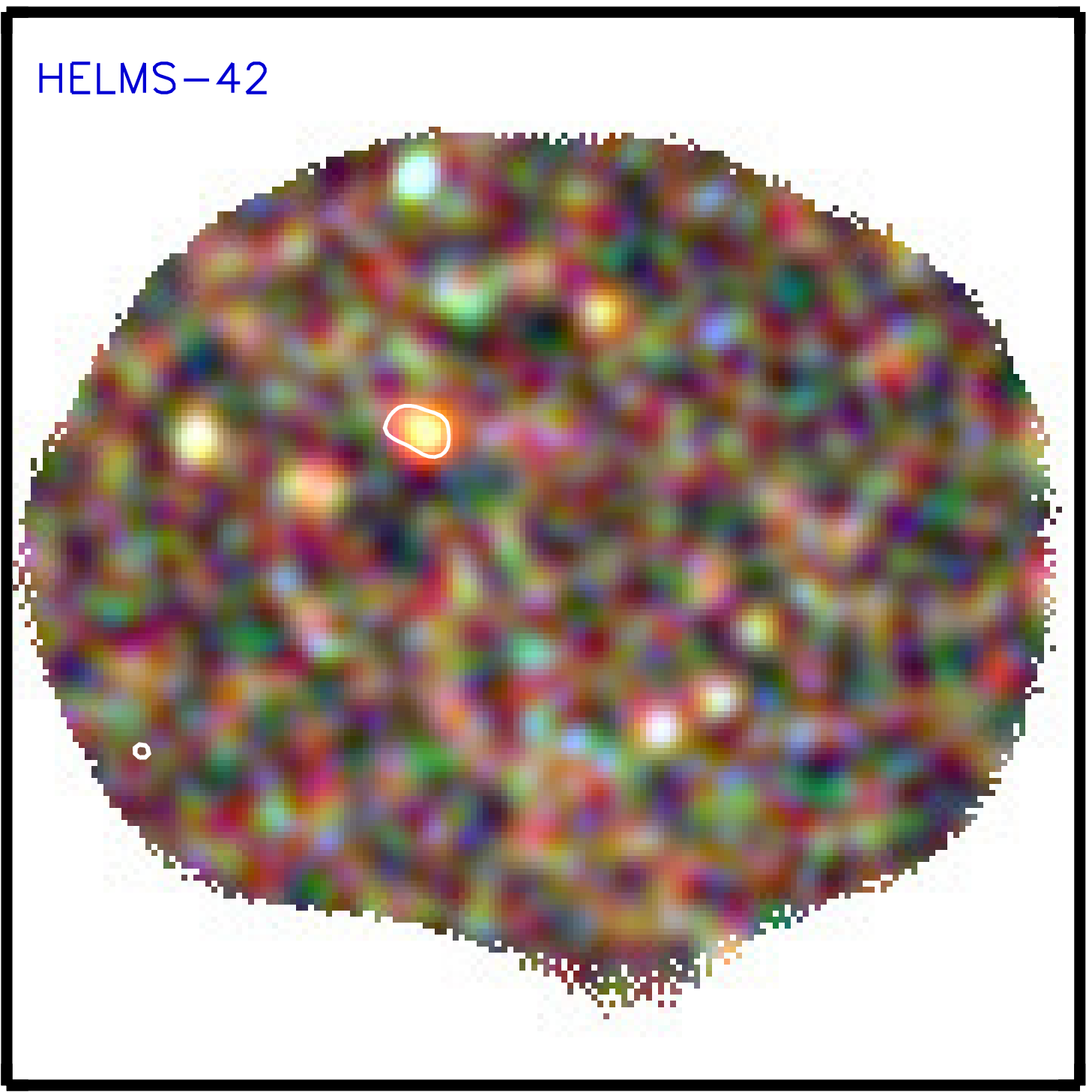}\\\vspace{0.1em}
        \includegraphics[height=0.22\textwidth]{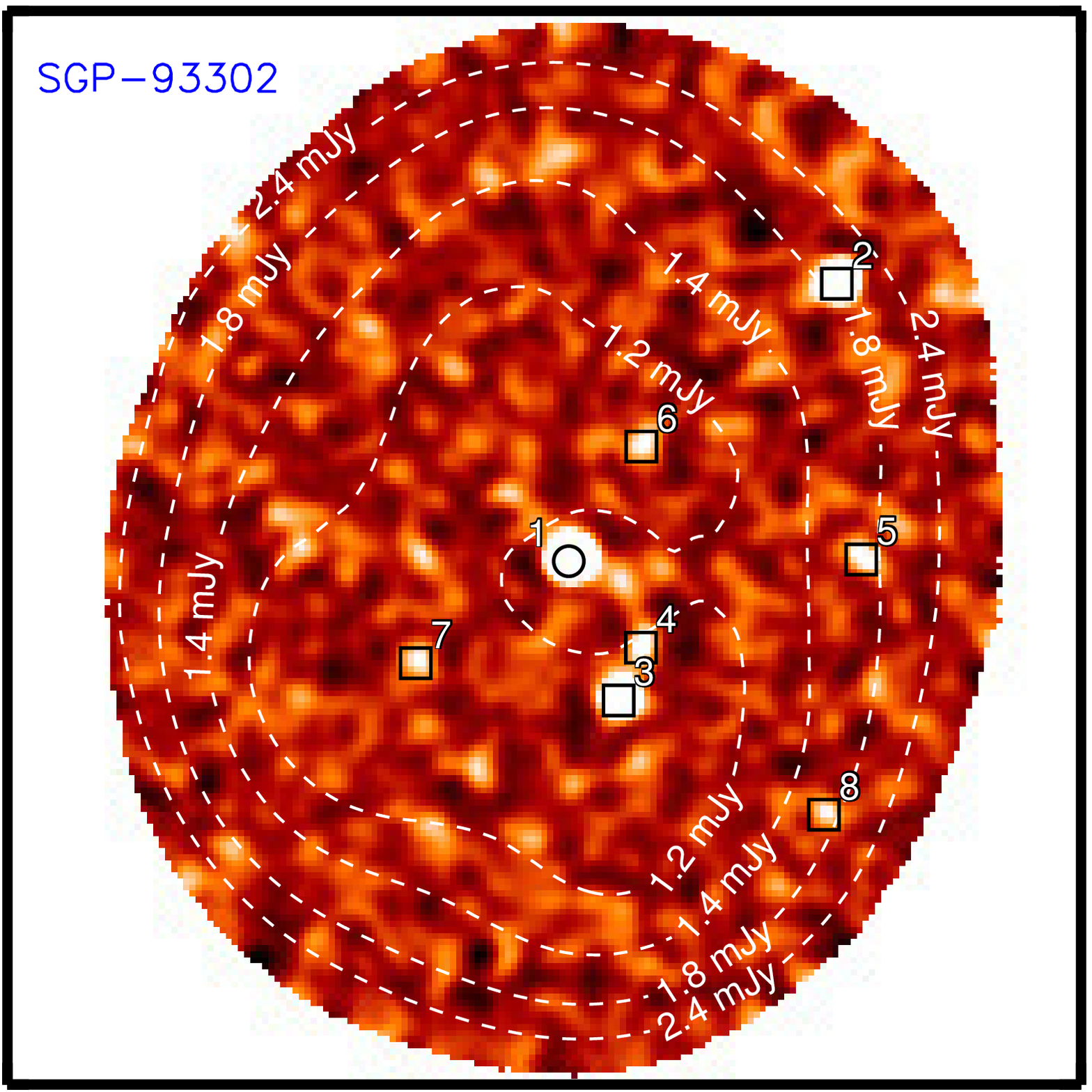}\hspace{-0.3em}
        \includegraphics[height=0.22\textwidth]{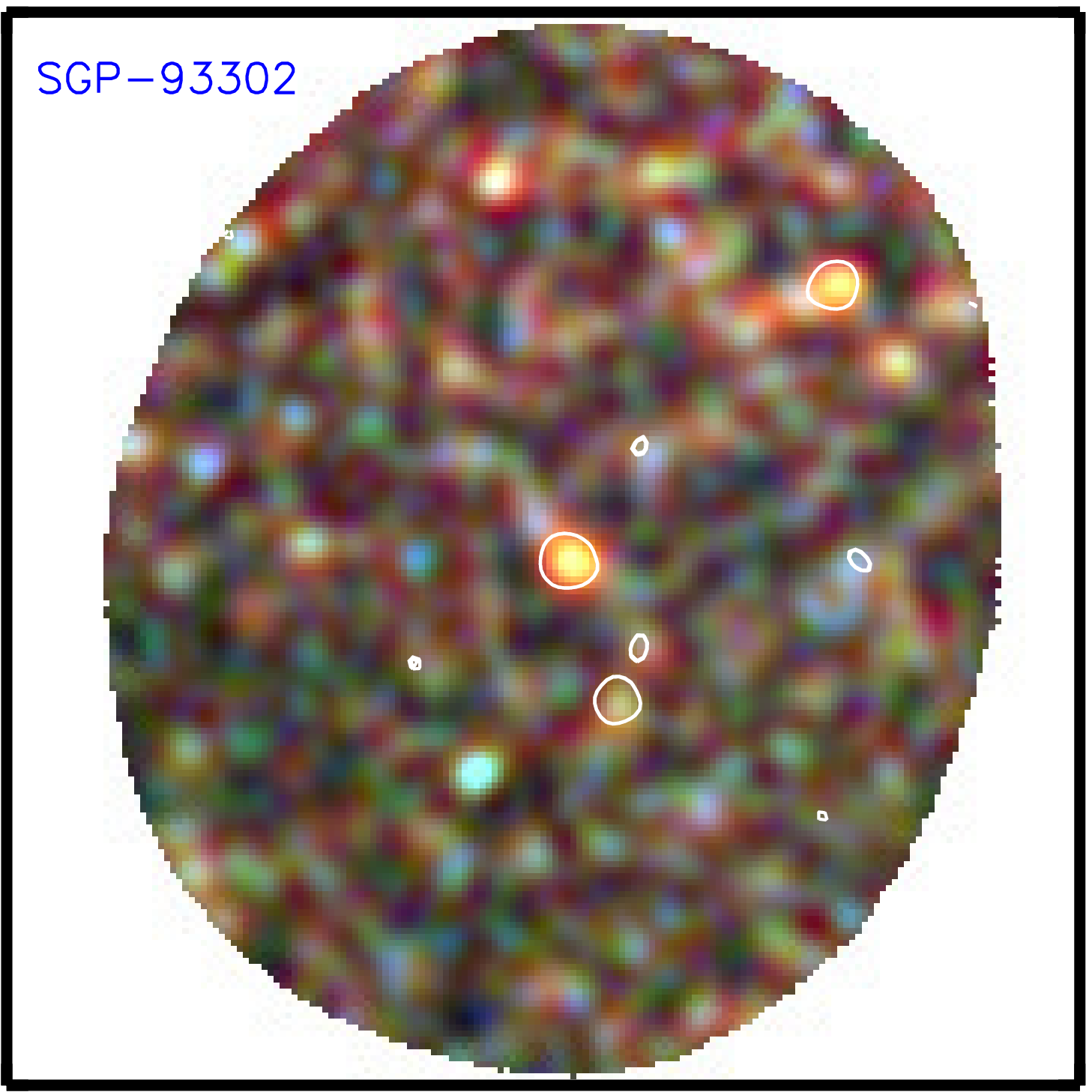}\hspace{1em}
        \includegraphics[height=0.22\textwidth]{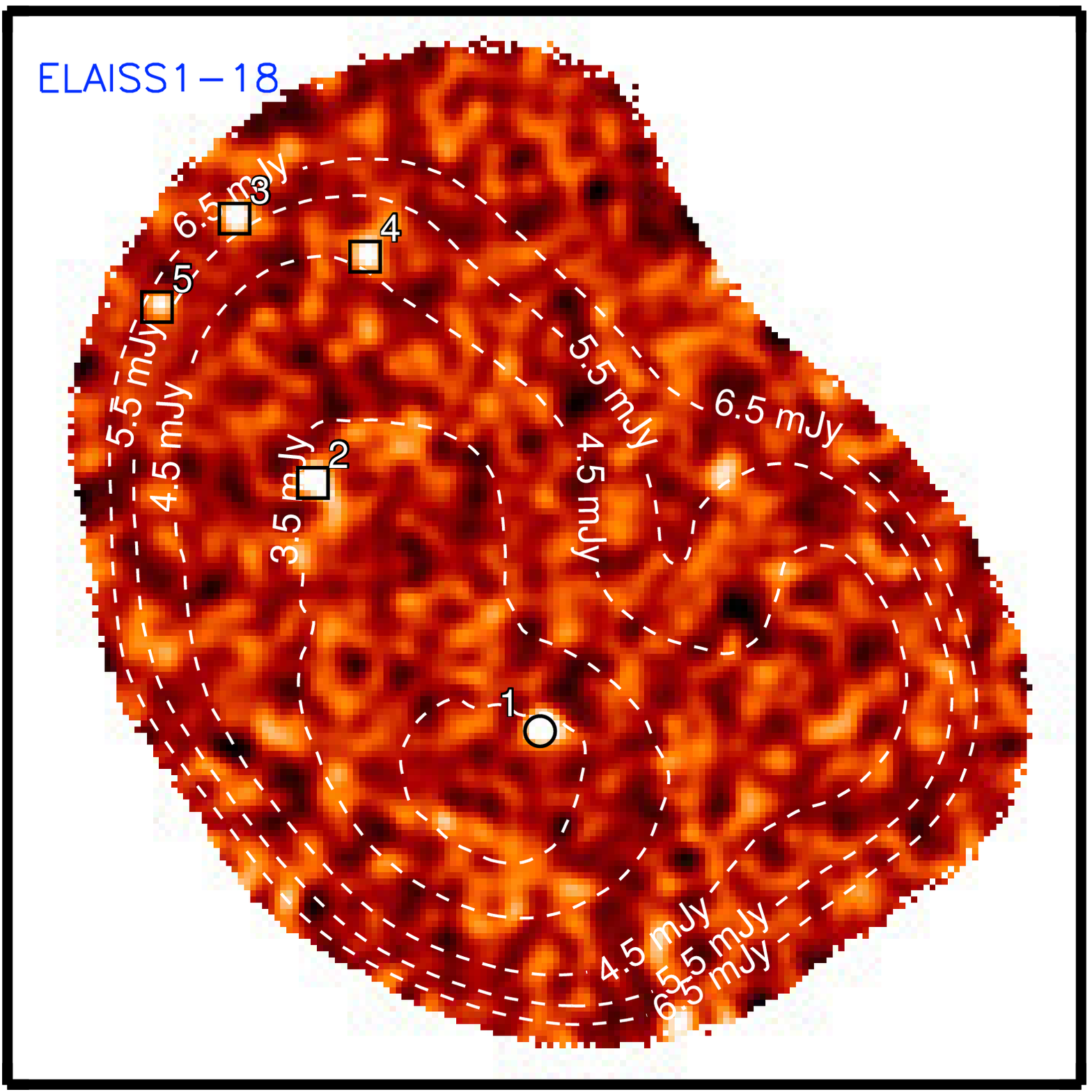}\hspace{-0.3em}
        \includegraphics[height=0.22\textwidth]{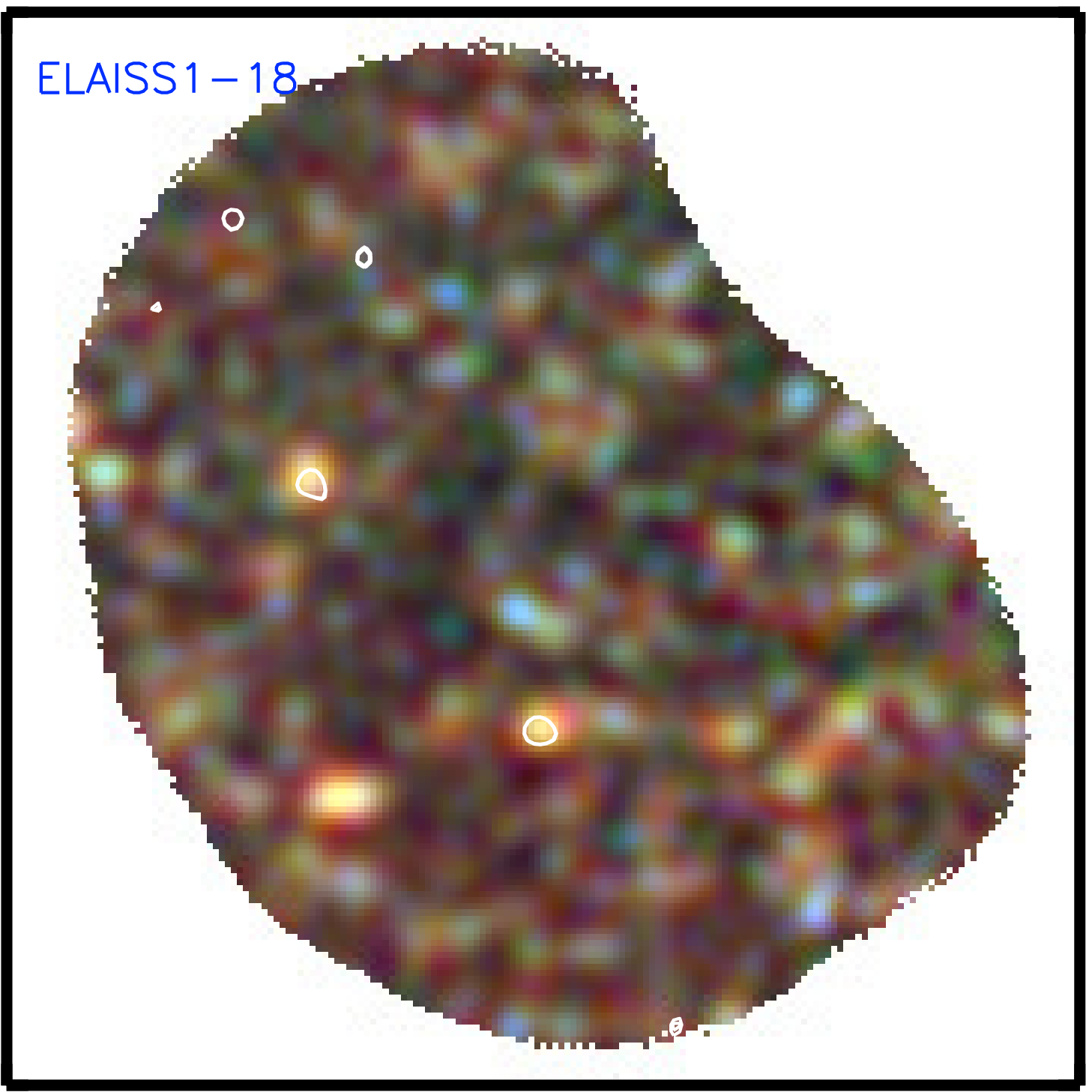}\\\vspace{0.1em}
        \includegraphics[height=0.22\textwidth]{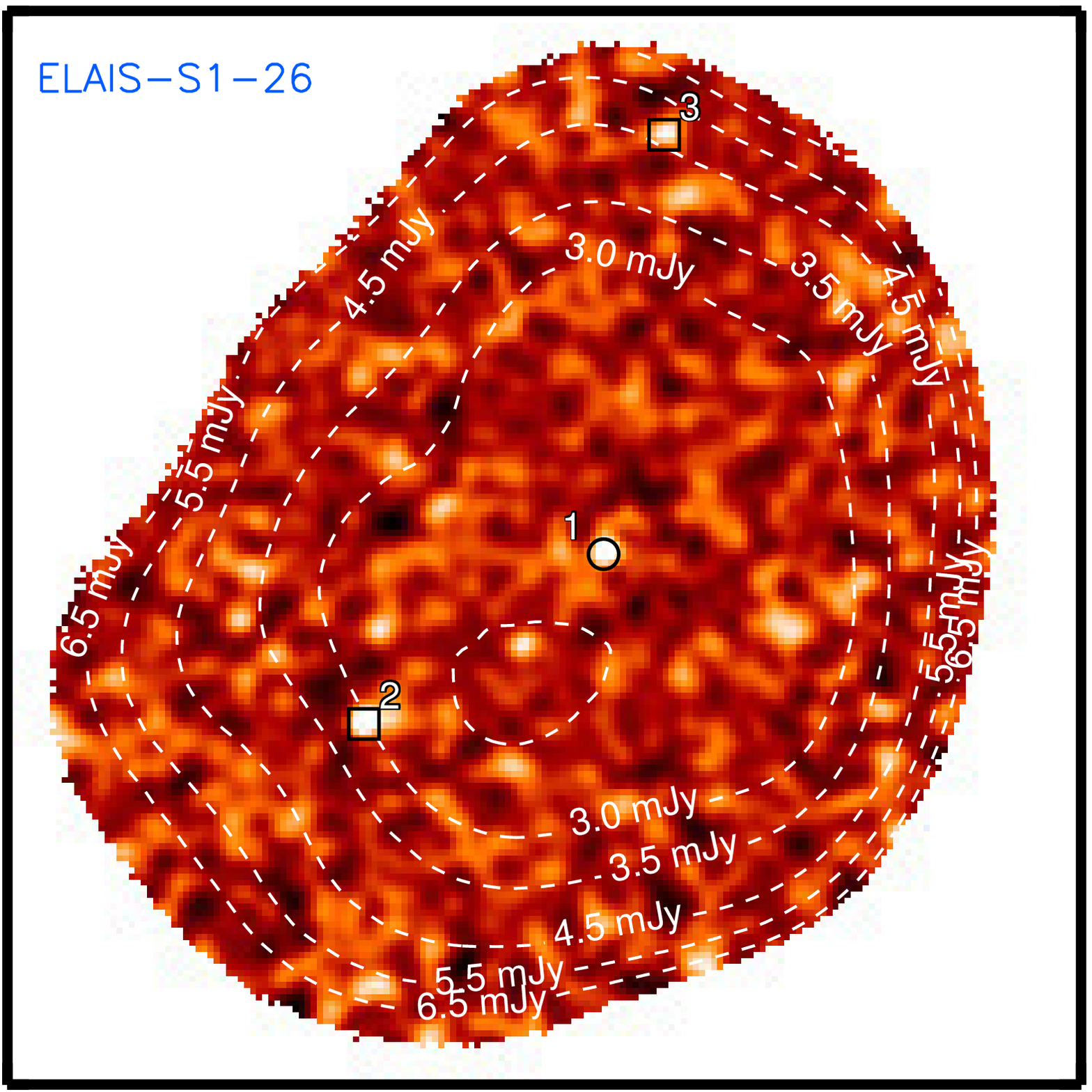}\hspace{-0.3em}
        \includegraphics[height=0.22\textwidth]{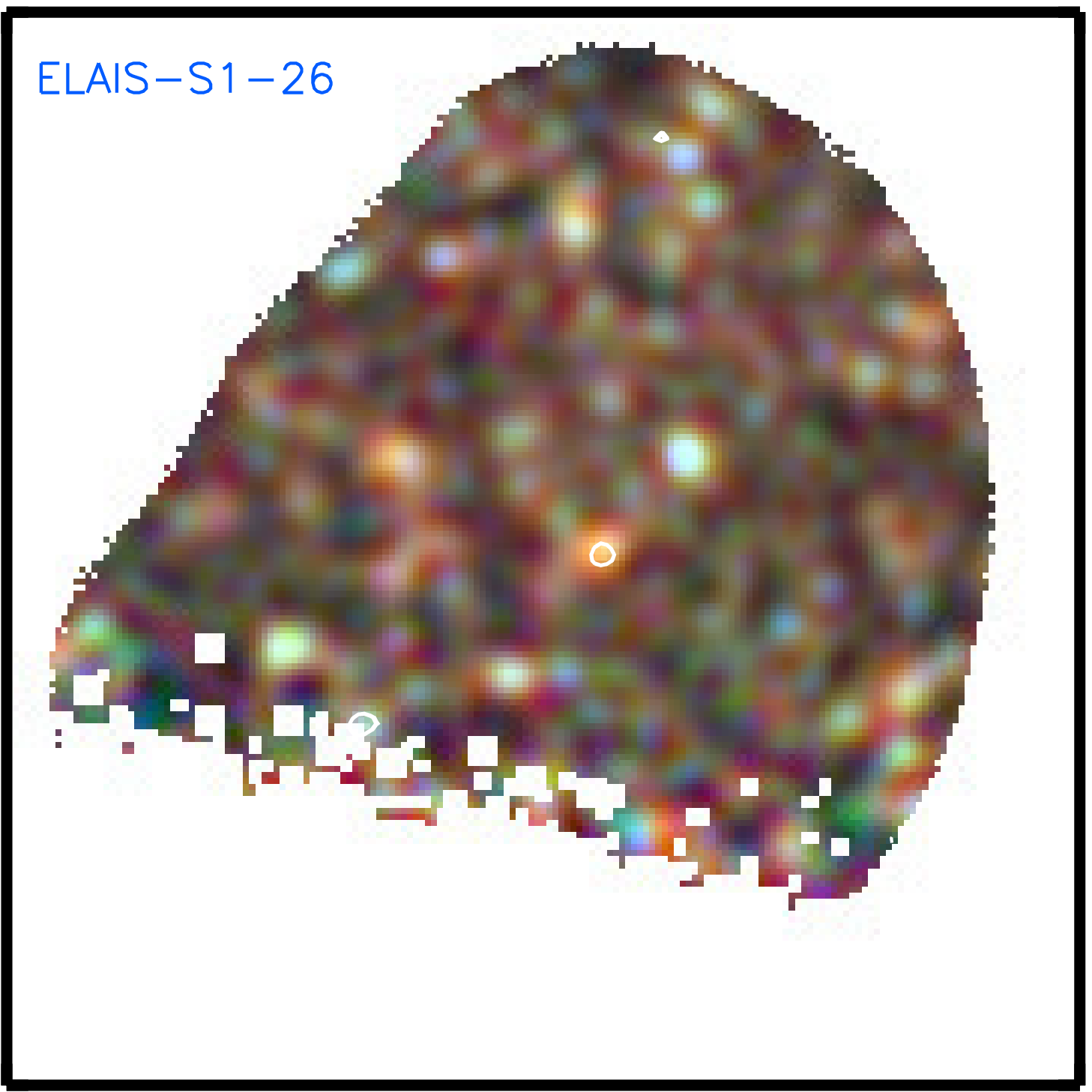}\hspace{1em}
        \includegraphics[height=0.22\textwidth]{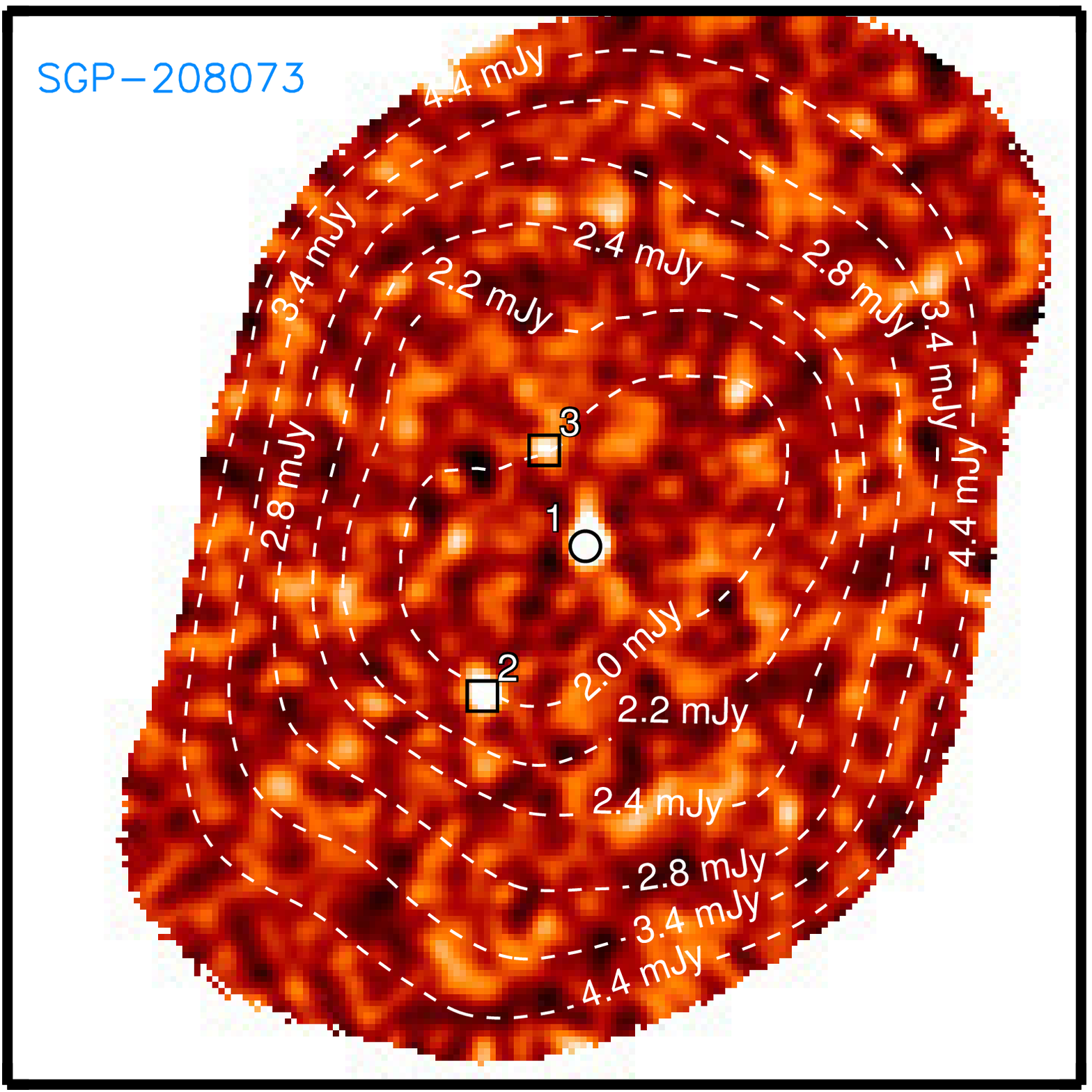}\hspace{-0.3em}
        \includegraphics[height=0.22\textwidth]{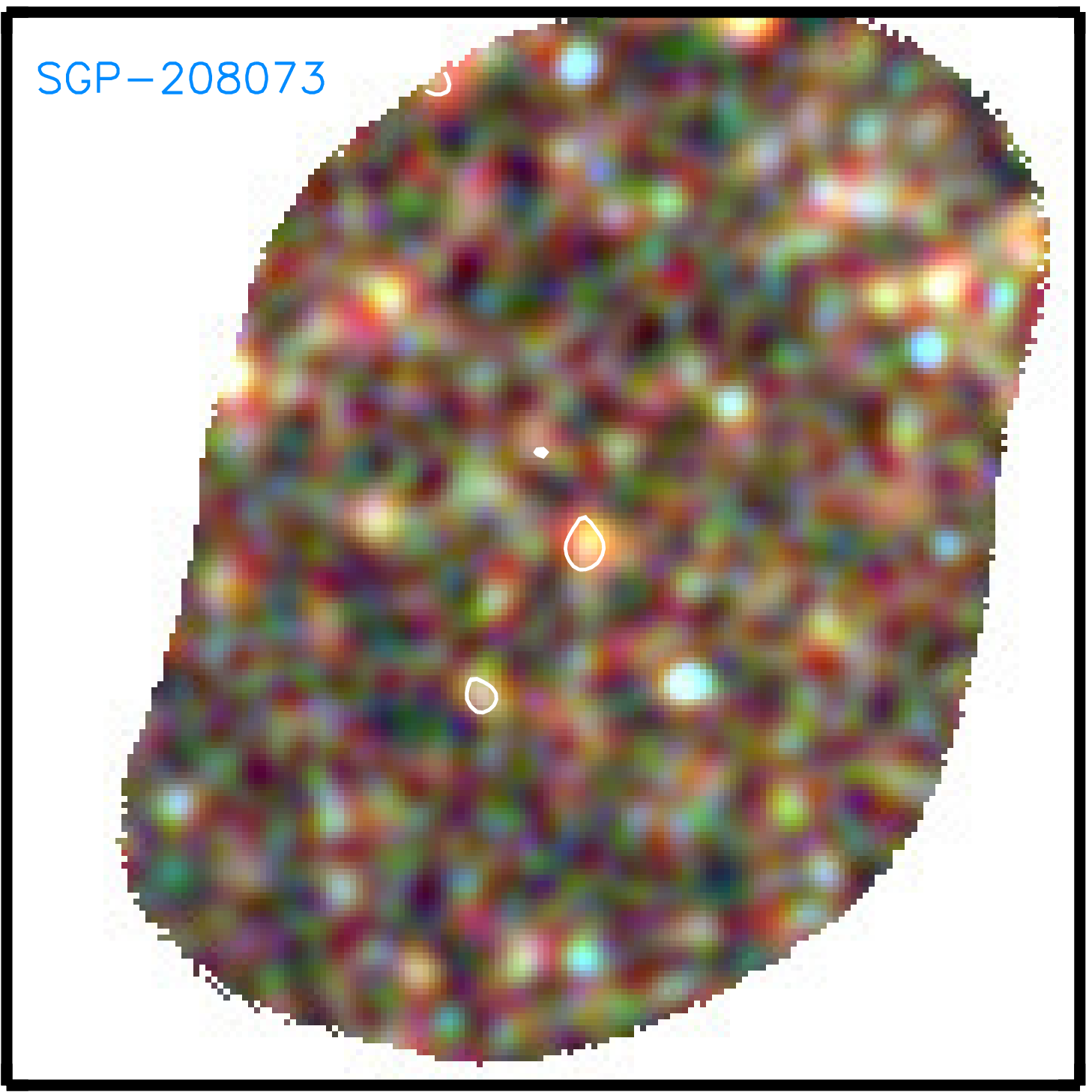}\\\vspace{0.1em}
        \includegraphics[height=0.22\textwidth]{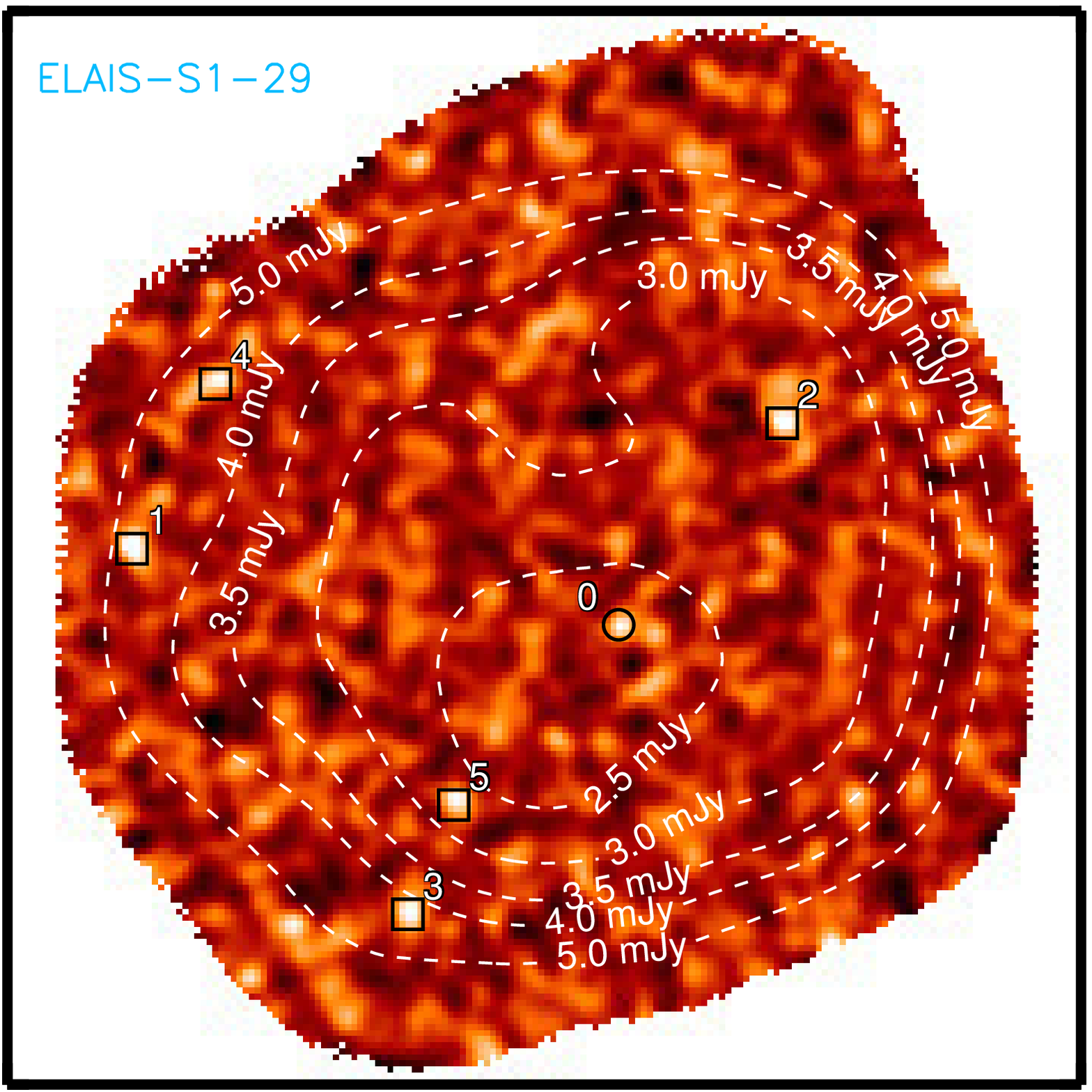}\hspace{-0.3em}
        \includegraphics[height=0.22\textwidth]{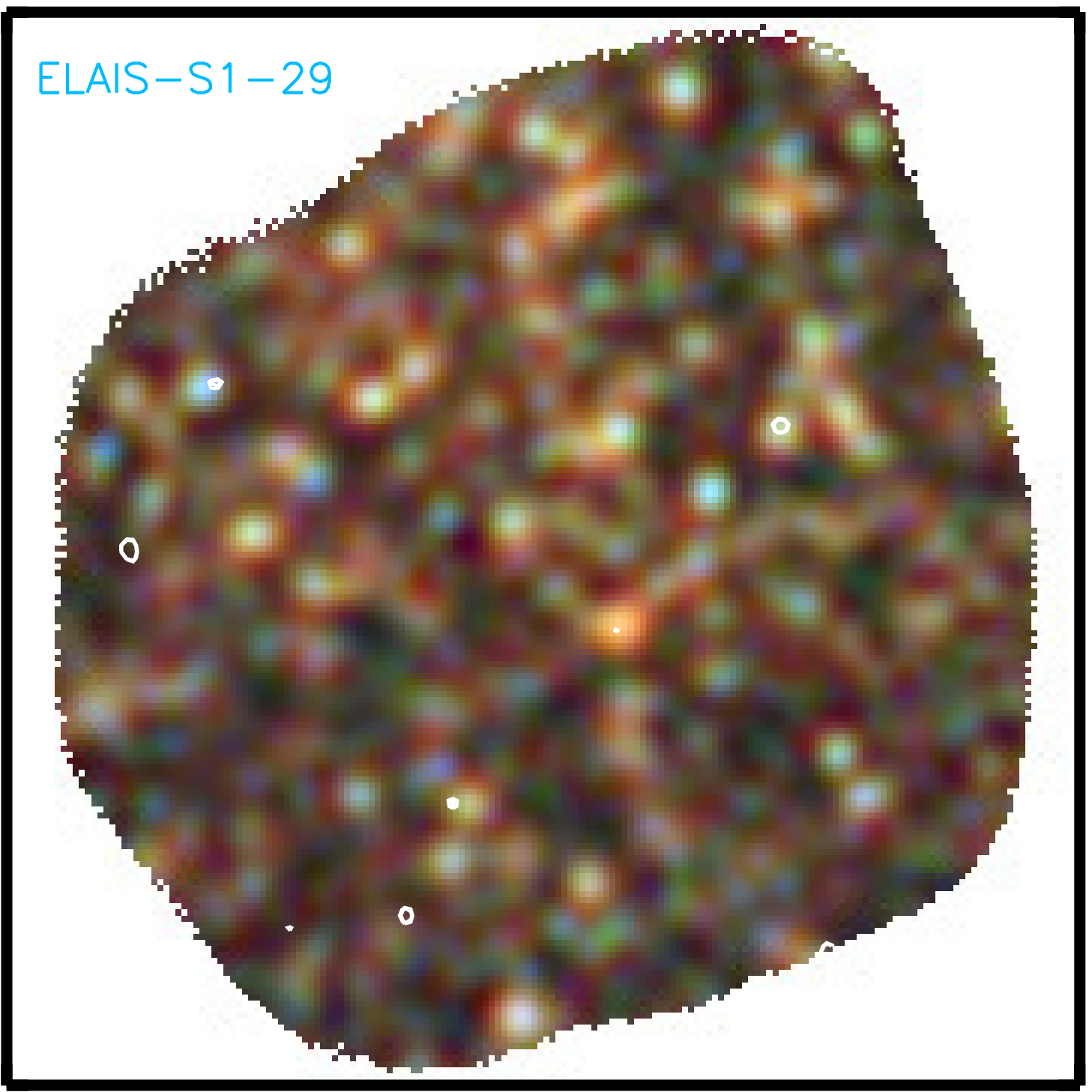}\hspace{1em}
        \includegraphics[height=0.22\textwidth]{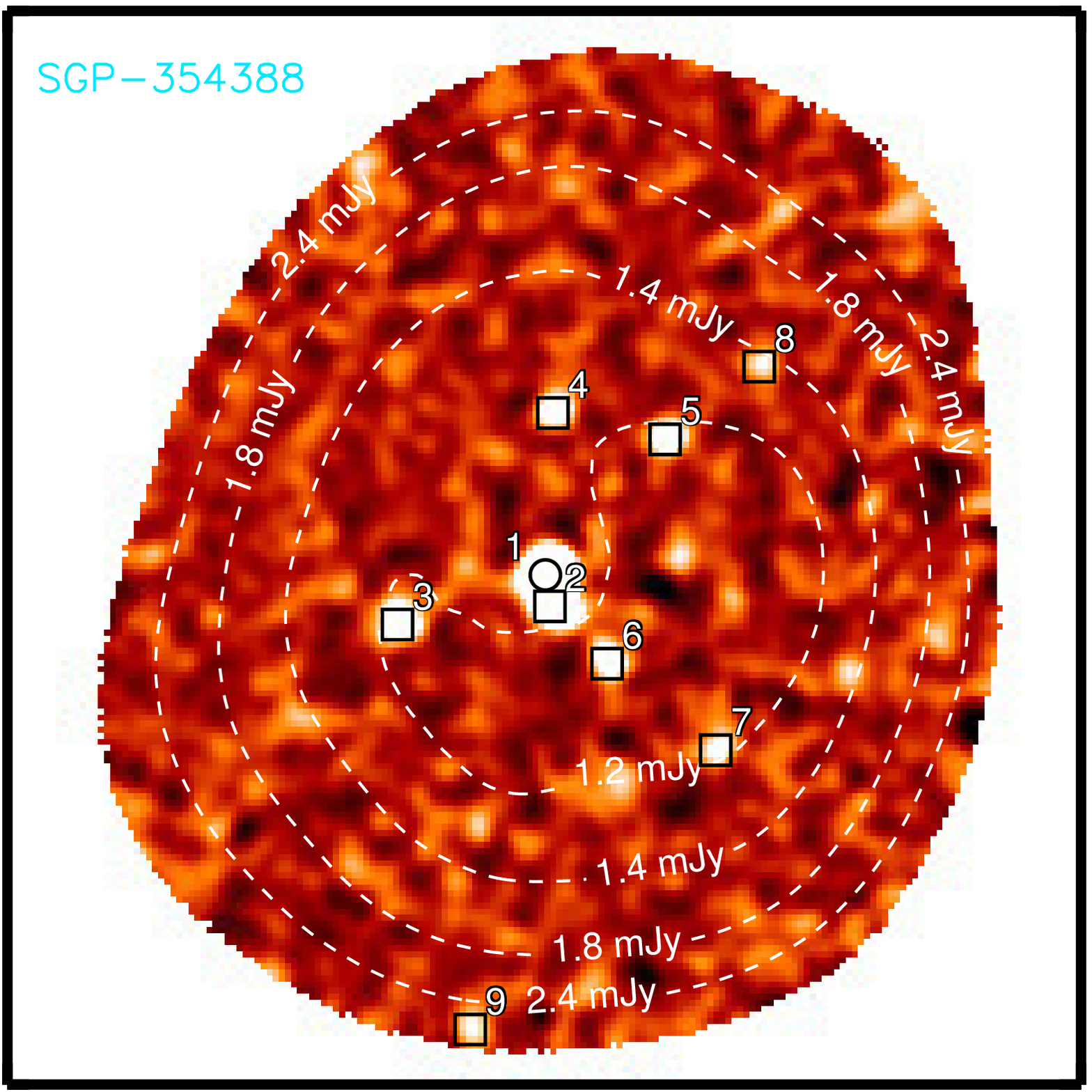}\hspace{-0.3em}
        \includegraphics[height=0.22\textwidth]{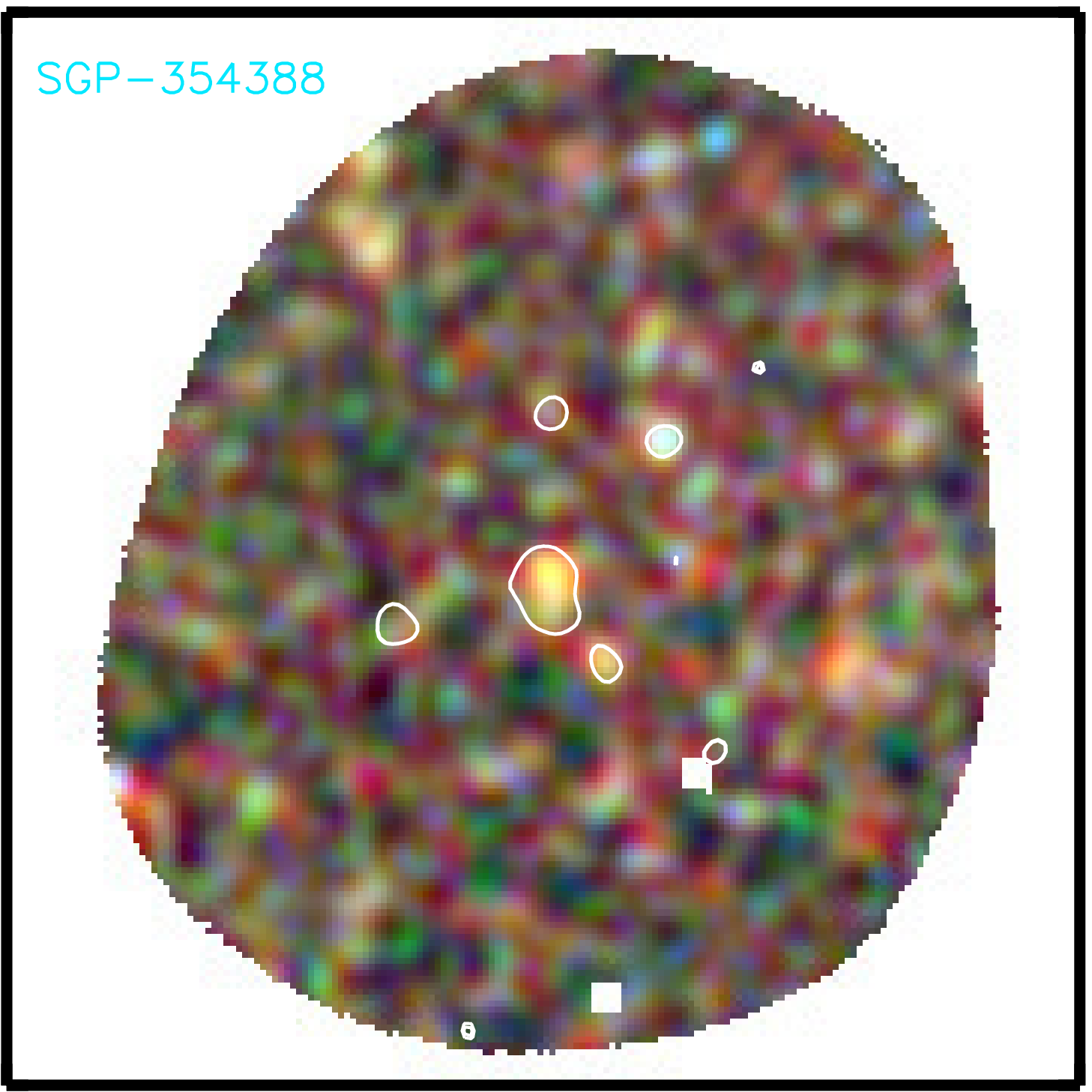}\\\vspace{0.1em}
        \includegraphics[height=0.22\textwidth]{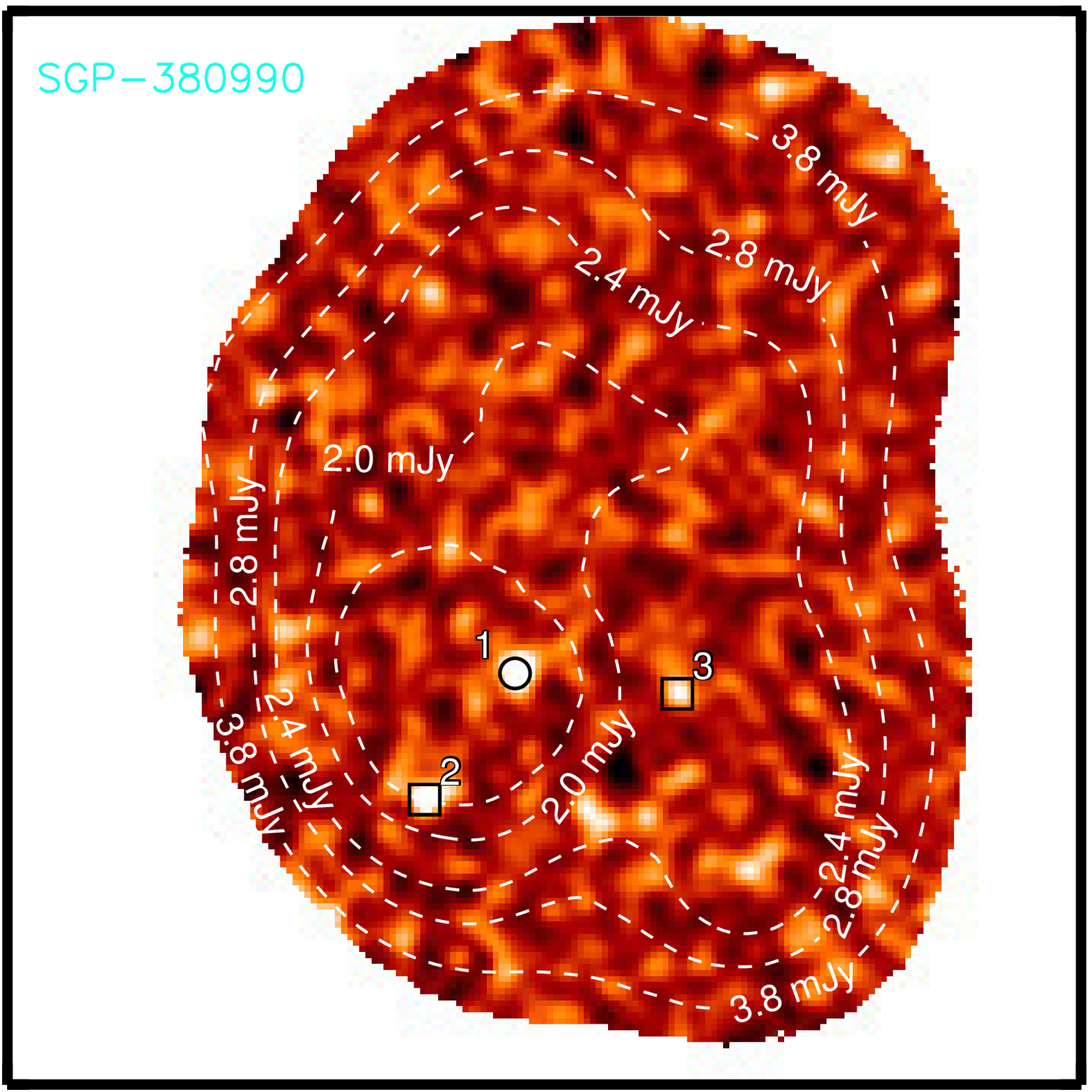}\hspace{-0.3em}
        \includegraphics[height=0.22\textwidth]{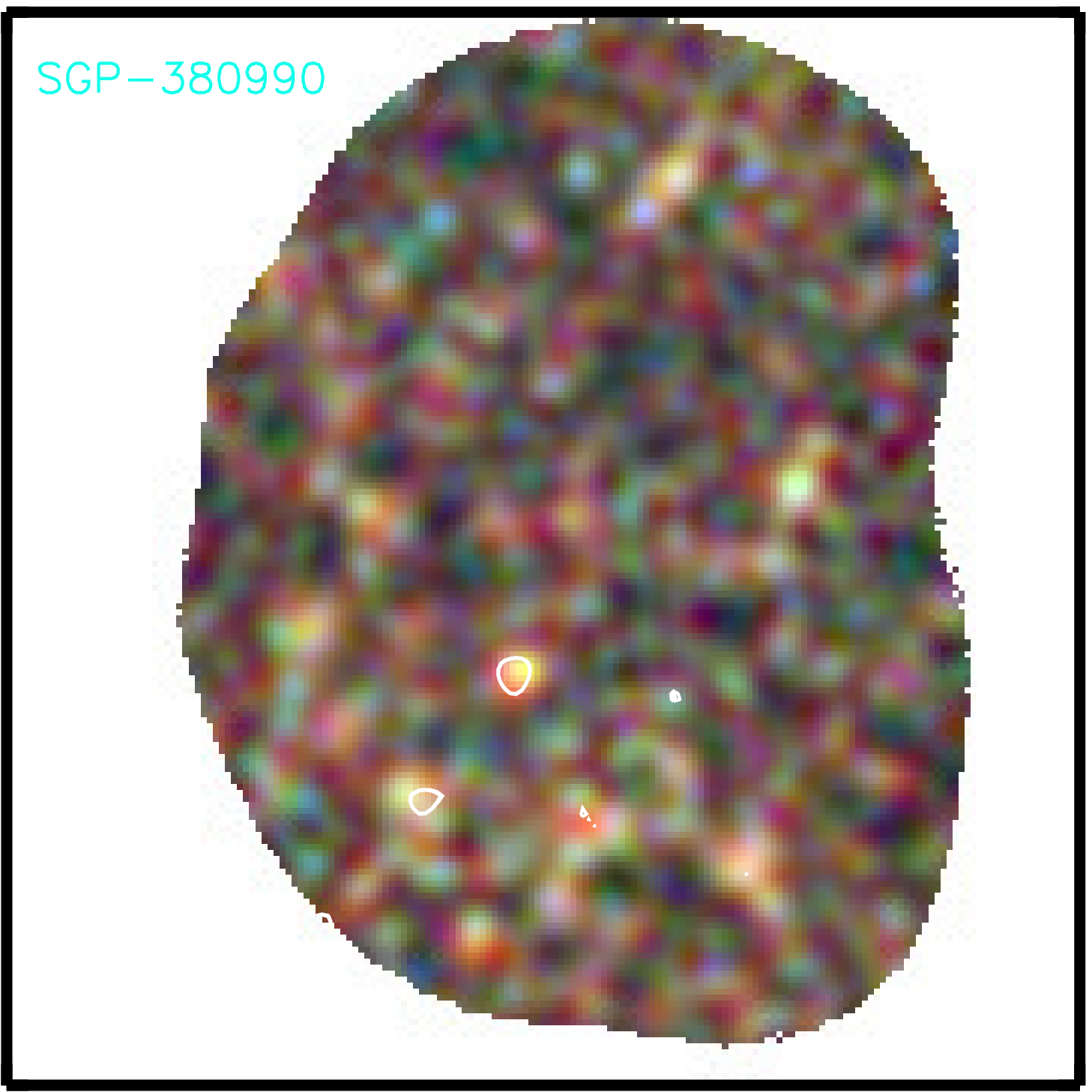}\hspace{1em}
        \includegraphics[height=0.22\textwidth]{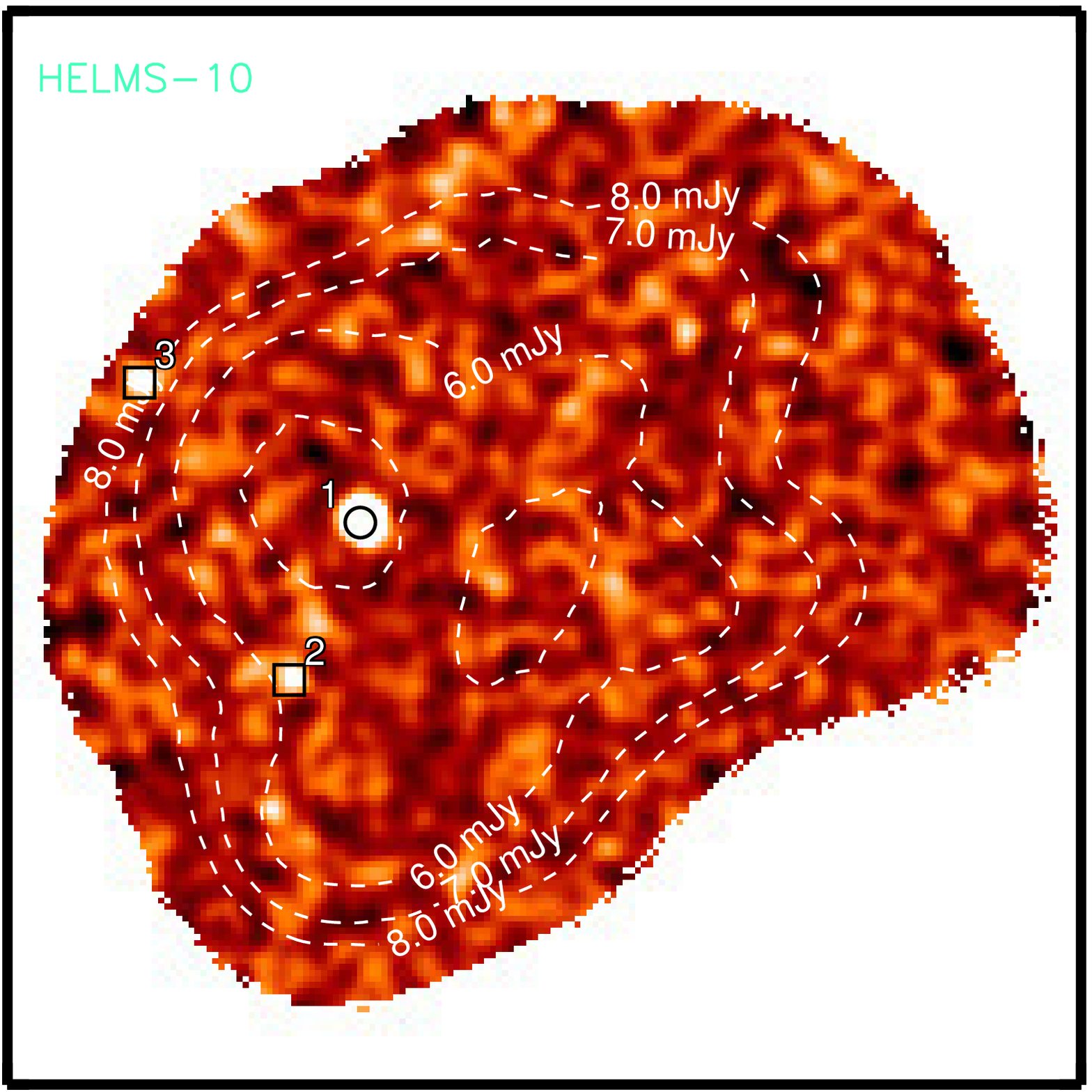}\hspace{-0.3em}
        \includegraphics[height=0.22\textwidth]{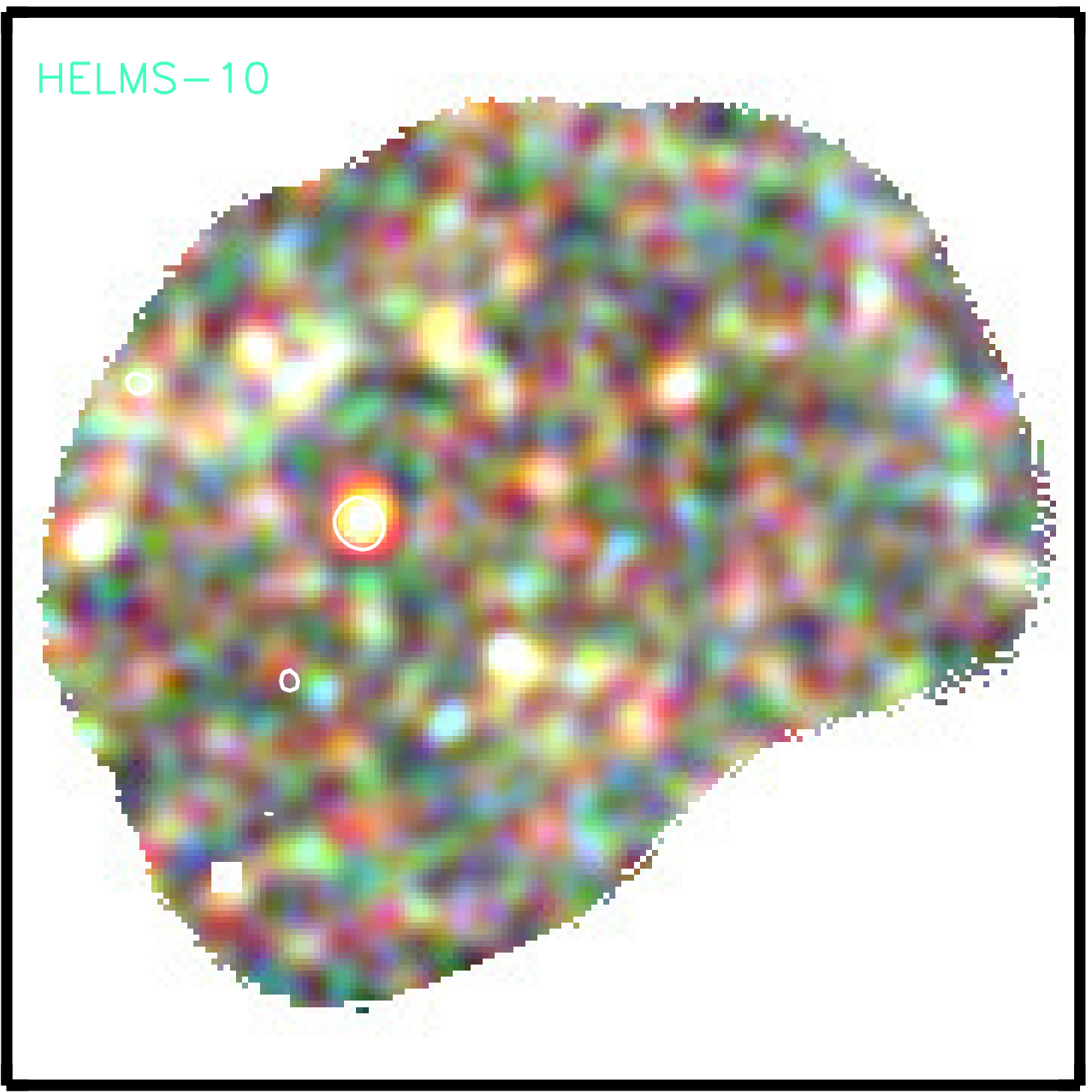}\\\vspace{0.1em}
        \caption{\textit{Left}: $14'\times14'$ cut-outs of our LABOCA $S/N$
          maps at a spatial resolution of $\approx27''$, stretched linearly
          between $\pm3.5\sigma$ (see beam inset and scale on top-left panel).
          North is up; East is left. Detections
          above $\Sigma_{\textrm{thresh}}=3.5$ are numbered in decreasing
          order of $S/N$ with hollow squares and stars representing
          signpost and field galaxies, respectively. Signposts numbered `0'
          are sources that we have been unable to detect above our
          $3.5\sigma$ threshold. We place dashed white contours at varying
          values of map noise. We show an arcminute scale and a LABOCA beam
          on the top row. \textit{Right}: false-color, matched-filtered
          \textit{Herschel} SPIRE $14'\times14'$ cut-out images, aligned
          with their LABOCA counterparts, which we use to measure the SPIRE
          photometry. White dashed contours are placed at LABOCA
          $3.5\sigma$ values. \textbf{Note}. Maps are presented in
          increasing order of right ascension, i.e.\ in
    the same order as they appear in Table~\ref{tab: photometry}, and their labels
    have been color-coded from blue to red.}
    \label{fig: reduced maps}
    \end{center}
\end{figure*}

\addtocounter{figure}{-1} 
\begin{figure*}
    \centering
    \begin{center}
        \includegraphics[height=0.22\textwidth]{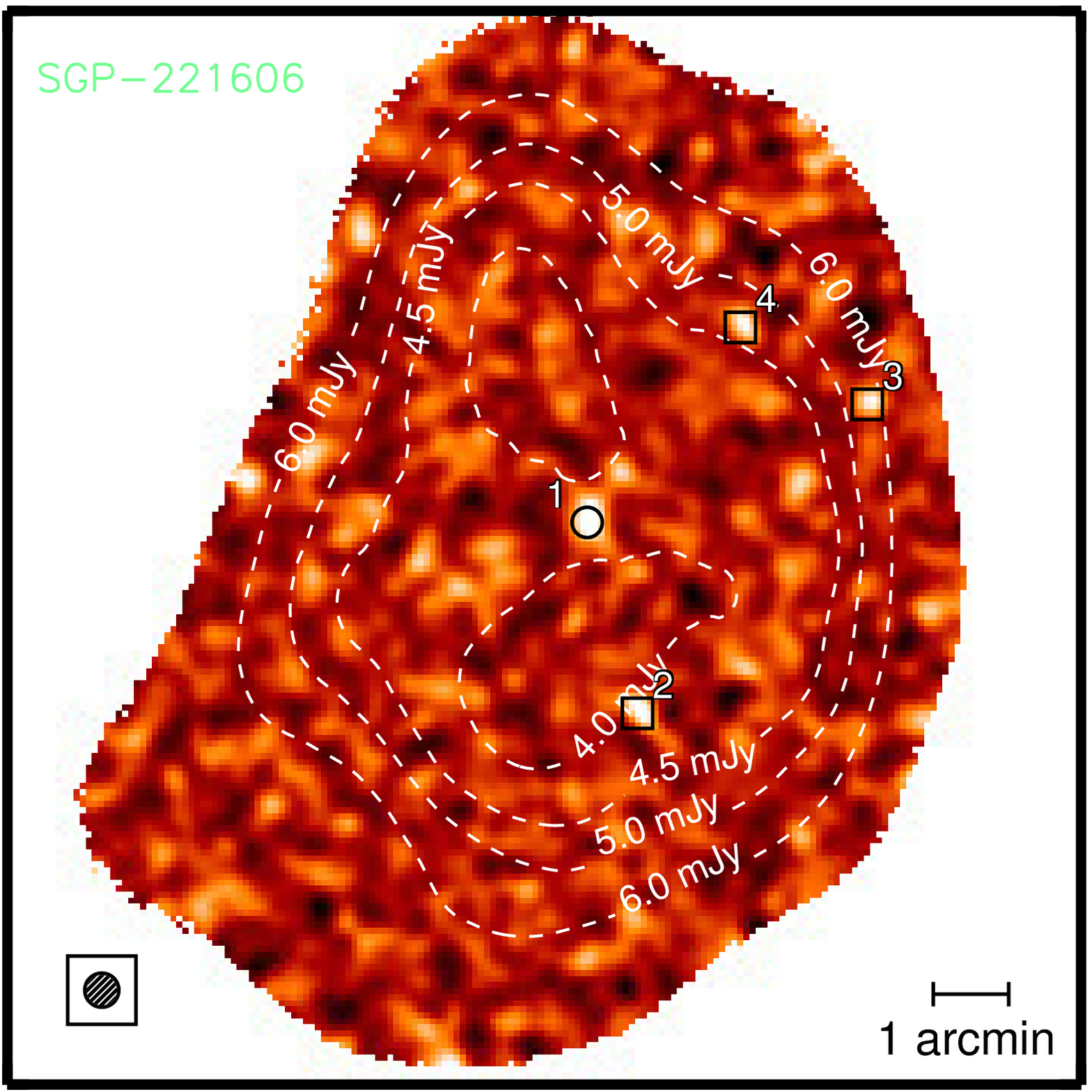}\hspace{-0.3em}
        \includegraphics[height=0.22\textwidth]{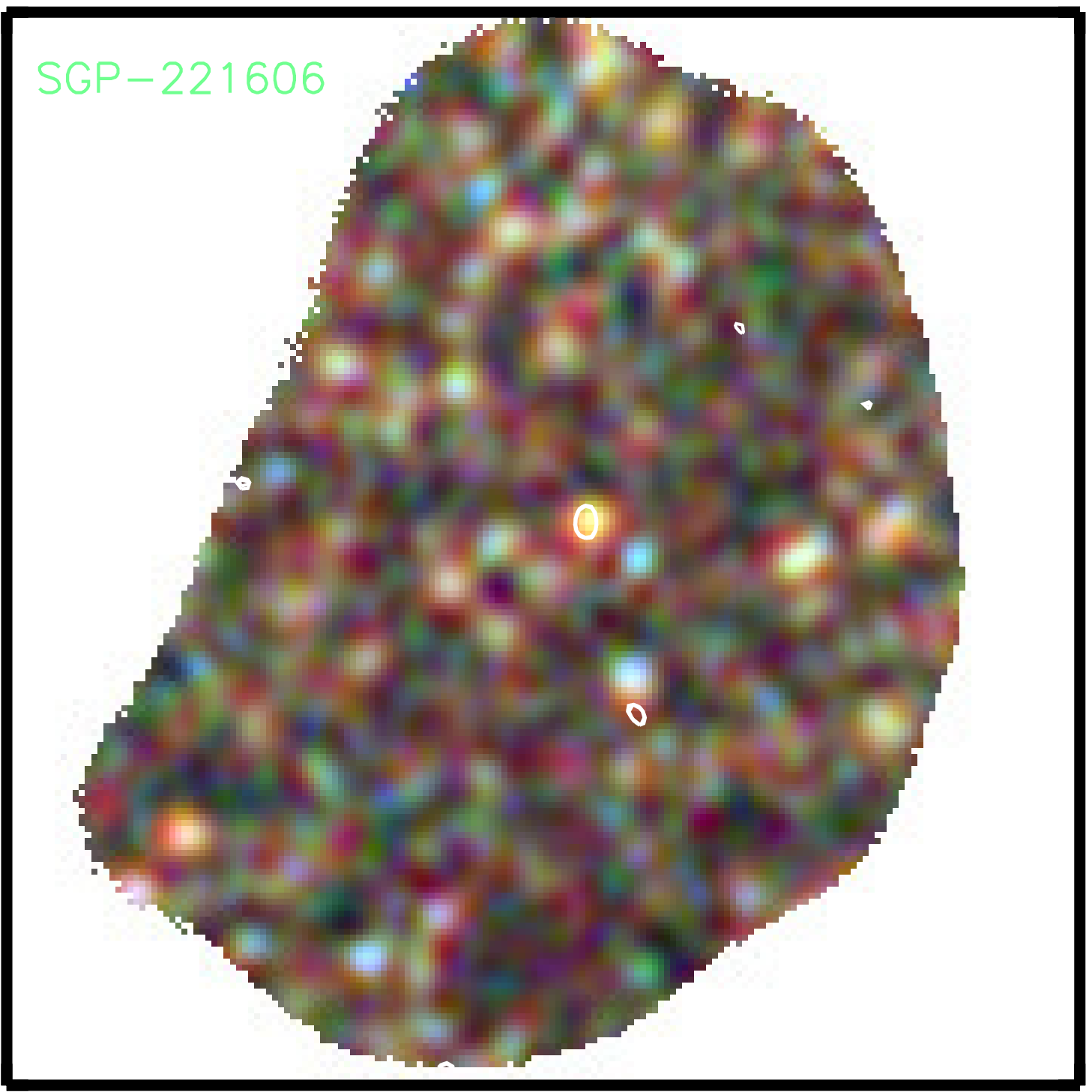}\hspace{1em}
        \includegraphics[height=0.22\textwidth]{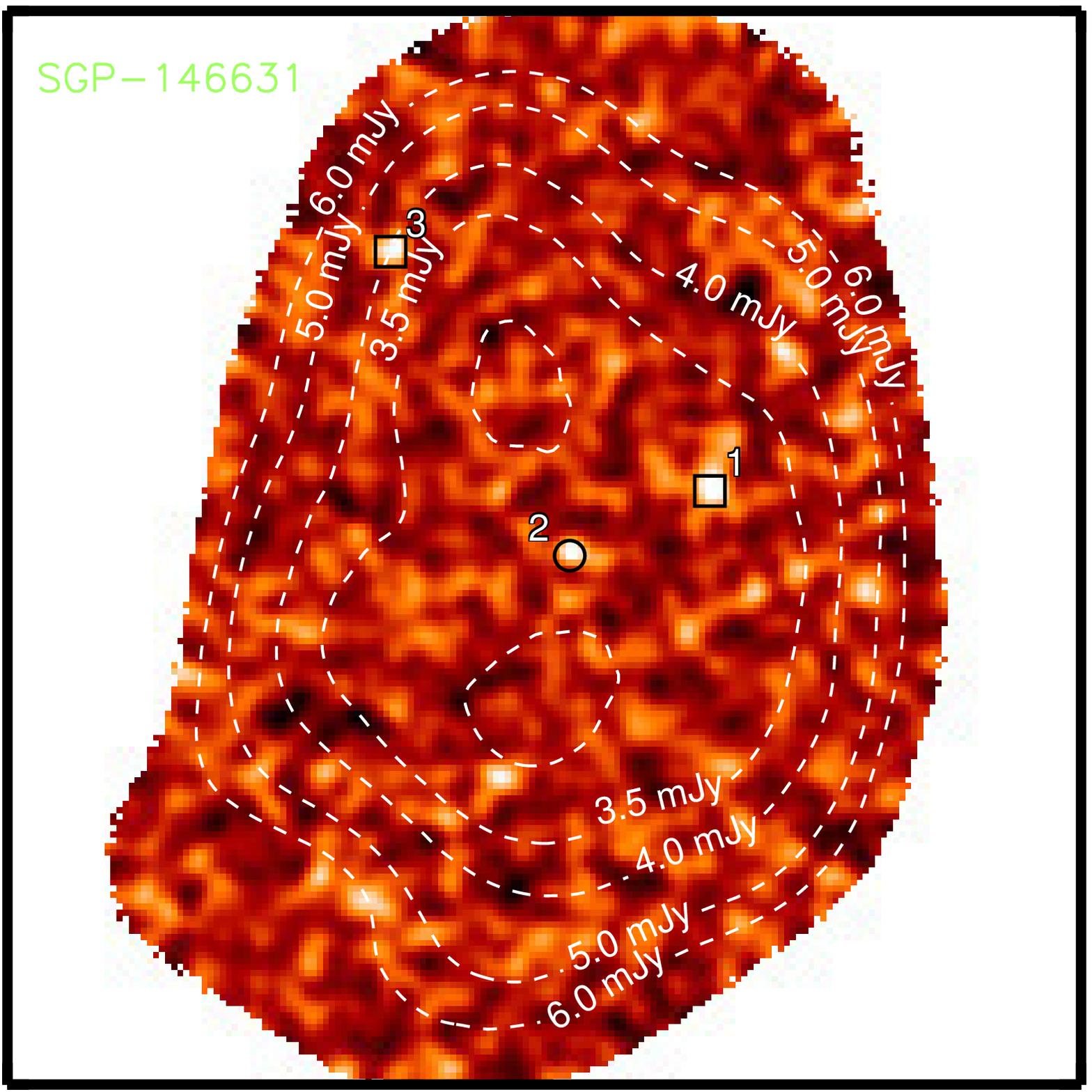}\hspace{-0.3em}
        \includegraphics[height=0.22\textwidth]{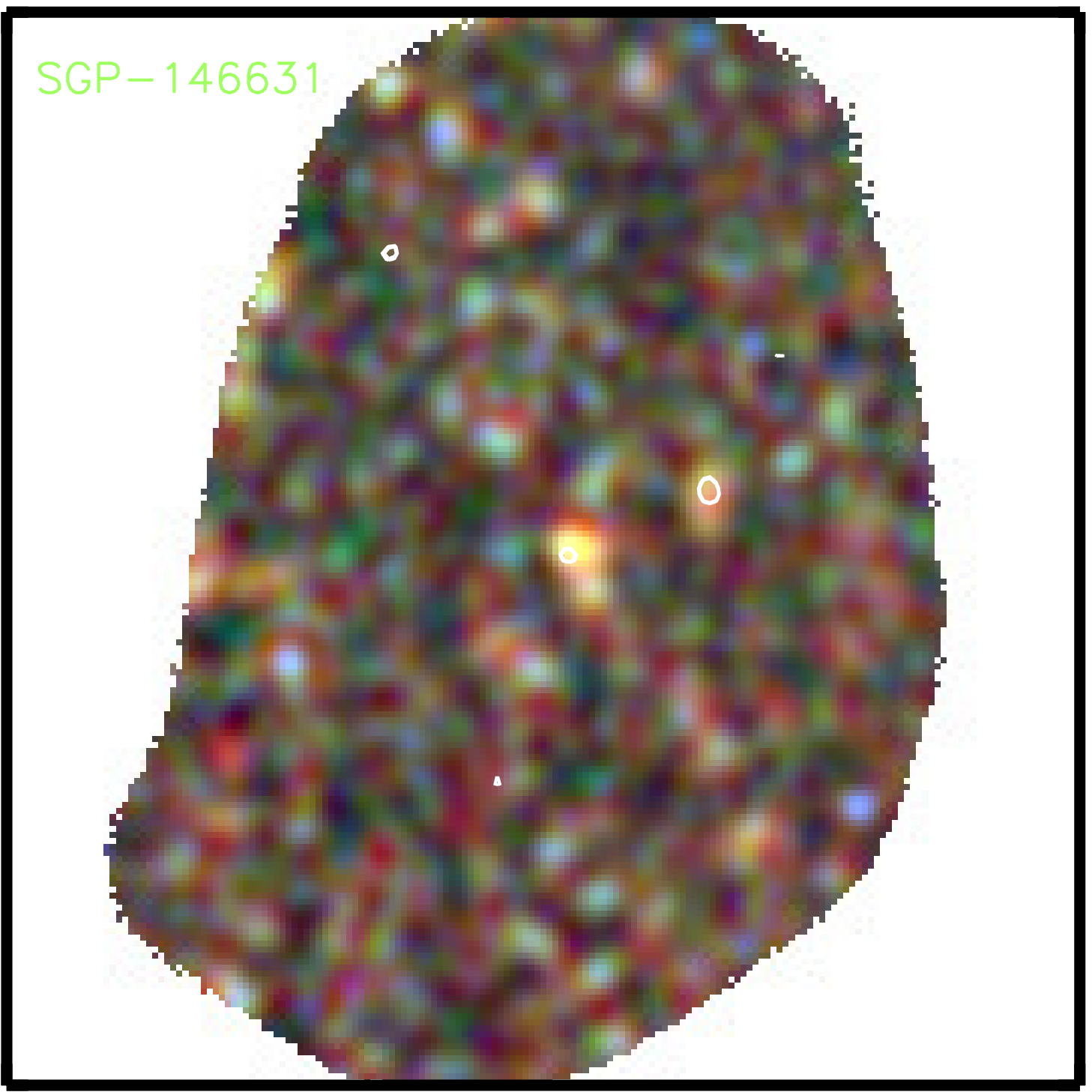}\\\vspace{0.1em}
        \includegraphics[height=0.22\textwidth]{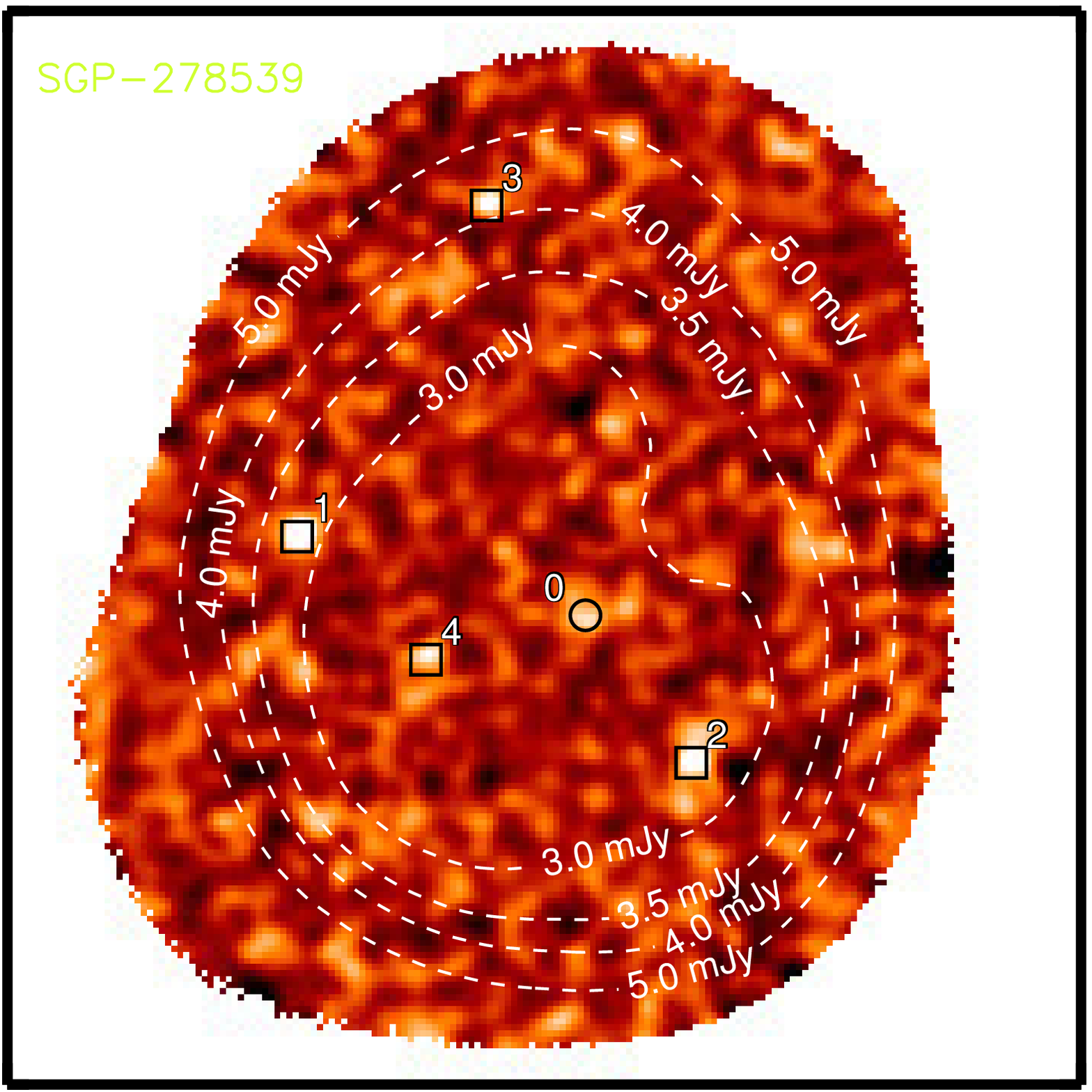}\hspace{-0.3em}
        \includegraphics[height=0.22\textwidth]{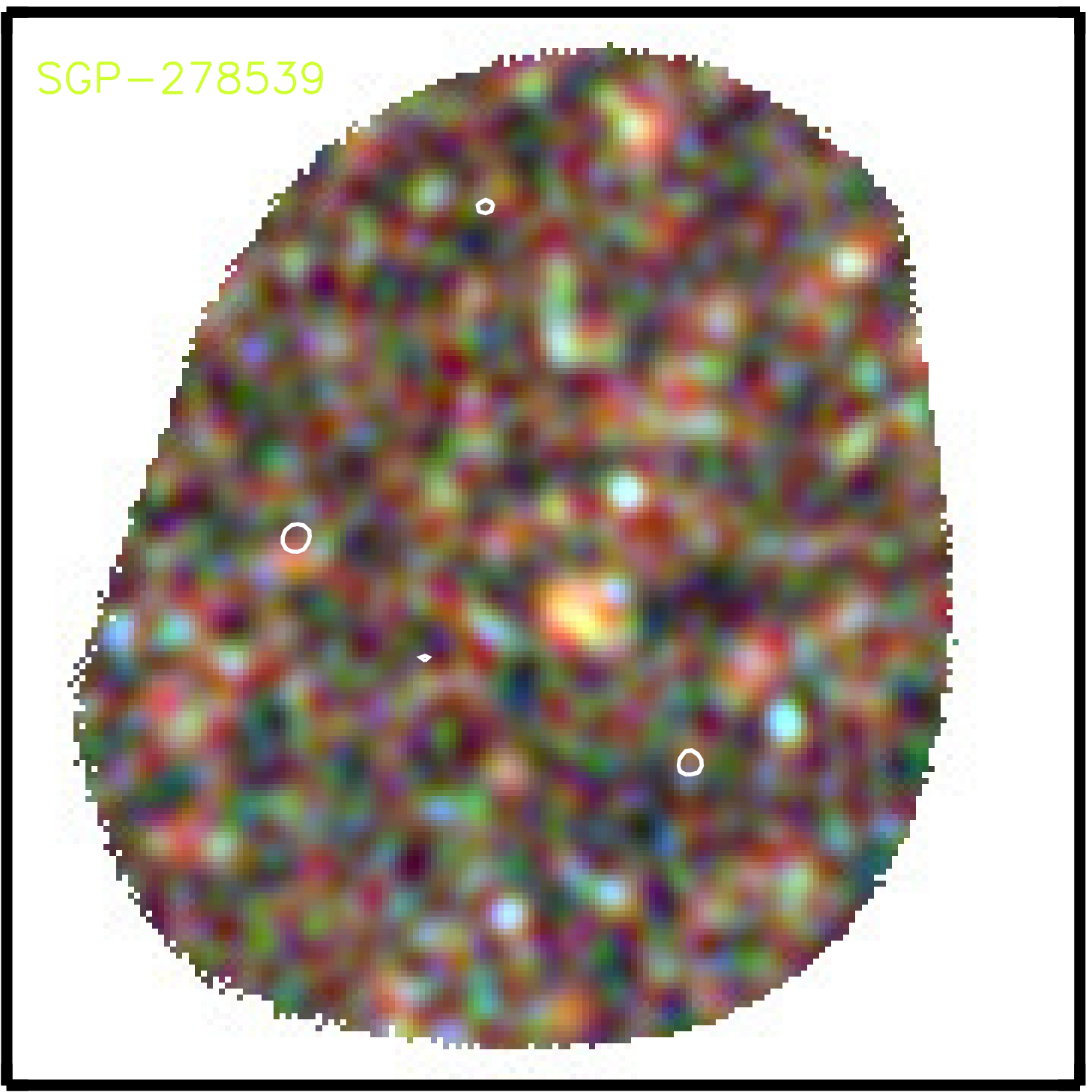}\hspace{1em}
        \includegraphics[height=0.22\textwidth]{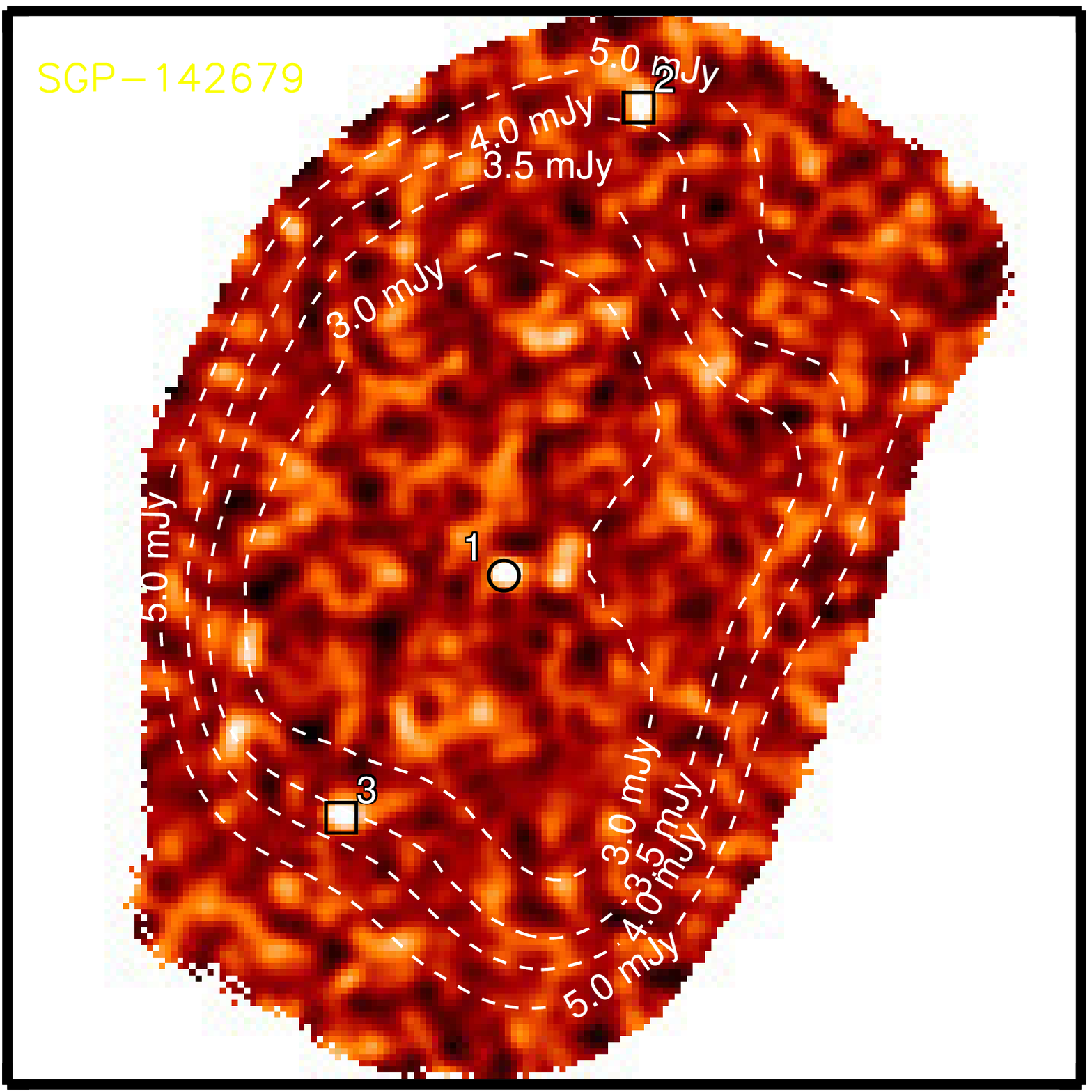}\hspace{-0.3em}
        \includegraphics[height=0.22\textwidth]{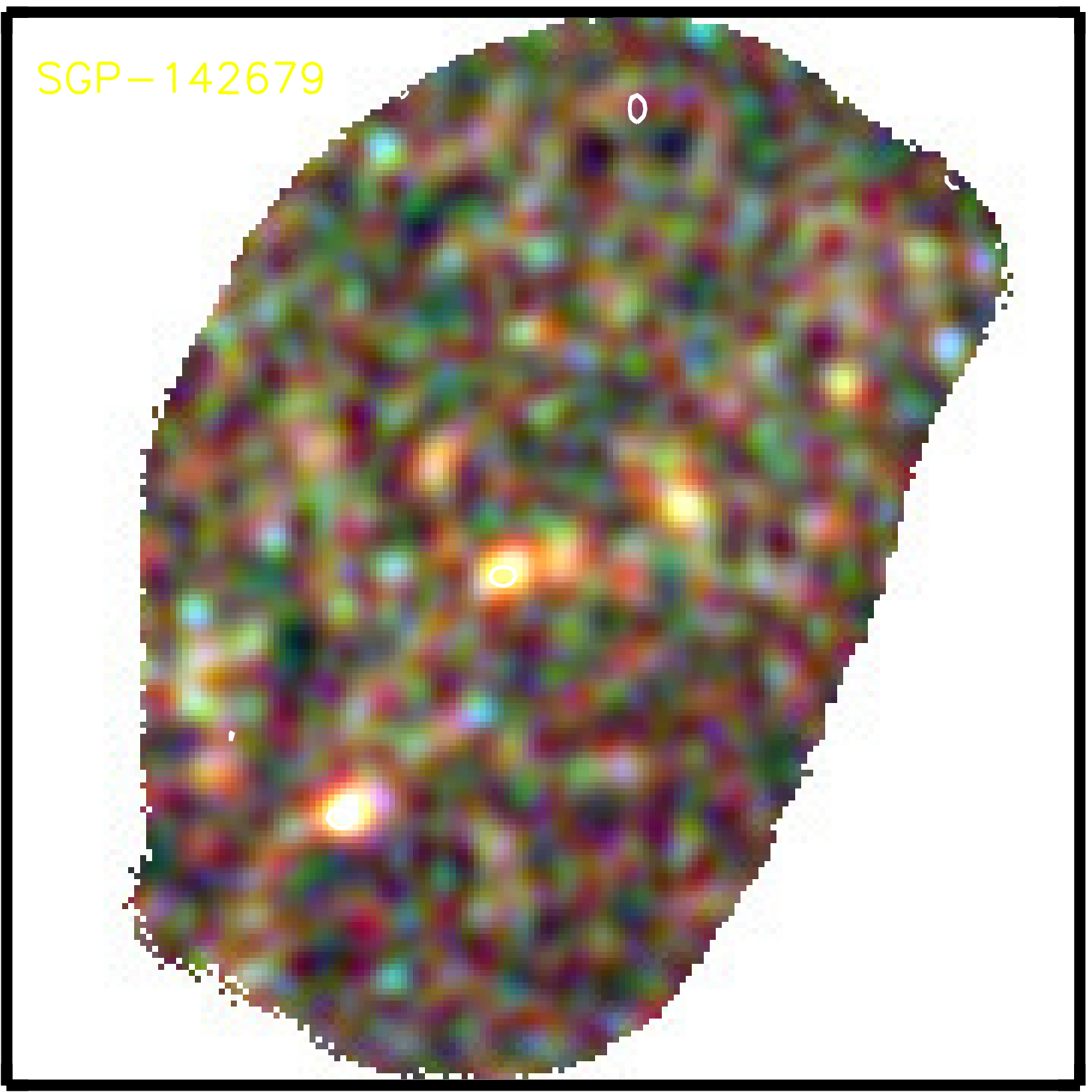}\\\vspace{0.1em}
        \includegraphics[height=0.22\textwidth]{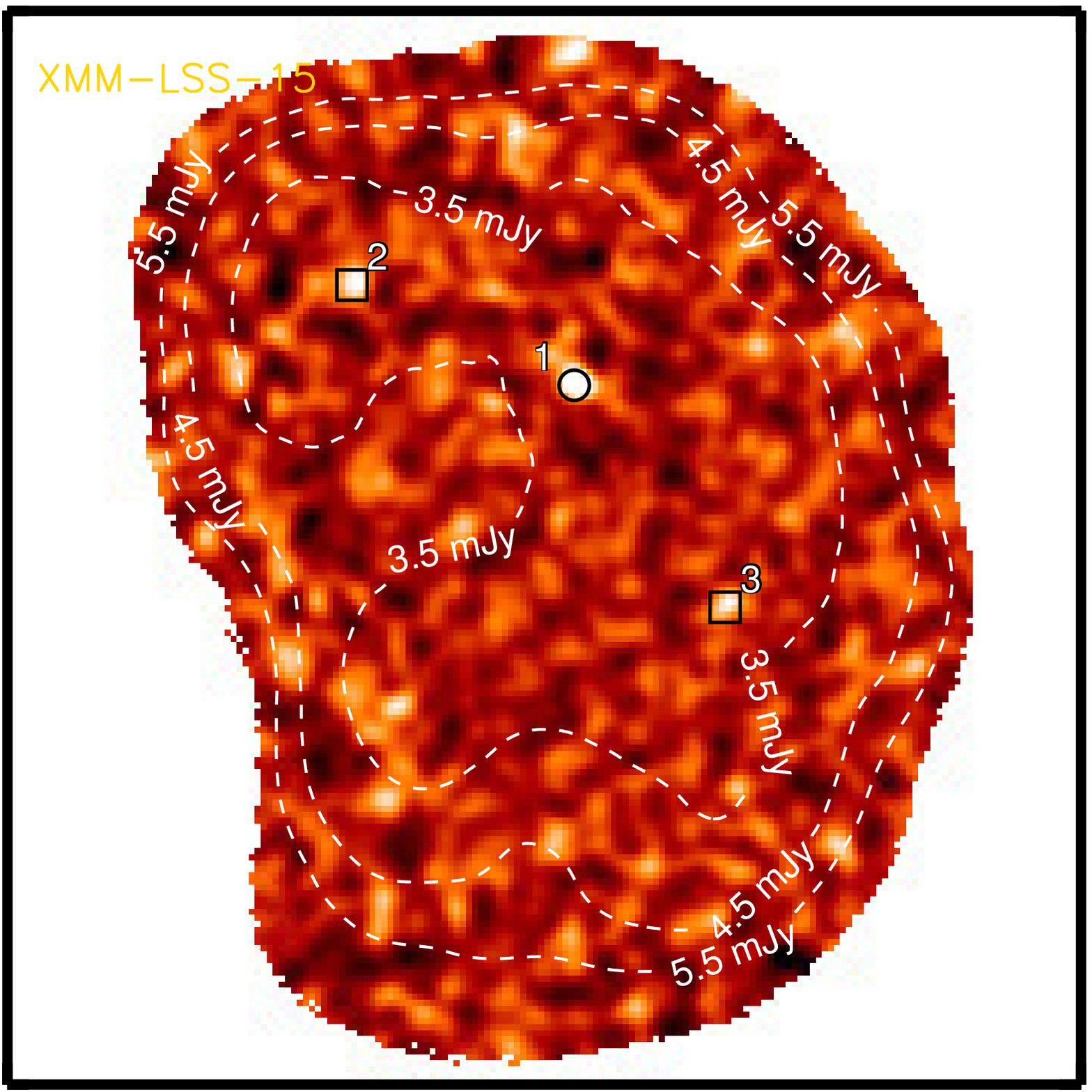}\hspace{-0.3em}
        \includegraphics[height=0.22\textwidth]{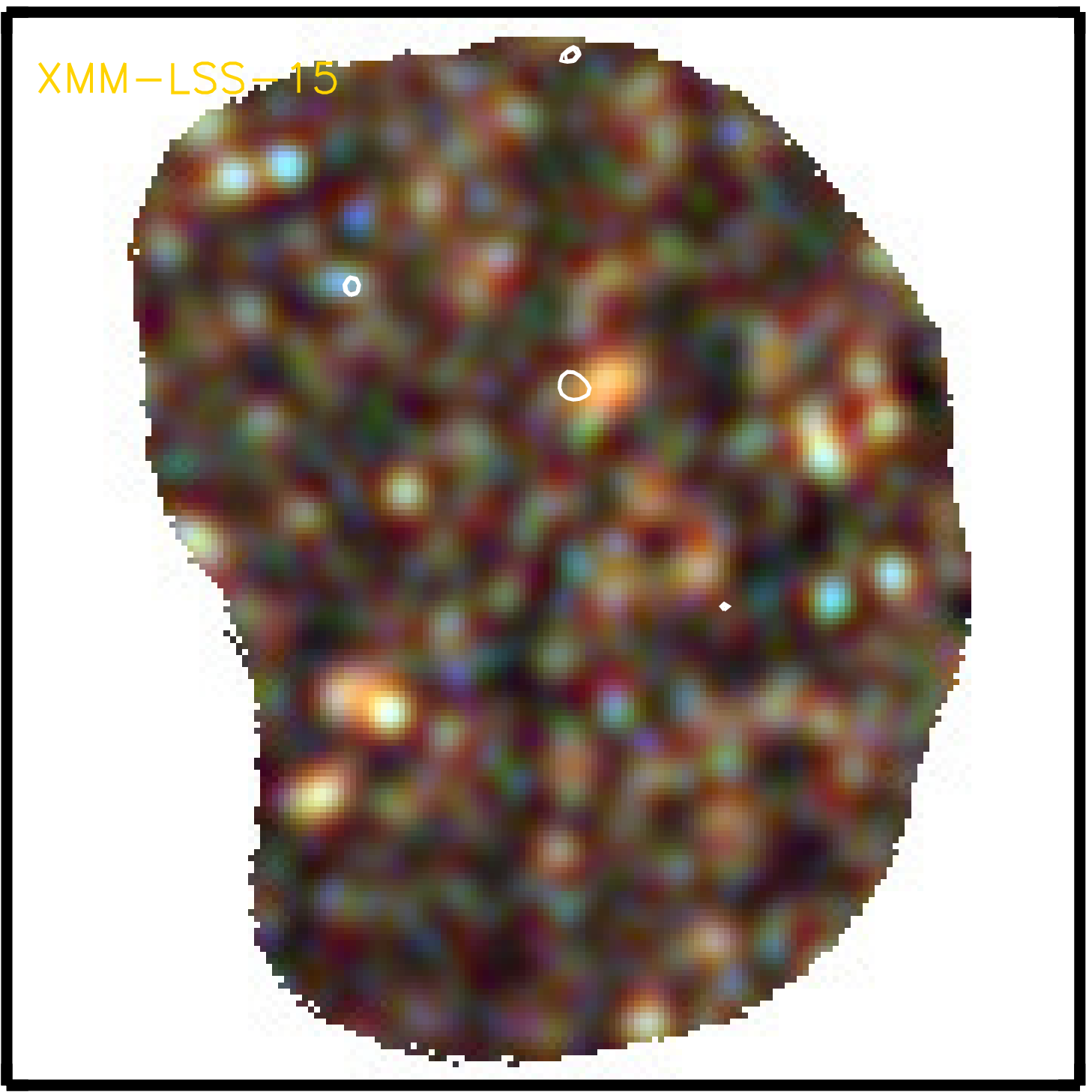}\hspace{1em}
        \includegraphics[height=0.22\textwidth]{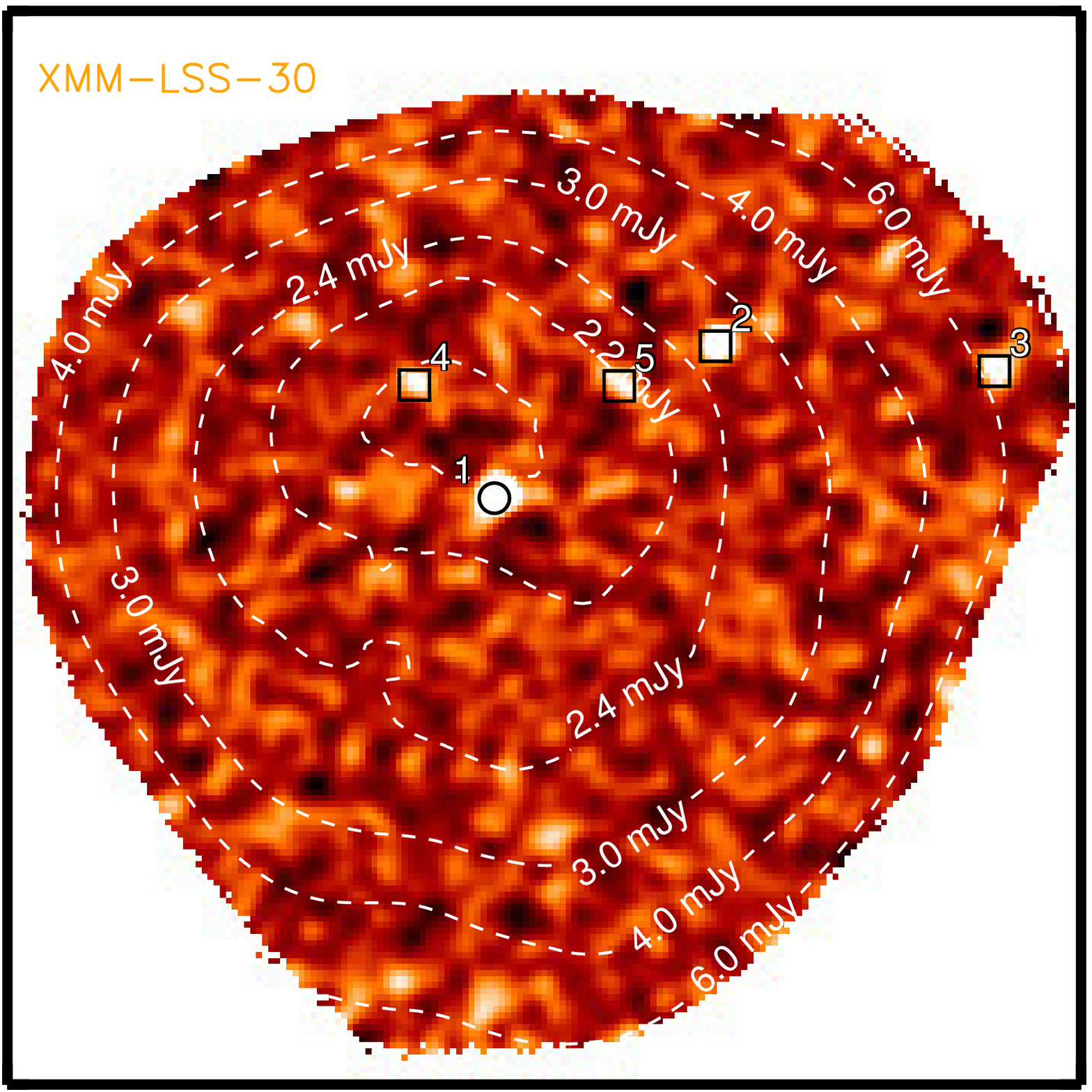}\hspace{-0.3em}
        \includegraphics[height=0.22\textwidth]{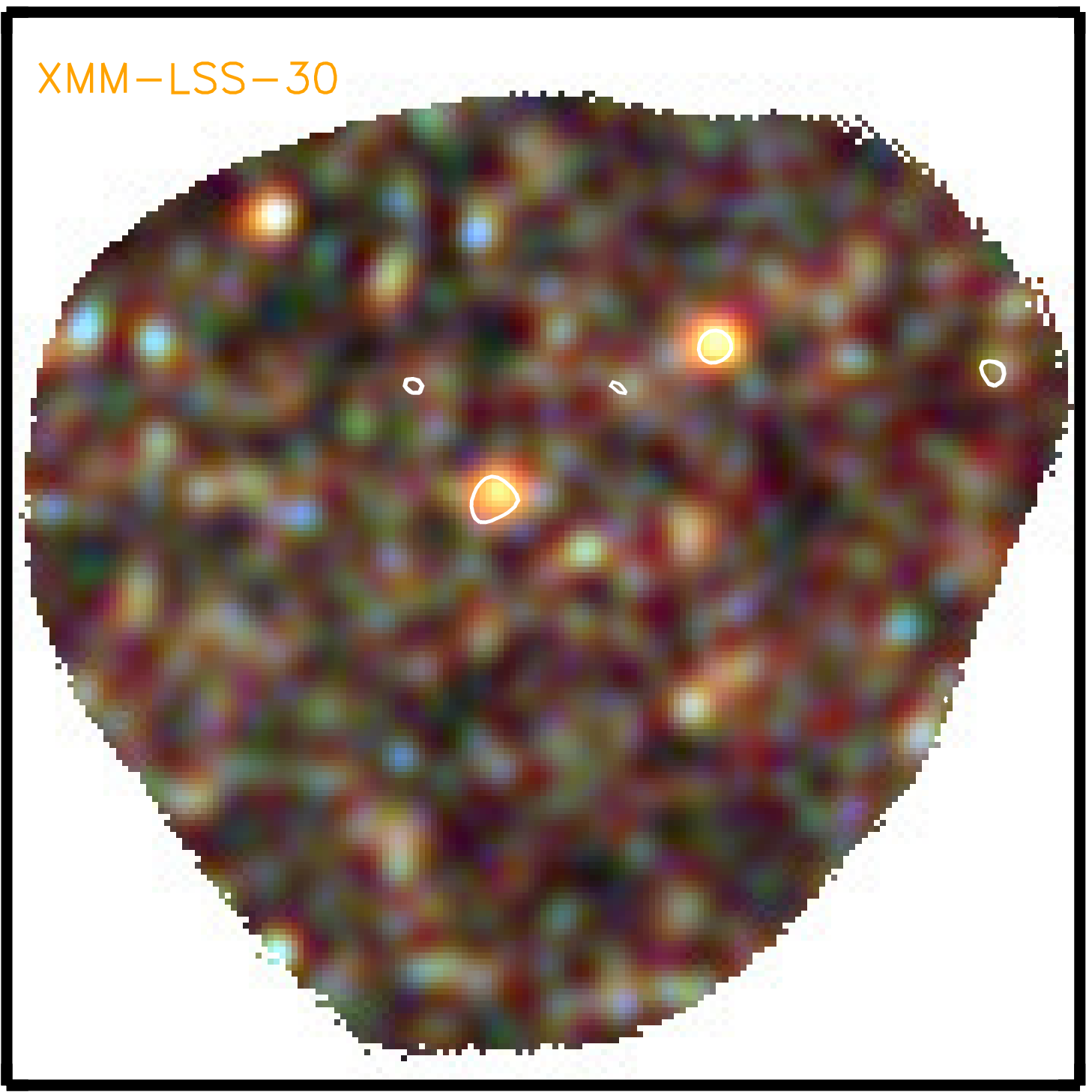}\\\vspace{0.1em}
        \includegraphics[height=0.22\textwidth]{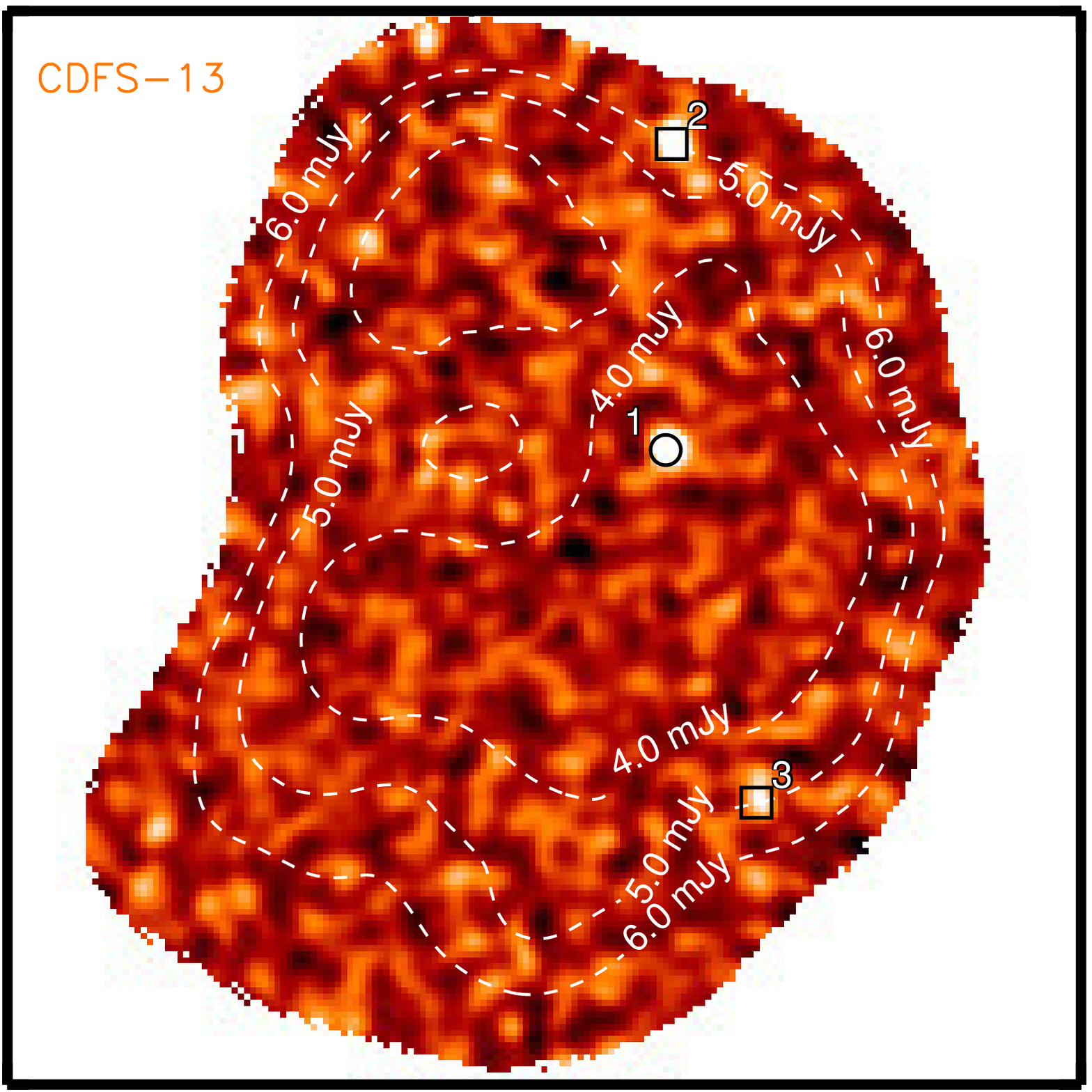}\hspace{-0.3em}
        \includegraphics[height=0.22\textwidth]{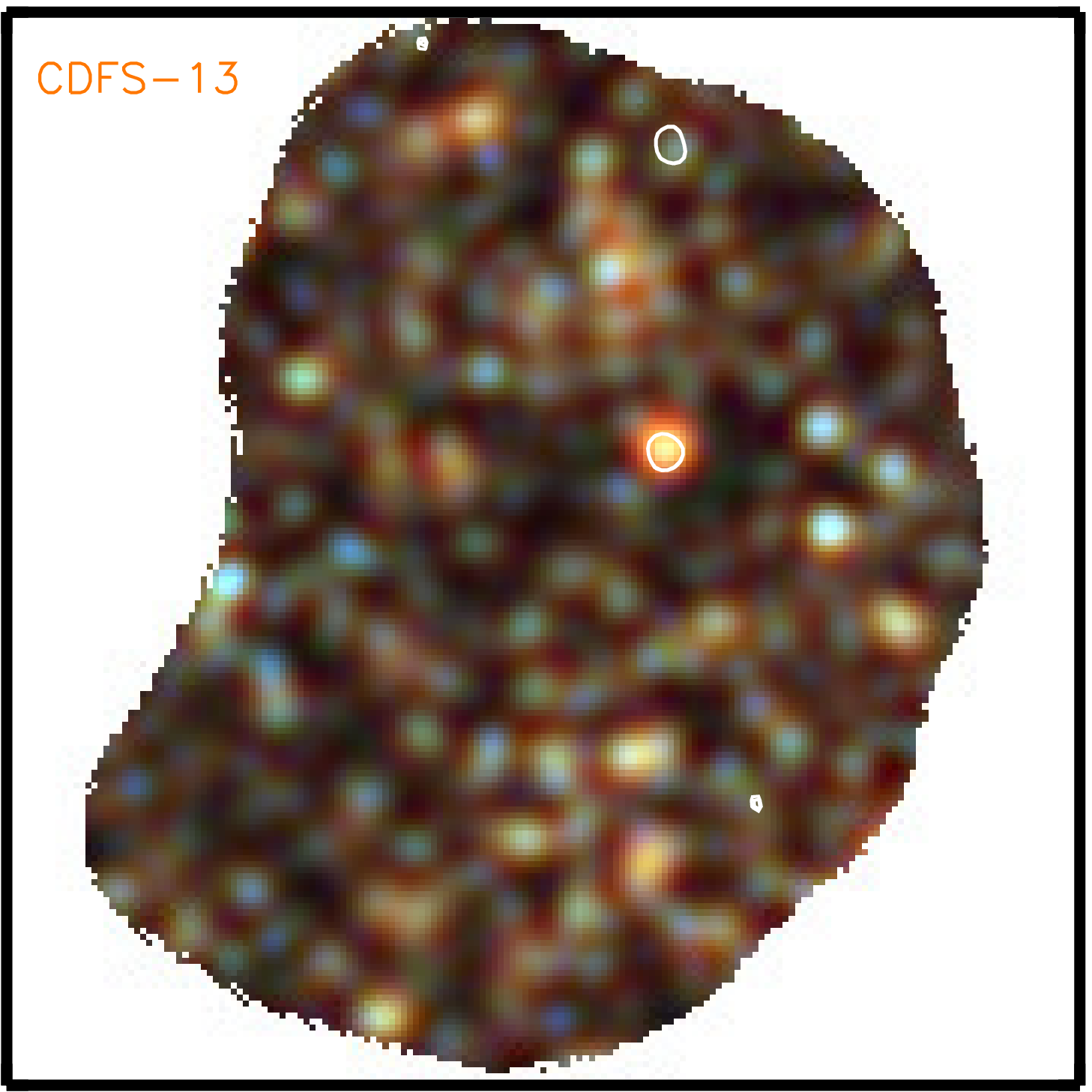}\hspace{1em}
        \includegraphics[height=0.22\textwidth]{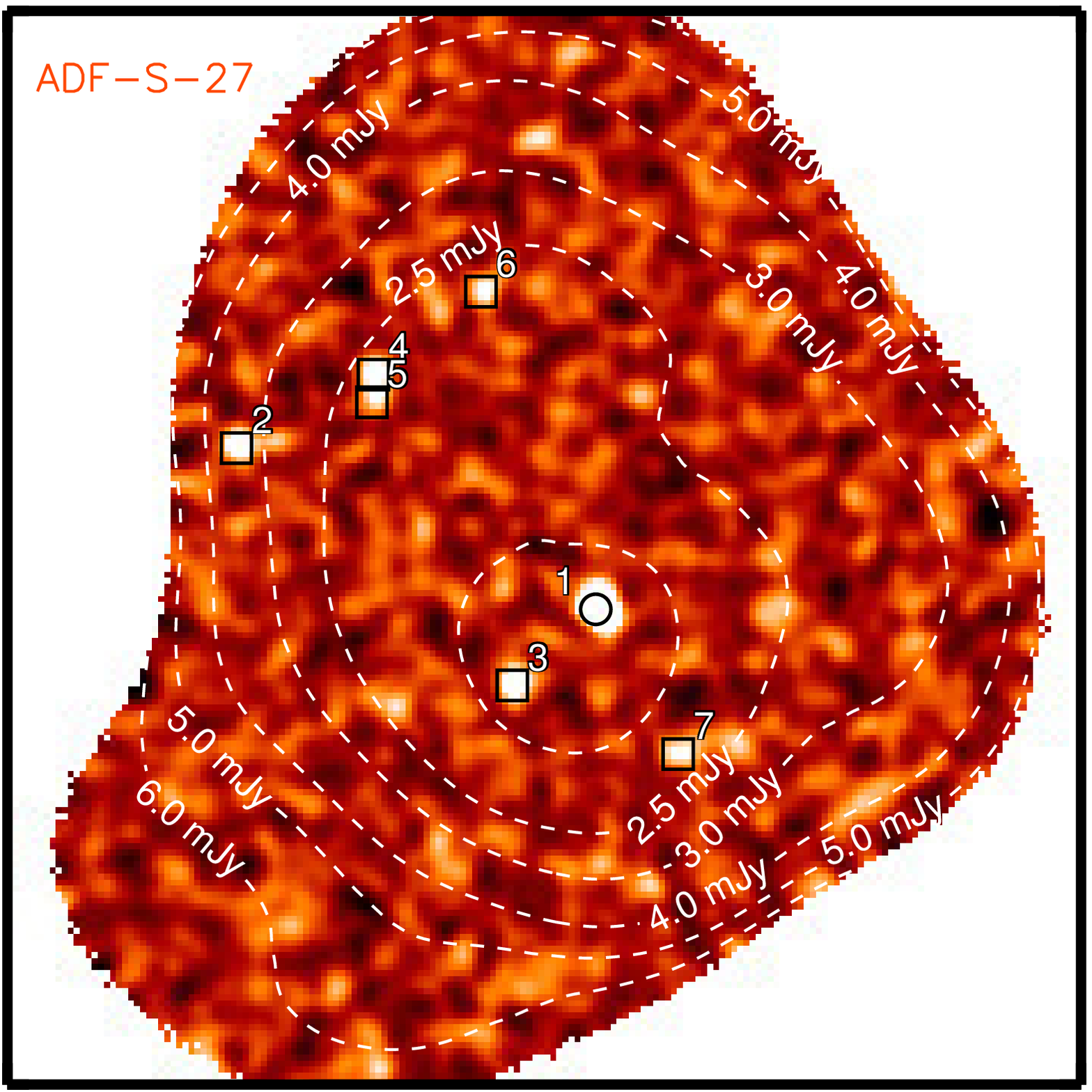}\hspace{-0.3em}
        \includegraphics[height=0.22\textwidth]{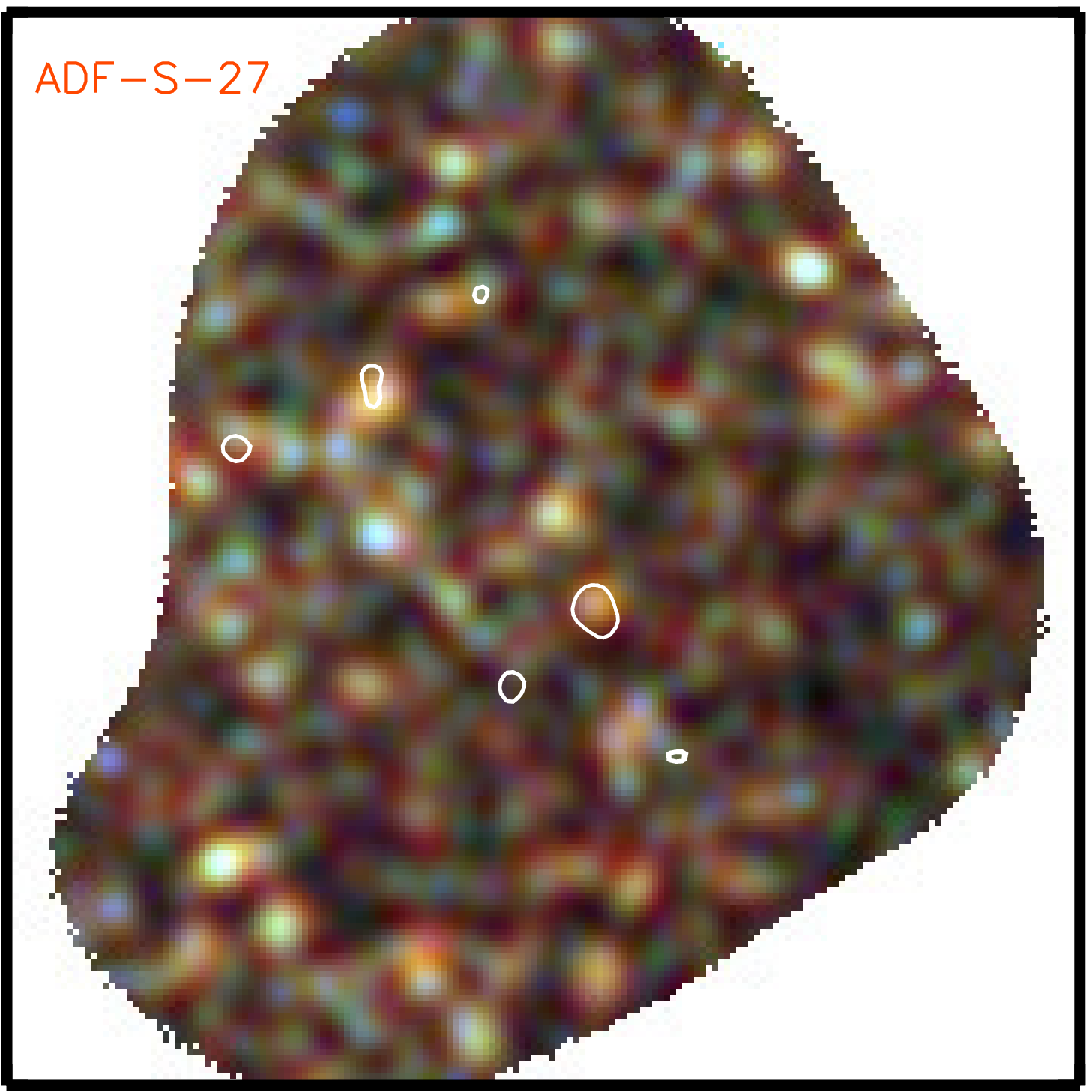}\\\vspace{0.1em}
        \includegraphics[height=0.22\textwidth]{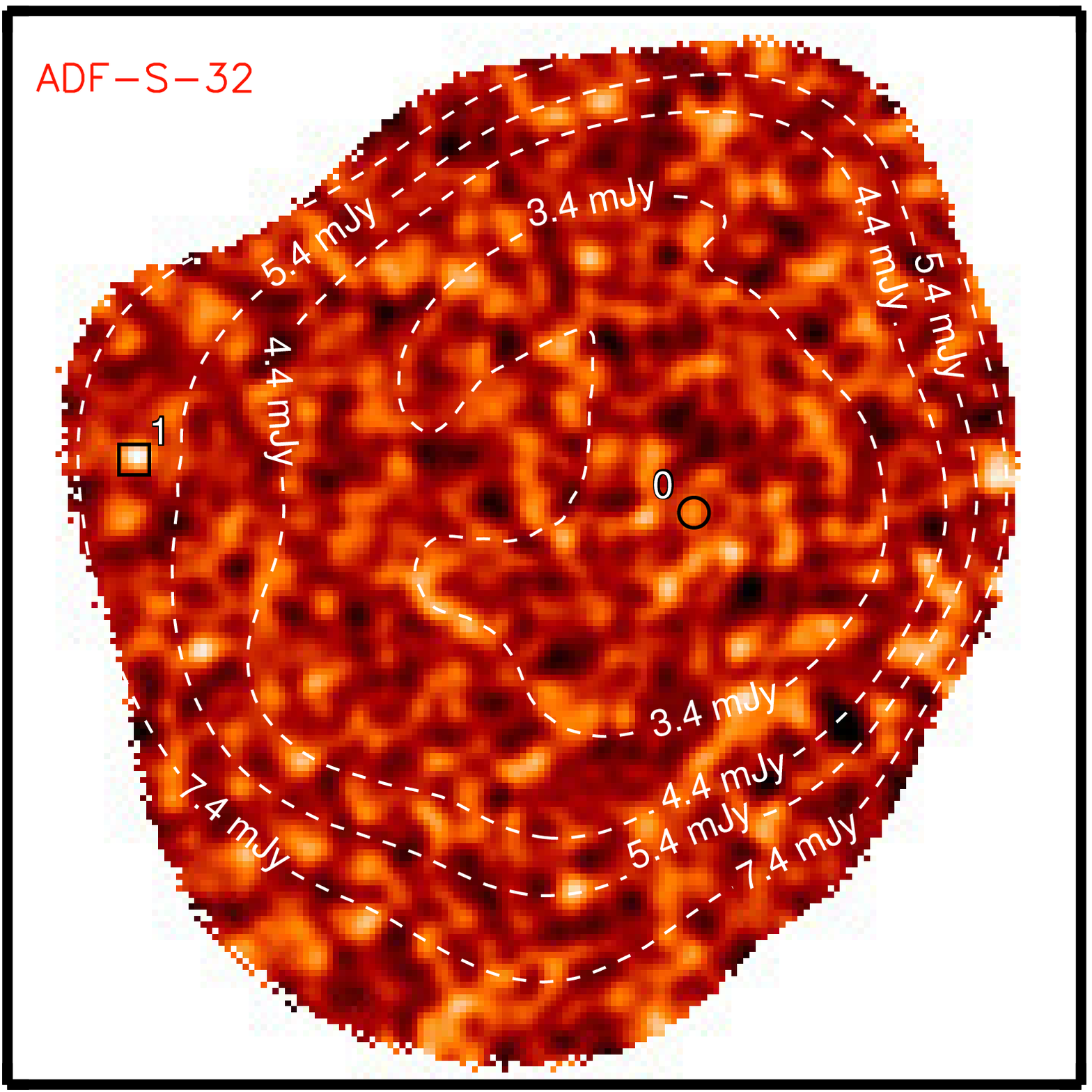}\hspace{-0.3em}
        \includegraphics[height=0.22\textwidth]{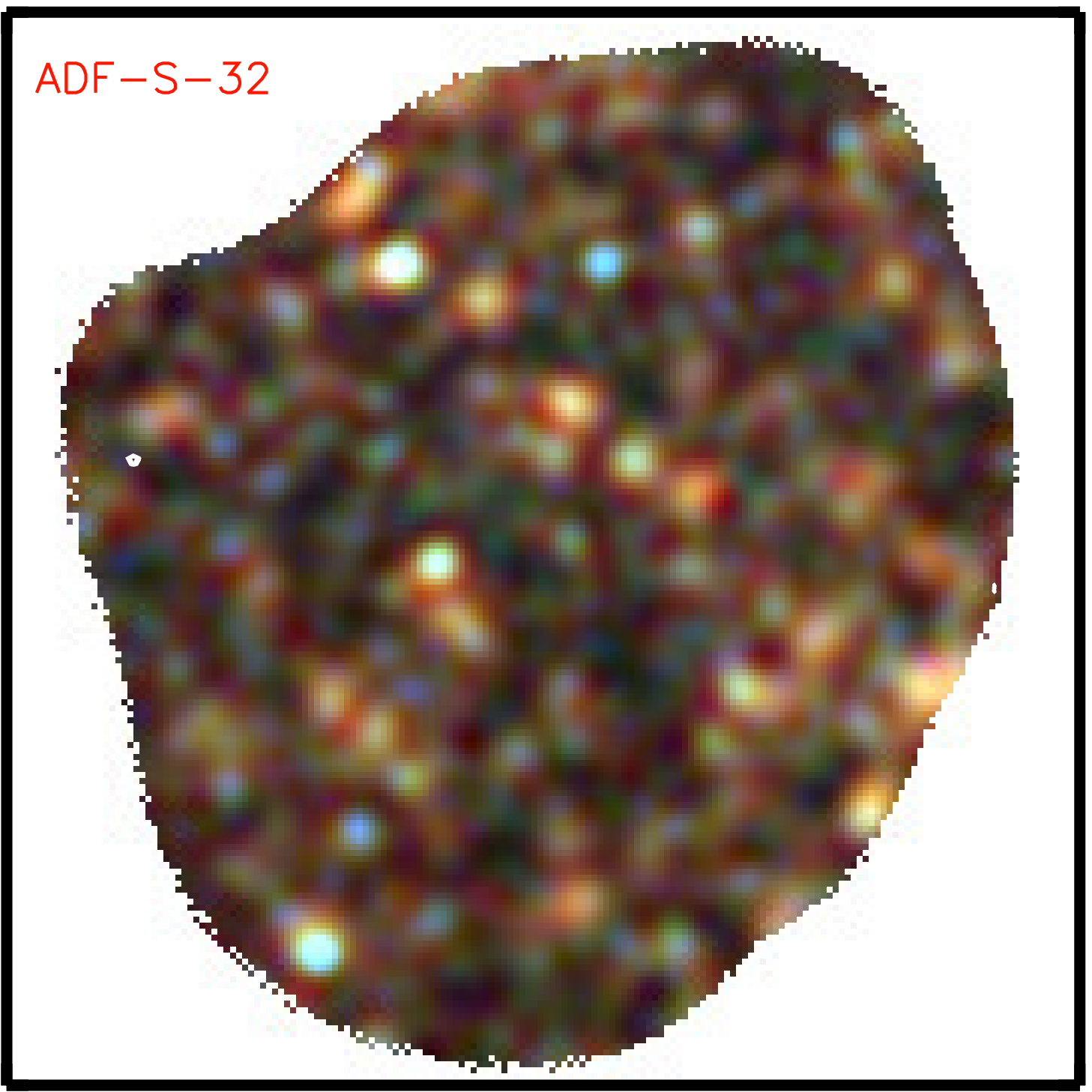}\hspace{1em}
        \includegraphics[height=0.22\textwidth]{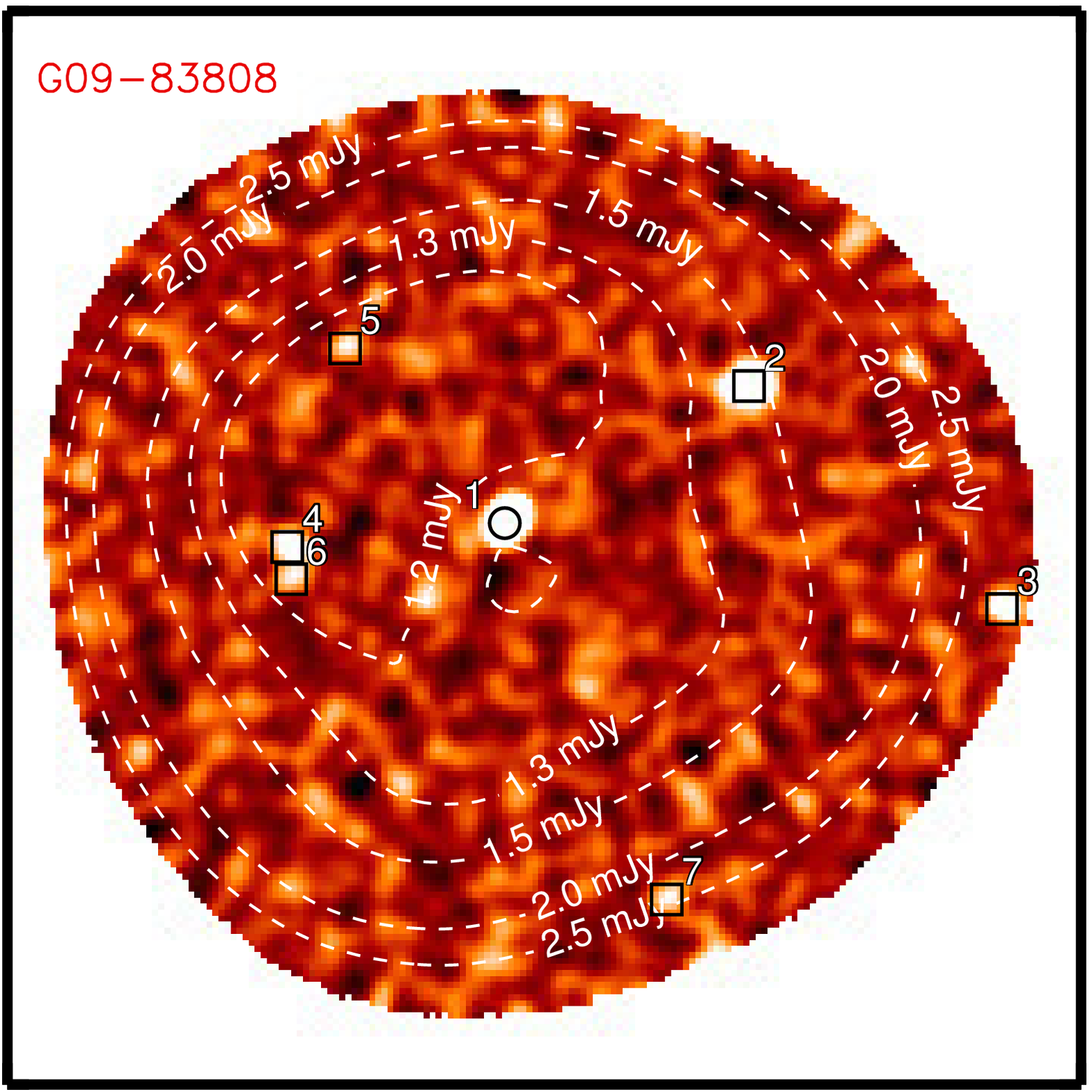}\hspace{-0.3em}
        \includegraphics[height=0.22\textwidth]{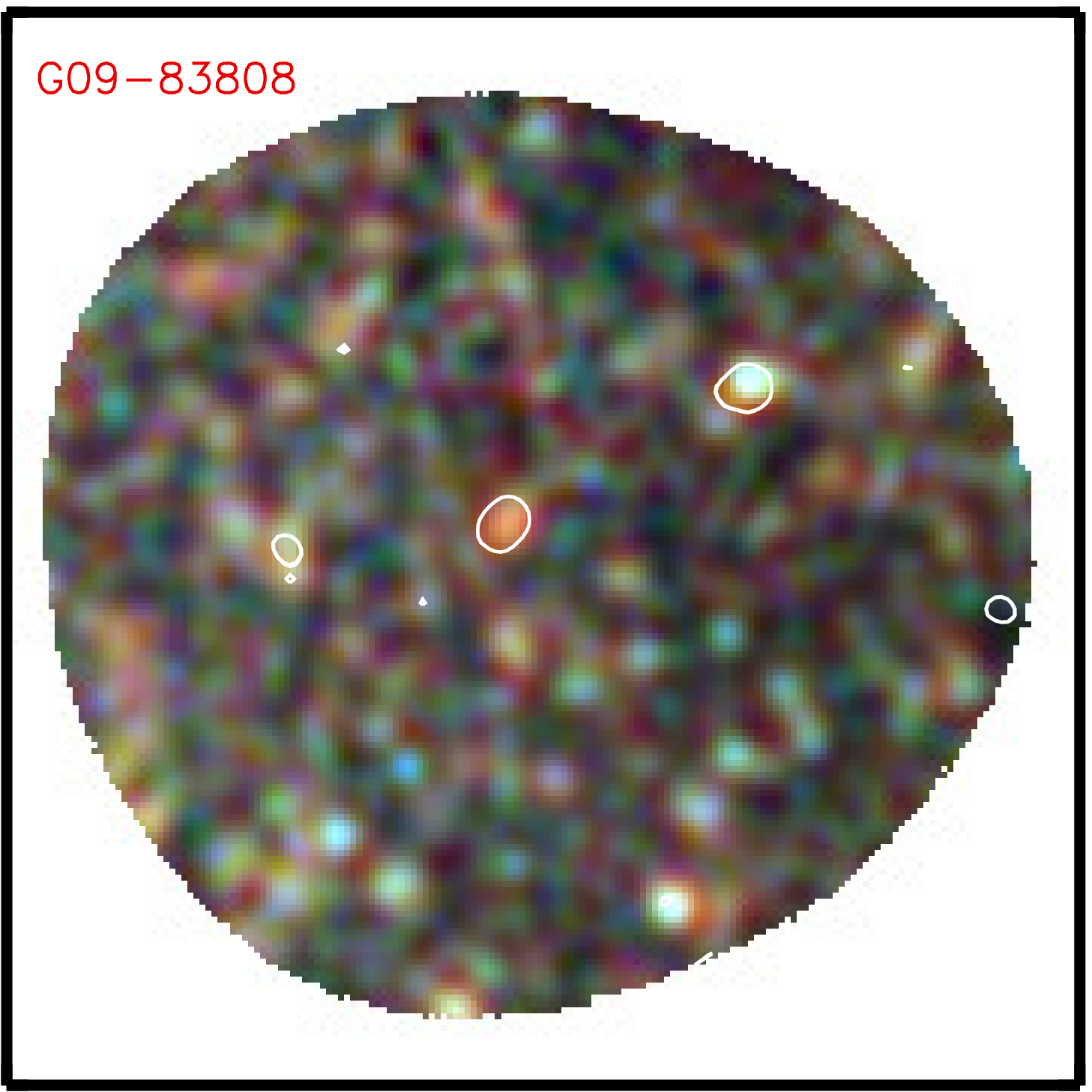}\\\vspace{0.1em}
        \includegraphics[height=0.22\textwidth]{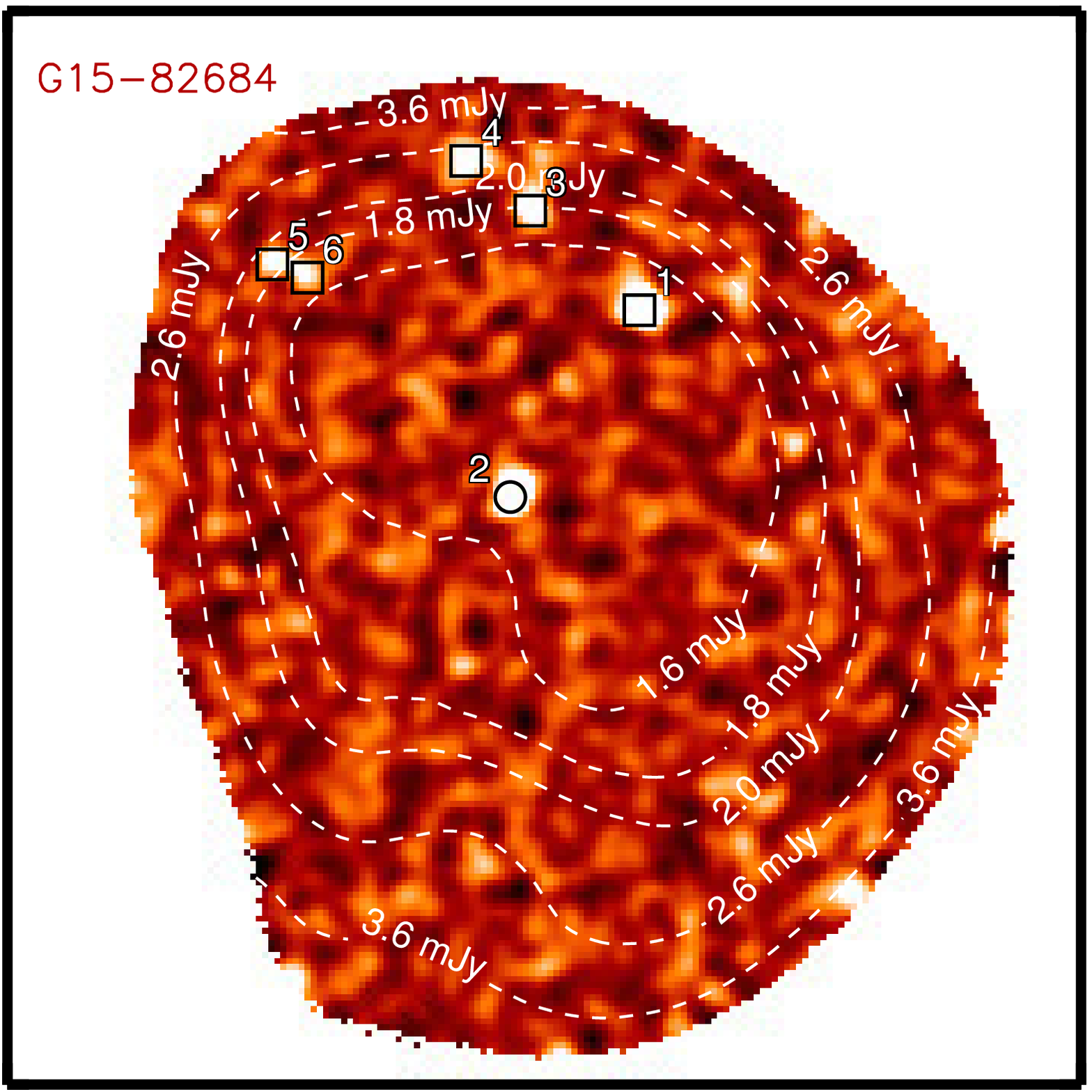}\hspace{-0.3em}
        \includegraphics[height=0.22\textwidth]{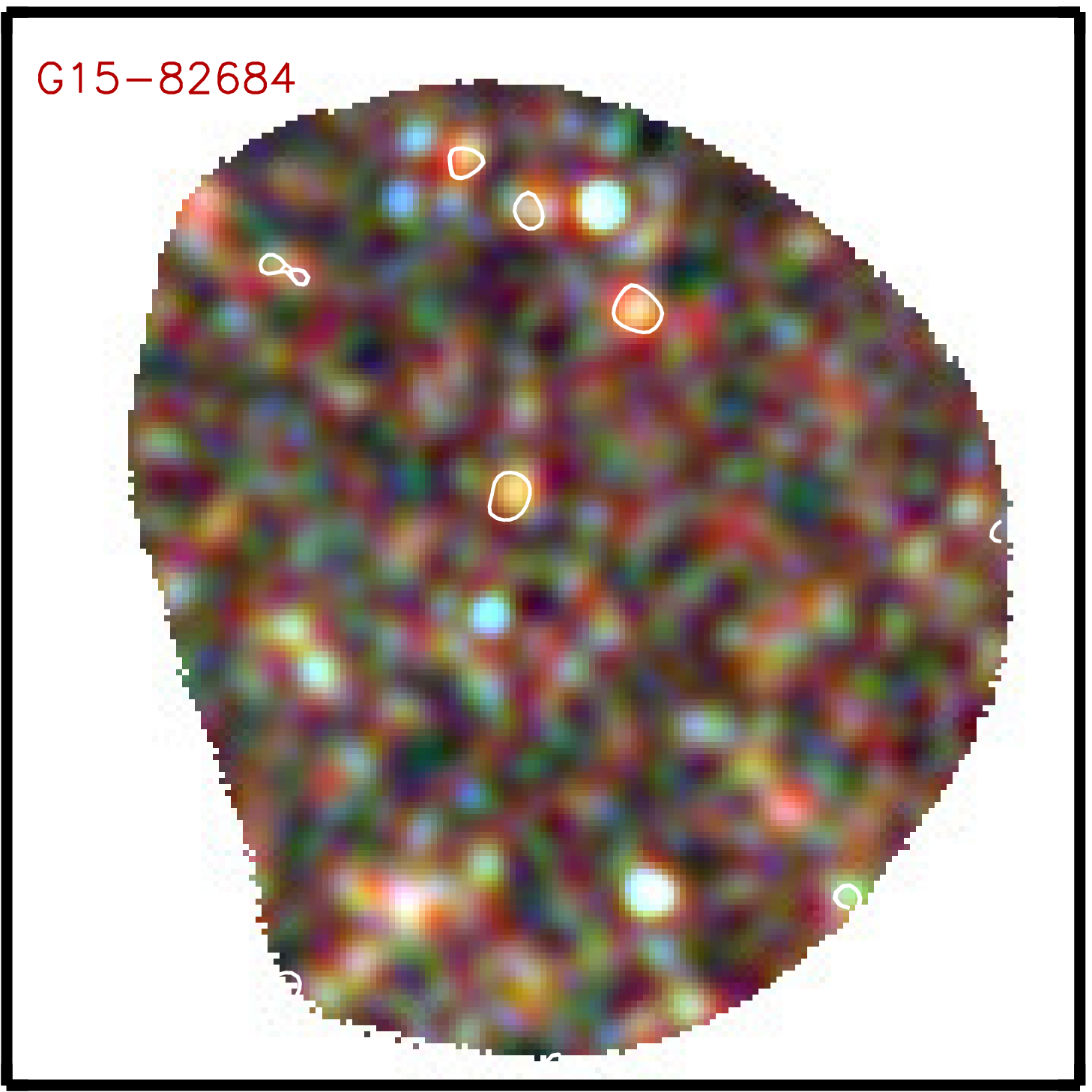}\hspace{1em}
        \includegraphics[height=0.22\textwidth]{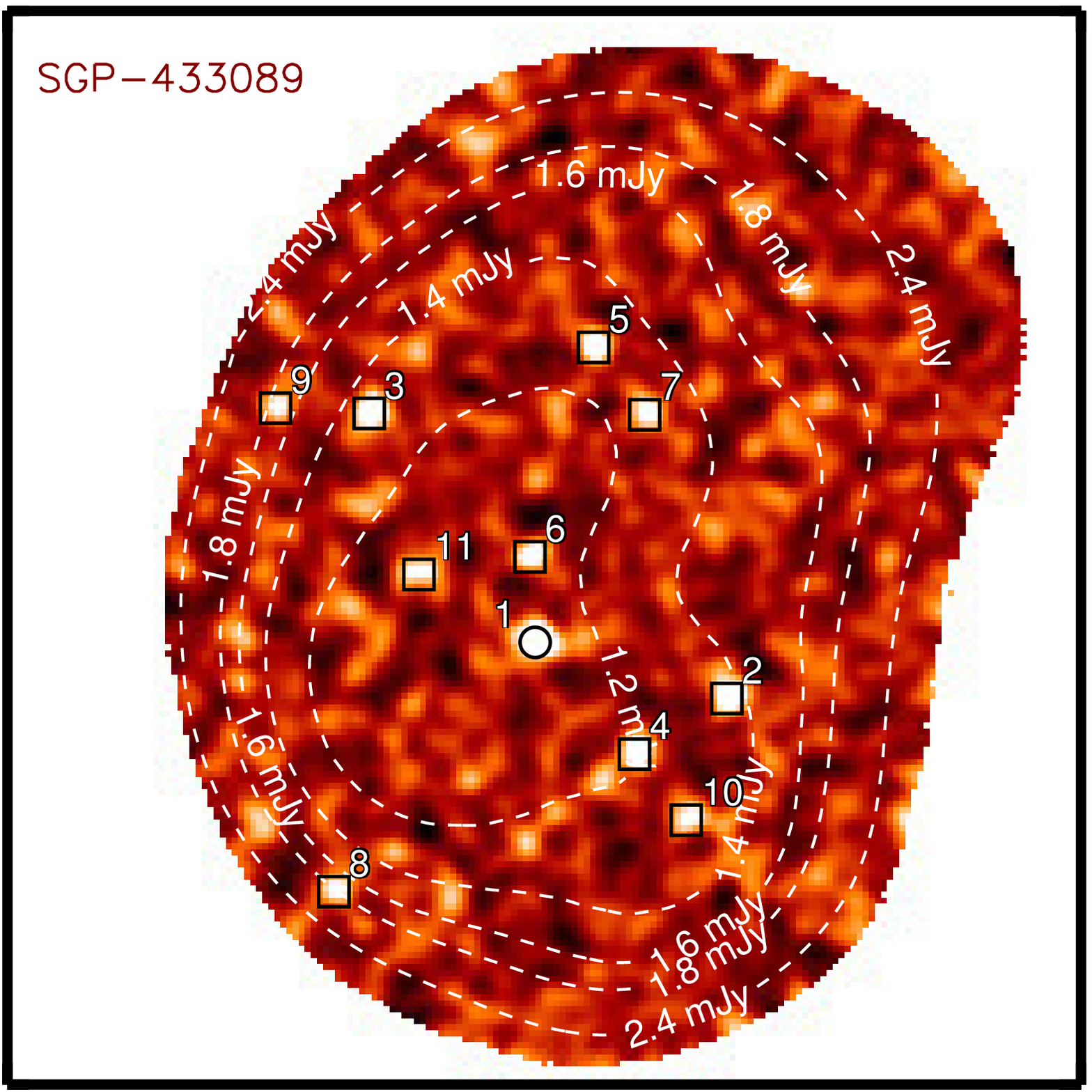}\hspace{-0.3em}
        \includegraphics[height=0.22\textwidth]{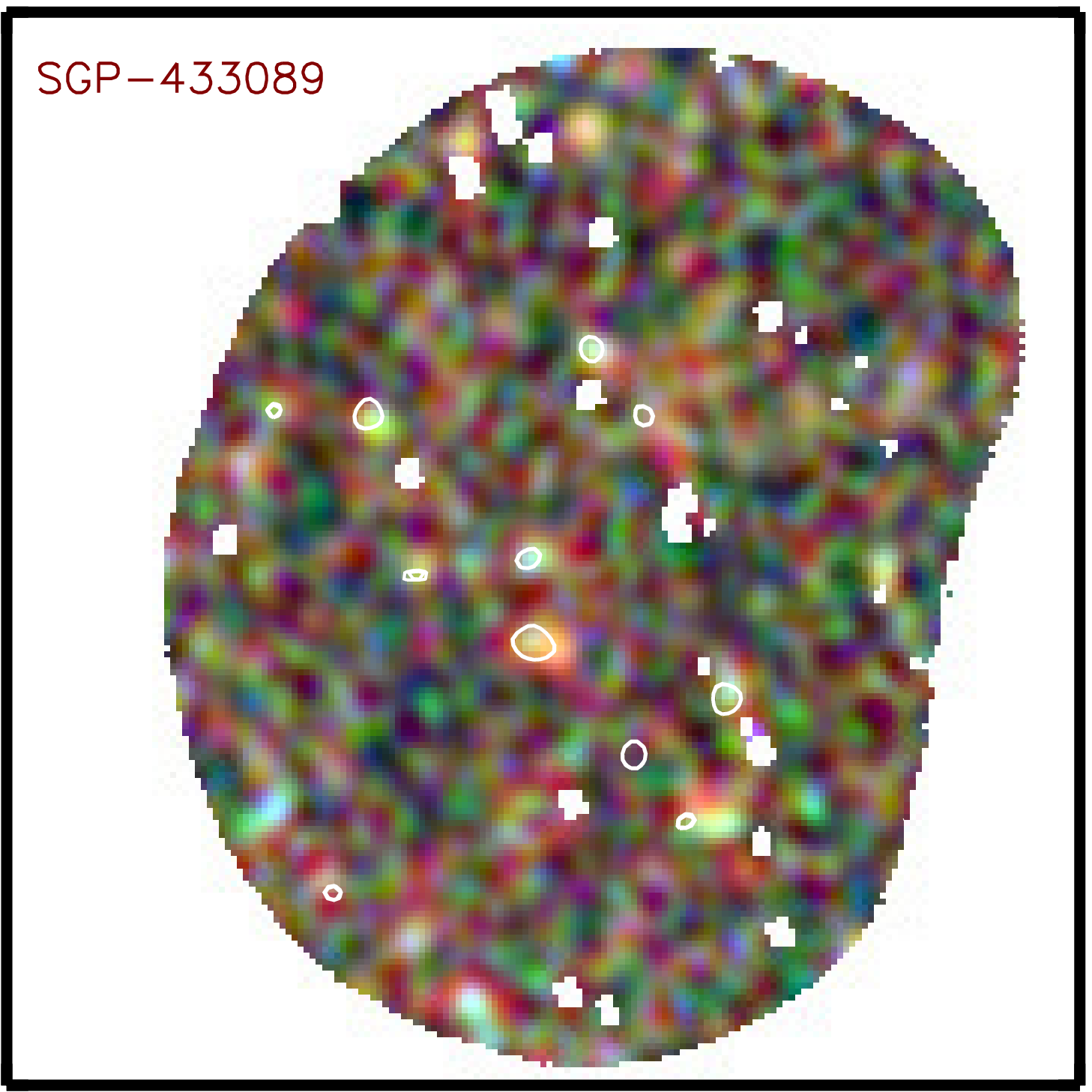}\\\vspace{0.1em}
    \caption{Cont...}
    \end{center}
\end{figure*}

\section{Photometry and redshift catalogs}
\label{sec:catalogues}

Here we present our photometry and photometric redshift catalogs for our sample
of ultra-red galaxies and their surrounding DSFGs.

\begin{table*}
    \begin{center}
    \caption{Signpost galaxies and their photometric properties.}
    \label{tab: photometry}
    \begin{scriptsize}
        \begin{tabular}{l lr r@{\,$\pm$\,}l r@{\,$\pm$\,}l r@{\,$\pm$\,}l r@{\,$\pm$\,}l c c}
            \hline\hline
            \multicolumn{1}{l}{IAU name} &
            \multicolumn{2}{c}{$\alpha$ \quad (J2000) \quad $\delta$} &
            \multicolumn{2}{c}{$S_{250}^{\dagger}$} &
            \multicolumn{2}{c}{$S_{350}^{\dagger}$} &
            \multicolumn{2}{c}{$S_{500}^{\dagger}$} &
            \multicolumn{2}{c}{$S_{870}^{\dagger}$} &
            \multicolumn{1}{c}{$\mathcal{B}$} &
            \multicolumn{1}{c}{$\mathcal{F}$}\\
            \multicolumn{1}{l}{} &
            \multicolumn{1}{c}{$^{\textrm{h}}$ \, $^{\textrm{m}}$ \, $^{\textrm{s}}$ } &
            \multicolumn{1}{c}{$\, ^{\circ} \quad ' \quad ''$} &
            \multicolumn{2}{c}{$\millijanksy{}\,\beam{}^{-1}$} &
            \multicolumn{2}{c}{$\millijanksy{}\,\beam{}^{-1}$} &
            \multicolumn{2}{c}{$\millijanksy{}\,\beam{}^{-1}$} &
            \multicolumn{2}{c}{$\millijanksy{}$} &
            \multicolumn{1}{c}{} \\
            \hline
            \multicolumn{13}{c}{\dotfill\,\textbf{SGP-28124}\,\dotfill}\\
            \textbf{URG\,J000124.9$-$354212} &  \textbf{00:01:24.88} & \textbf{$-$35:42:12.2} & \textbf{62.2} & \textbf{9.1} & \textbf{89.8} & \textbf{8.8} & \textbf{119.9} & \textbf{9.3} & \textbf{44.3} & \textbf{1.4} & \textbf{1.04} & \textbf{1.00}\\
            URG\,J000145.0$-$353822 &  00:01:44.95 & $-$35:38:22.1 & 55.9 & 7.9 & 67.4 & 8.6 & 52.4 & 9.4 & 15.9 & 2.6 & 1.15 & 1.00\\
            URG\,J00014.2$-$354123 &  00:01:04.20 & $-$35:41:23.0 & 5.9 & 7.5 & 11.7 & 8.8 & 4.7 & 9.7 & 6.4 & 1.5 & 1.35 & 0.97\\
            URG\,J000122.9$-$354211 &  00:01:22.91 & $-$35:42:11.2 & 31.9 & 9.0 & 47.9 & 8.7 & 87.8 & 9.4 & 10.2 & 1.4 & 1.11 & 0.92\\
            URG\,J000138.5$-$35442 &  00:01:38.50 & $-$35:44:02.3 & 4.0 & 9.2 & 9.2 & 9.2 & $-$3.6 & 10.3 & 4.7 & 1.2 & 1.55 & 0.85\\
            URG\,J000115.9$-$35411 &  00:01:15.90 & $-$35:41:01.3 & 28.4 & 8.1 & 27.4 & 8.6 & 6.2 & 9.3 & 4.4 & 1.2 & 1.59 & 0.85\\
            URG\,J000129.4$-$354416 &  00:01:29.39 & $-$35:44:15.7 & 30.0 & 9.6 & 23.6 & 9.0 & 26.7 & 10.2 & 3.5 & 1.2 & 1.65 & 0.57\\
            \multicolumn{13}{c}{\dotfill\,\textbf{HeLMS-42}\,\dotfill}\\
            \textbf{URG\,J00034.2$+$024114} &  \textbf{00:03:04.17} & \textbf{$+$02:41:13.7} & \textbf{39.8} & \textbf{9.2} & \textbf{60.3} & \textbf{9.9} & \textbf{81.0} & \textbf{11.3} & \textbf{42.6} & \textbf{3.6} & \textbf{1.89} & \textbf{1.00}\\
            URG\,J000319.2$+$02371 &  00:03:19.16 & $+$02:37:00.7 & 1.3 & 8.6 & 3.6 & 8.9 & $-$1.1 & 11.0 & 24.5 & 6.5 & 5.06 & 0.87\\
            \multicolumn{13}{c}{\dotfill\,\textbf{SGP-93302}\,\dotfill}\\
            \textbf{URG\,J000624.4$-$323018} &  \textbf{00:06:24.44} & \textbf{$-$32:30:17.7} & \textbf{32.1} & \textbf{7.1} & \textbf{59.6} & \textbf{8.3} & \textbf{59.6} & \textbf{8.9} & \textbf{32.0} & \textbf{1.3} & \textbf{1.03} & \textbf{1.00}\\
            URG\,J00067.7$-$322638 &  00:06:07.68 & $-$32:26:38.0 & 24.0 & 7.7 & 49.7 & 9.3 & 60.9 & 9.1 & 32.4 & 1.9 & 1.03 & 1.00\\
            URG\,J000621.3$-$32328 &  00:06:21.31 & $-$32:32:07.9 & 15.8 & 7.5 & 27.3 & 7.8 & 22.9 & 8.5 & 13.3 & 1.1 & 1.05 & 1.00\\
            URG\,J000619.9$-$323126 &  00:06:19.92 & $-$32:31:26.2 & 23.2 & 7.6 & 21.6 & 8.0 & 21.2 & 8.4 & 5.3 & 1.2 & 1.43 & 0.99\\
            URG\,J00066.1$-$323016 &  00:06:06.14 & $-$32:30:16.1 & 40.1 & 7.2 & 23.2 & 8.8 & 13.9 & 8.7 & 7.3 & 1.7 & 1.48 & 0.96\\
            URG\,J000619.9$-$322847 &  00:06:19.91 & $-$32:28:46.8 & 23.7 & 7.8 & 23.3 & 8.5 & 18.6 & 8.8 & 4.7 & 1.2 & 1.57 & 0.85\\
            URG\,J000634.0$-$323138 &  00:06:34.00 & $-$32:31:38.1 & 11.8 & 7.2 & 10.7 & 7.7 & 10.8 & 8.1 & 4.0 & 1.0 & 1.67 & 0.75\\
            URG\,J00068.5$-$323338 &  00:06:08.47 & $-$32:33:38.2 & 6.7 & 7.4 & 6.3 & 8.1 & 5.3 & 8.0 & 5.7 & 1.7 & 1.79 & 0.61\\
            \multicolumn{13}{c}{\dotfill\,\textbf{ELAIS-S1-18}\,\dotfill}\\
            \textbf{URG\,J002851.3$-$431353} &  \textbf{00:28:51.31} & \textbf{$-$43:13:52.8} & \textbf{33.4} & \textbf{5.7} & \textbf{48.8} & \textbf{7.0} & \textbf{46.5} & \textbf{7.3} & \textbf{17.8} & \textbf{2.9} & \textbf{1.44} & \textbf{1.00}\\
            URG\,J00297.7$-$431036 &  00:29:07.74 & $-$43:10:36.2 & 35.7 & 5.6 & 43.5 & 6.6 & 42.4 & 7.4 & 18.9 & 3.4 & 1.66 & 1.00\\
            URG\,J002913.4$-$43077 &  00:29:13.39 & $-$43:07:07.0 & 6.7 & 5.1 & $-$0.2 & 6.2 & 6.5 & 7.1 & 25.1 & 5.9 & 3.20 & 0.99\\
            URG\,J00294.0$-$430737 &  00:29:03.95 & $-$43:07:37.2 & 17.7 & 5.8 & 11.1 & 6.6 & 4.2 & 7.2 & 18.0 & 4.6 & 4.60 & 0.87\\
            URG\,J002919.0$-$430817 &  00:29:19.01 & $-$43:08:16.8 & $-$1.6 & 5.3 & $-$1.8 & 6.2 & 7.7 & 7.5 & 17.5 & 5.9 & 5.29 & 0.69\\
            \multicolumn{13}{c}{\dotfill\,\textbf{ELAIS-S1-26}\,\dotfill}\\
            \textbf{URG\,J003352.4$-$452015} &  \textbf{00:33:52.39} & \textbf{$-$45:20:14.6} & \textbf{24.5} & \textbf{6.6} & \textbf{37.0} & \textbf{8.3} & \textbf{43.1} & \textbf{9.6} & \textbf{12.6} & \textbf{2.6} & \textbf{1.57} & \textbf{1.00}\\
            URG\,J003410.4$-$452230 &  00:34:10.40 & $-$45:22:29.7 & 45.7 & 9.2 & 37.6 & 9.1 & 18.6 & 10.2 & 14.8 & 3.1 & 1.55 & 1.00\\
            URG\,J003347.9$-$451441 &  00:33:47.86 & $-$45:14:40.8 & 11.6 & 6.1 & 20.6 & 6.9 & 13.8 & 7.3 & 15.9 & 4.6 & 3.11 & 0.78\\
            \multicolumn{13}{c}{\dotfill\,\textbf{SGP-208073}\,\dotfill}\\
            \textbf{URG\,J003533.9$-$280260} &  \textbf{00:35:33.90} & \textbf{$-$28:02:59.5} & \textbf{27.7} & \textbf{7.7} & \textbf{37.4} & \textbf{8.8} & \textbf{47.6} & \textbf{9.7} & \textbf{19.2} & \textbf{1.8} & \textbf{1.16} & \textbf{1.00}\\
            URG\,J003540.1$-$280459 &  00:35:40.07 & $-$28:04:58.7 & 32.3 & 7.6 & 31.2 & 8.5 & 28.1 & 9.8 & 12.4 & 2.0 & 1.22 & 1.00\\
            URG\,J003536.4$-$280143 &  00:35:36.37 & $-$28:01:43.3 & 14.7 & 7.9 & 16.8 & 9.0 & 23.4 & 9.7 & 7.1 & 2.0 & 2.23 & 0.72\\
            \multicolumn{13}{c}{\dotfill\,\textbf{ELAIS-S1-29}\,\dotfill}\\
            \textbf{URG\,J003756.6$-$421519$^{\dagger}$} &  \textbf{00:37:56.62} & \textbf{$-$42:15:19.0} & \textbf{24.9} & \textbf{6.2} & \textbf{35.1} & \textbf{7.5} & \textbf{43.5} & \textbf{8.0} & \textbf{7.7} & \textbf{2.3} & \textbf{\textemdash} & \textbf{\textemdash}\\
            URG\,J003831.5$-$421418 &  00:38:31.49 & $-$42:14:18.4 & $-$2.3 & 5.7 & 1.8 & 6.6 & $-$1.4 & 7.3 & 20.0 & 4.8 & 2.02 & 0.95\\
            URG\,J003744.9$-$421240 &  00:37:44.90 & $-$42:12:39.6 & 41.7 & 6.7 & 45.8 & 7.7 & 27.8 & 8.3 & 10.3 & 2.7 & 2.59 & 0.90\\
            URG\,J003811.7$-$42198 &  00:38:11.74 & $-$42:19:08.0 & 0.5 & 5.5 & $-$0.5 & 6.1 & 0.2 & 7.2 & 16.4 & 4.3 & 2.73 & 0.87\\
            URG\,J003825.5$-$42128 &  00:38:25.48 & $-$42:12:08.1 & 59.5 & 6.0 & 29.6 & 6.9 & 15.3 & 8.0 & 15.7 & 4.5 & 3.14 & 0.78\\
            URG\,J00388.4$-$421742 &  00:38:08.44 & $-$42:17:41.7 & 23.8 & 5.7 & 33.7 & 6.4 & 22.8 & 7.7 & 9.3 & 2.7 & 3.22 & 0.72\\
            \multicolumn{13}{c}{\dotfill\,\textbf{SGP-354388}\,\dotfill}\\
            \textbf{URG\,J004223.7$-$334325} &  \textbf{00:42:23.73} & \textbf{$-$33:43:25.0} & \textbf{15.4} & \textbf{8.6} & \textbf{47.6} & \textbf{8.8} & \textbf{59.7} & \textbf{9.8} & \textbf{34.3} & \textbf{1.2} & \textbf{1.04} & \textbf{1.00}\\
            URG\,J004223.5$-$334350 &  00:42:23.46 & $-$33:43:49.6 & 23.4 & 8.5 & 35.3 & 8.9 & 33.8 & 9.9 & 17.5 & 1.2 & 1.05 & 1.00\\
            URG\,J004233.2$-$33444 &  00:42:33.16 & $-$33:44:04.2 & 12.8 & 8.1 & 14.3 & 8.9 & 14.8 & 9.5 & 9.4 & 1.2 & 1.09 & 1.00\\
            URG\,J004223.2$-$334117 &  00:42:23.25 & $-$33:41:16.9 & 18.8 & 8.0 & 13.8 & 9.0 & 17.6 & 9.6 & 8.7 & 1.2 & 1.11 & 1.00\\
            URG\,J004216.1$-$334138 &  00:42:16.11 & $-$33:41:37.8 & 63.5 & 8.2 & 56.3 & 9.2 & 28.9 & 9.7 & 7.9 & 1.2 & 1.13 & 1.00\\
            URG\,J004219.8$-$334435 &  00:42:19.79 & $-$33:44:35.2 & 16.8 & 8.7 & 34.0 & 8.9 & 34.1 & 10.0 & 7.2 & 1.2 & 1.16 & 1.00\\
            URG\,J004212.9$-$334544 &  00:42:12.86 & $-$33:45:43.5 & 5.5 & 8.6 & 8.7 & 9.0 & 3.8 & 10.3 & 5.5 & 1.2 & 1.30 & 0.99\\
            URG\,J004210.1$-$334040 &  00:42:10.09 & $-$33:40:40.0 & 1.8 & 8.6 & $-$1.1 & 8.6 & $-$9.0 & 9.6 & 4.9 & 1.4 & 1.57 & 0.75\\
            URG\,J004228.5$-$334925 &  00:42:28.53 & $-$33:49:24.6 & $-$4.0 & 8.6 & $-$1.1 & 9.2 & $-$15.2 & 10.3 & 10.9 & 2.8 & 1.49 & 0.72\\
            \multicolumn{13}{c}{\dotfill\,\textbf{SGP-380990}\,\dotfill}\\
            \textbf{URG\,J004614.6$-$321828} &  \textbf{00:46:14.55} & \textbf{$-$32:18:28.1} & \textbf{20.4} & \textbf{8.2} & \textbf{43.1} & \textbf{8.9} & \textbf{46.6} & \textbf{9.3} & \textbf{10.4} & \textbf{1.6} & \textbf{1.18} & \textbf{1.00}\\
            URG\,J004620.2$-$32209 &  00:46:20.19 & $-$32:20:08.5 & 24.3 & 8.5 & 29.2 & 9.0 & 34.3 & 9.3 & 9.2 & 1.8 & 1.31 & 1.00\\
            URG\,J00464.4$-$321844 &  00:46:04.41 & $-$32:18:44.2 & 23.2 & 8.0 & 17.4 & 8.6 & 8.3 & 9.3 & 7.6 & 2.2 & 2.18 & 0.69\\
            \multicolumn{13}{c}{\dotfill\,\textbf{HeLMS-10}\,\dotfill}\\
            \textbf{URG\,J005258.6$+$061318} &  \textbf{00:52:58.61} & \textbf{$+$06:13:18.2} & \textbf{68.9} & \textbf{11.5} & \textbf{105.4} & \textbf{11.2} & \textbf{124.3} & \textbf{11.7} & \textbf{81.7} & \textbf{4.7} & \textbf{2.19} & \textbf{1.00}\\
            URG\,J00532.4$+$061113 &  00:53:02.41 & $+$06:11:12.9 & 7.3 & 9.8 & $-$3.7 & 10.7 & 6.7 & 12.3 & 23.8 & 5.8 & 7.62 & 0.98\\
            URG\,J005310.4$+$061510 &  00:53:10.40 & $+$06:15:09.5 & 45.3 & 11.4 & 51.6 & 11.8 & 29.5 & 12.5 & 38.3 & 8.4 & 3.59 & 0.98\\
            \multicolumn{13}{c}{\dotfill\,\textbf{SGP-221606}\,\dotfill}\\
            \textbf{URG\,J011918.9$-$294516} &  \textbf{01:19:18.93} & \textbf{$-$29:45:15.7} & \textbf{34.9} & \textbf{7.7} & \textbf{53.6} & \textbf{8.8} & \textbf{52.1} & \textbf{9.9} & \textbf{20.3} & \textbf{3.9} & \textbf{1.82} & \textbf{1.00}\\
            URG\,J011915.9$-$294748 &  01:19:15.86 & $-$29:47:47.6 & 1.2 & 8.0 & 0.0 & 9.0 & 22.6 & 9.1 & 16.2 & 4.1 & 3.80 & 0.94\\
            URG\,J01191.8$-$294342 &  01:19:01.83 & $-$29:43:42.0 & 7.9 & 7.6 & 7.2 & 9.1 & $-$3.1 & 9.9 & 17.9 & 5.5 & 5.92 & 0.69\\
            URG\,J01199.6$-$294241 &  01:19:09.59 & $-$29:42:40.6 & $-$0.1 & 7.7 & $-$0.9 & 9.6 & 0.5 & 9.8 & 15.5 & 4.6 & 5.87 & 0.61\\
            \multicolumn{13}{c}{\dotfill\,\textbf{SGP-146631}\,\dotfill}\\
            URG\,J013155.8$-$311147 &  01:31:55.82 & $-$31:11:47.0 & 26.1 & 7.4 & 32.7 & 7.5 & 39.9 & 8.0 & 15.0 & 3.3 & 1.87 & 0.98\\
            \textbf{URG\,J01324.5$-$311239} &  \textbf{01:32:04.46} & \textbf{$-$31:12:38.5} & \textbf{47.2} & \textbf{7.9} & \textbf{78.7} & \textbf{7.6} & \textbf{67.9} & \textbf{8.5} & \textbf{11.5} & \textbf{3.2} & \textbf{3.92} & \textbf{0.94}\\
            URG\,J013215.5$-$310837 &  01:32:15.51 & $-$31:08:36.6 & 5.7 & 8.5 & 8.6 & 8.8 & 6.4 & 9.4 & 14.9 & 4.0 & 3.73 & 0.85\\
            \hline
            \multicolumn{13}{r}{Continued on next page ...}\\
        \end{tabular}
    \end{scriptsize}
    \end{center}
\end{table*}

\addtocounter{table}{-1} 

\begin{table*}
    \begin{center}
    \caption{Cont...}
    \begin{scriptsize}
        \begin{tabular}{l lr r@{\,$\pm$\,}l r@{\,$\pm$\,}l r@{\,$\pm$\,}l r@{\,$\pm$\,}l c c}
            \hline\hline
            \multicolumn{1}{l}{IAU name} &
            \multicolumn{2}{c}{$\alpha$ \quad (J2000) \quad $\delta$} &
            \multicolumn{2}{c}{$S_{250}$} &
            \multicolumn{2}{c}{$S_{350}$} &
            \multicolumn{2}{c}{$S_{500}$} &
            \multicolumn{2}{c}{$S_{870}$} &
            \multicolumn{1}{c}{$\mathcal{B}$} &
            \multicolumn{1}{c}{$\mathcal{F}$}\\
            \multicolumn{1}{l}{} &
            \multicolumn{1}{c}{$^{\textrm{h}}$ \, $^{\textrm{m}}$ \, $^{\textrm{s}}$ } &
            \multicolumn{1}{c}{$\, ^{\circ} \quad ' \quad ''$} &
            \multicolumn{2}{c}{$\millijanksy{}\,\beam{}^{-1}$} &
            \multicolumn{2}{c}{$\millijanksy{}\,\beam{}^{-1}$} &
            \multicolumn{2}{c}{$\millijanksy{}\,\beam{}^{-1}$} &
            \multicolumn{2}{c}{$\millijanksy{}$} &
            \multicolumn{1}{c}{} \\
            \hline
            \multicolumn{13}{c}{\dotfill\,\textbf{SGP-278539}\,\dotfill}\\
            \textbf{URG\,J01428.2$-$323426$^{\dagger}$} &  \textbf{01:42:08.20} & \textbf{$-$32:34:26.3} & \textbf{22.7} & \textbf{8.3} & \textbf{39.0} & \textbf{9.2} & \textbf{50.7} & \textbf{9.5} & \textbf{8.7} & \textbf{2.8} & \textbf{\textemdash} & \textbf{\textemdash}\\
            URG\,J014226.2$-$323324 &  01:42:26.25 & $-$32:33:23.8 & 7.0 & 8.4 & 2.6 & 8.5 & 8.2 & 9.2 & 17.2 & 3.2 & 1.40 & 1.00\\
            URG\,J01421.6$-$323624 &  01:42:01.58 & $-$32:36:23.8 & 6.7 & 8.7 & 7.4 & 9.0 & 9.3 & 9.0 & 14.1 & 2.9 & 1.49 & 0.99\\
            URG\,J014214.4$-$32290 &  01:42:14.41 & $-$32:29:00.2 & 6.1 & 8.1 & 9.5 & 8.6 & 8.6 & 9.6 & 15.7 & 4.2 & 2.83 & 0.92\\
            URG\,J014218.2$-$32352 &  01:42:18.19 & $-$32:35:01.5 & $-$0.1 & 8.3 & $-$7.2 & 8.7 & $-$2.8 & 9.2 & 9.6 & 2.8 & 3.26 & 0.65\\
            \multicolumn{13}{c}{\dotfill\,\textbf{SGP-142679}\,\dotfill}\\
            \textbf{URG\,J014456.9$-$284146} &  \textbf{01:44:56.88} & \textbf{$-$28:41:46.0} & \textbf{29.9} & \textbf{8.1} & \textbf{65.0} & \textbf{9.8} & \textbf{71.7} & \textbf{9.9} & \textbf{12.9} & \textbf{2.8} & \textbf{1.59} & \textbf{1.00}\\
            URG\,J014448.8$-$283535 &  01:44:48.78 & $-$28:35:35.4 & 7.5 & 7.7 & $-$9.0 & 8.5 & 10.5 & 8.9 & 18.3 & 4.2 & 1.88 & 0.97\\
            URG\,J01456.7$-$284457 &  01:45:06.66 & $-$28:44:57.3 & 97.2 & 8.5 & 101.8 & 9.8 & 82.2 & 9.8 & 15.6 & 3.5 & 1.70 & 0.96\\
            \multicolumn{13}{c}{\dotfill\,\textbf{XMM-LSS-15}\,\dotfill}\\
            \textbf{URG\,J021745.3$-$030912} &  \textbf{02:17:45.30} & \textbf{$-$03:09:12.3} & \textbf{12.6} & \textbf{6.2} & \textbf{22.2} & \textbf{7.2} & \textbf{24.0} & \textbf{7.8} & \textbf{17.6} & \textbf{3.0} & \textbf{1.47} & \textbf{1.00}\\
            URG\,J021757.1$-$030753 &  02:17:57.12 & $-$03:07:53.0 & 56.8 & 6.5 & 34.5 & 7.4 & 14.6 & 7.6 & 11.5 & 2.9 & 2.67 & 0.90\\
            URG\,J021737.3$-$03128 &  02:17:37.29 & $-$03:12:08.0 & 0.5 & 6.7 & $-$0.3 & 7.5 & 4.6 & 8.2 & 10.8 & 3.2 & 3.55 & 0.69\\
            \multicolumn{13}{c}{\dotfill\,\textbf{XMM-LSS-30}\,\dotfill}\\
            \textbf{URG\,J022656.6$-$032711} &  \textbf{02:26:56.60} & \textbf{$-$03:27:11.1} & \textbf{25.6} & \textbf{6.3} & \textbf{44.8} & \textbf{7.0} & \textbf{61.6} & \textbf{7.1} & \textbf{23.3} & \textbf{2.0} & \textbf{1.16} & \textbf{1.00}\\
            URG\,J022644.9$-$032510 &  02:26:44.90 & $-$03:25:10.1 & 44.2 & 6.3 & 65.6 & 6.8 & 63.9 & 7.5 & 18.8 & 2.6 & 1.23 & 1.00\\
            URG\,J022630.2$-$032530 &  02:26:30.16 & $-$03:25:30.0 & 20.7 & 5.7 & 24.3 & 7.0 & 18.4 & 7.7 & 29.8 & 6.4 & 2.04 & 0.97\\
            URG\,J02270.8$-$032541 &  02:27:00.81 & $-$03:25:41.0 & 10.3 & 6.5 & 10.3 & 7.1 & 13.9 & 7.8 & 7.6 & 2.0 & 3.38 & 0.93\\
            URG\,J022650.0$-$032542 &  02:26:50.00 & $-$03:25:41.9 & 28.9 & 6.5 & 28.6 & 6.7 & 18.0 & 7.3 & 7.6 & 2.1 & 3.53 & 0.61\\
            \multicolumn{13}{c}{\dotfill\,\textbf{CDFS-13}\,\dotfill}\\
            \textbf{URG\,J03370.7$-$292148} &  \textbf{03:37:00.72} & \textbf{$-$29:21:48.0} & \textbf{41.1} & \textbf{5.9} & \textbf{51.0} & \textbf{7.1} & \textbf{55.4} & \textbf{7.2} & \textbf{26.2} & \textbf{3.5} & \textbf{1.45} & \textbf{1.00}\\
            URG\,J03370.3$-$291746 &  03:37:00.35 & $-$29:17:45.8 & 23.3 & 5.8 & 20.6 & 6.8 & 10.5 & 6.8 & 37.6 & 5.9 & 1.45 & 1.00\\
            URG\,J033655.2$-$292627 &  03:36:55.23 & $-$29:26:26.9 & 11.6 & 7.3 & 15.7 & 7.3 & 7.6 & 7.0 & 17.8 & 5.0 & 5.46 & 0.75\\
            \multicolumn{13}{c}{\dotfill\,\textbf{ADF-S-27}\,\dotfill}\\
            \textbf{URG\,J043657.0$-$543813} &  \textbf{04:36:57.01} & \textbf{$-$54:38:13.2} & \textbf{16.5} & \textbf{6.0} & \textbf{24.0} & \textbf{7.1} & \textbf{28.2} & \textbf{7.8} & \textbf{25.3} & \textbf{1.8} & \textbf{1.24} & \textbf{1.00}\\
            URG\,J043729.9$-$54365 &  04:37:29.90 & $-$54:36:04.5 & 14.9 & 6.8 & 17.9 & 7.9 & 19.9 & 7.7 & 18.0 & 3.3 & 1.34 & 1.00\\
            URG\,J04374.7$-$543914 &  04:37:04.65 & $-$54:39:13.7 & 3.7 & 6.0 & 2.4 & 8.0 & 0.4 & 7.8 & 10.2 & 1.9 & 1.35 & 1.00\\
            URG\,J043717.4$-$54356 &  04:37:17.35 & $-$54:35:06.2 & 13.5 & 7.1 & 21.7 & 7.9 & 25.5 & 7.6 & 8.8 & 2.4 & 2.35 & 0.98\\
            URG\,J043717.5$-$543528 &  04:37:17.49 & $-$54:35:28.3 & 48.7 & 7.1 & 54.5 & 7.8 & 49.0 & 7.6 & 6.2 & 2.3 & 2.59 & 0.93\\
            URG\,J04377.5$-$54341 &  04:37:07.51 & $-$54:34:00.6 & 34.2 & 6.6 & 27.3 & 7.9 & 13.6 & 7.9 & 8.9 & 2.3 & 2.18 & 0.93\\
            URG\,J043649.4$-$54408 &  04:36:49.44 & $-$54:40:08.4 & 7.9 & 5.4 & 13.9 & 6.9 & 5.2 & 8.2 & 9.0 & 2.2 & 2.00 & 0.78\\
            \multicolumn{13}{c}{\dotfill\,\textbf{ADF-S-32}\,\dotfill}\\
            \textbf{URG\,J044410.1$-$534949$^{\dagger}$} &  \textbf{04:44:10.13} & \textbf{$-$53:49:49.1} & \textbf{13.1} & \textbf{6.0} & \textbf{16.6} & \textbf{6.8} & \textbf{20.8} & \textbf{8.0} & \textbf{5.5} & \textbf{2.8} & \textbf{\textemdash} & \textbf{\textemdash}\\
            URG\,J04450.4$-$53496 &  04:45:00.43 & $-$53:49:06.2 & 9.3 & 5.6 & 0.9 & 6.8 & $-$0.6 & 8.0 & 20.0 & 6.0 & 3.81 & 0.78\\
            \multicolumn{13}{c}{\dotfill\,\textbf{G09-83808}\,\dotfill}\\
            \textbf{URG\,J090045.7$+$004124} &  \textbf{09:00:45.74} & \textbf{$+$00:41:24.1} & \textbf{10.9} & \textbf{7.5} & \textbf{24.1} & \textbf{8.3} & \textbf{42.4} & \textbf{8.7} & \textbf{26.3} & \textbf{1.3} & \textbf{1.06} & \textbf{1.00}\\
            URG\,J090032.8$+$004313 &  09:00:32.77 & $+$00:43:13.0 & 79.5 & 6.6 & 69.2 & 7.7 & 40.9 & 8.1 & 18.5 & 1.4 & 1.06 & 1.00\\
            URG\,J090019.4$+$004016 &  09:00:19.37 & $+$00:40:15.7 & 5.6 & 6.4 & $-$2.3 & 7.4 & $-$8.1 & 7.3 & 18.3 & 3.3 & 1.18 & 1.00\\
            URG\,J090057.3$+$00415 &  09:00:57.28 & $+$00:41:04.8 & 30.1 & 7.3 & 32.5 & 8.2 & 28.1 & 9.0 & 5.5 & 1.1 & 1.25 & 1.00\\
            URG\,J090054.2$+$004343 &  09:00:54.21 & $+$00:43:43.1 & 19.2 & 7.5 & 18.8 & 8.2 & 19.9 & 8.9 & 3.7 & 1.1 & 1.66 & 0.75\\
            URG\,J090057.1$+$004039 &  09:00:57.08 & $+$00:40:39.4 & 26.9 & 7.4 & 33.6 & 8.4 & 32.7 & 9.0 & 3.2 & 1.2 & 1.66 & 0.61\\
            URG\,J090037.1$+$003624 &  09:00:37.14 & $+$00:36:24.3 & 72.9 & 6.6 & 65.4 & 7.4 & 43.8 & 8.3 & 8.6 & 2.4 & 1.60 & 0.61\\
            \multicolumn{13}{c}{\dotfill\,\textbf{G15-82684}\,\dotfill}\\
            URG\,J14506.3$+$015038 &  14:50:06.29 & $+$01:50:38.4 & 31.5 & 7.1 & 37.9 & 7.4 & 45.4 & 8.9 & 17.4 & 1.5 & 1.07 & 1.00\\
            \textbf{URG\,J145013.1$+$014810} &  \textbf{14:50:13.10} & \textbf{$+$01:48:09.8} & \textbf{17.7} & \textbf{7.5} & \textbf{36.4} & \textbf{8.1} & \textbf{39.0} & \textbf{9.2} & \textbf{17.2} & \textbf{1.5} & \textbf{1.08} & \textbf{1.00}\\
            URG\,J145012.1$+$015158 &  14:50:12.06 & $+$01:51:57.5 & 30.5 & 7.3 & 34.0 & 7.2 & 34.4 & 8.7 & 11.2 & 1.8 & 1.17 & 1.00\\
            URG\,J145015.4$+$015237 &  14:50:15.43 & $+$01:52:37.1 & 18.5 & 7.3 & 33.9 & 7.6 & 37.9 & 8.5 & 13.2 & 2.3 & 1.21 & 1.00\\
            URG\,J145025.7$+$015115 &  14:50:25.66 & $+$01:51:14.8 & 21.9 & 7.8 & 31.7 & 7.7 & 22.8 & 9.1 & 7.1 & 1.9 & 1.68 & 1.00\\
            URG\,J145023.8$+$01514 &  14:50:23.82 & $+$01:51:04.4 & 13.7 & 7.6 & 9.8 & 7.7 & 23.9 & 8.9 & 5.4 & 1.7 & 1.92 & 0.92\\
            \multicolumn{13}{c}{\dotfill\,\textbf{SGP-433089}\,\dotfill}\\
            \textbf{URG\,J222737.4$-$333835} &  \textbf{22:27:37.37} & \textbf{$-$33:38:34.7} & \textbf{28.3} & \textbf{9.2} & \textbf{36.8} & \textbf{10.0} & \textbf{35.1} & \textbf{10.8} & \textbf{8.1} & \textbf{1.1} & \textbf{1.12} & \textbf{1.00}\\
            URG\,J222725.2$-$333920 &  22:27:25.22 & $-$33:39:19.5 & 35.3 & 9.4 & 38.8 & 10.4 & 20.2 & 11.3 & 8.1 & 1.4 & 1.16 & 1.00\\
            URG\,J222747.9$-$333533 &  22:27:47.89 & $-$33:35:32.7 & 21.7 & 9.4 & 32.0 & 9.8 & 25.1 & 10.9 & 7.5 & 1.3 & 1.17 & 1.00\\
            URG\,J222731.1$-$33404 &  22:27:31.09 & $-$33:40:03.7 & 5.0 & 9.1 & $-$8.0 & 10.4 & $-$1.1 & 11.1 & 6.3 & 1.2 & 1.21 & 1.00\\
            URG\,J222733.7$-$333440 &  22:27:33.67 & $-$33:34:40.2 & 40.2 & 9.7 & 43.8 & 10.0 & 28.8 & 10.7 & 6.4 & 1.3 & 1.24 & 1.00\\
            URG\,J222737.7$-$333727 &  22:27:37.70 & $-$33:37:26.8 & 49.7 & 9.5 & 47.2 & 9.9 & 23.2 & 10.5 & 5.1 & 1.1 & 1.31 & 0.99\\
            URG\,J222730.4$-$333534 &  22:27:30.44 & $-$33:35:33.6 & 18.5 & 9.5 & 18.8 & 9.9 & 18.2 & 11.0 & 5.5 & 1.3 & 1.35 & 0.96\\
            URG\,J222750.1$-$334153 &  22:27:50.14 & $-$33:41:53.2 & 10.3 & 9.9 & 11.5 & 10.3 & 19.9 & 10.8 & 7.0 & 1.8 & 1.50 & 0.93\\
            URG\,J222753.8$-$333529 &  22:27:53.81 & $-$33:35:28.5 & 4.3 & 9.7 & 38.1 & 10.2 & 16.2 & 10.9 & 6.4 & 1.7 & 1.55 & 0.90\\
            URG\,J222727.8$-$334056 &  22:27:27.79 & $-$33:40:56.3 & 17.5 & 9.6 & 27.9 & 10.5 & 25.9 & 11.1 & 5.2 & 1.3 & 1.44 & 0.85\\
            URG\,J222744.7$-$333741 &  22:27:44.74 & $-$33:37:40.8 & 5.5 & 9.4 & 37.0 & 9.9 & 27.6 & 10.8 & 4.5 & 1.1 & 1.46 & 0.75\\
            \hline
        \end{tabular}
    \end{scriptsize}
    \end{center}
\vspace{-\baselineskip}

\noindent
$^{\ddagger}$ SPIRE flux densities have been boosted to reflect the radial offset
of a LABOCA source.
Additionally, $870\textrm{-}\micron{}$ flux densities have been deboosted.

\noindent
$^{\dagger}$ Signpost ultra-red galaxies that are undetected. We report the
peak flux density and r.m.s.\
values for these sources within a $45''$ aperture centered on the
telescope pointing position. We do not provide flux boosting ($\mathcal{B}$) or fidelity ($\mathcal{F}$) values.

\noindent
\textbf{Note}. Targets are listed in order of increasing right ascension and are
highlighted in \textbf{bold}. Each source detected in a given field is
subsequently listed in increasing order of detected $S/N$.
\end{table*}

\begin{table*}
    \begin{center}
    \caption{Targets and their photometric redshift properties.}
    \label{tab: photometric redshifts}
    \begin{scriptsize}
        \begin{tabular}{l c c c l c c c}
        \hline
        \hline
        \multicolumn{1}{l}{ID} &
        \multicolumn{1}{c}{$z_\textrm{phot}^{\dagger}$} &
        \multicolumn{1}{c}{$\chi^2$} &
        \multicolumn{1}{c}{$\log_{10}\left(\lfir{}\right)$} &
        \multicolumn{1}{l}{ID} &
        \multicolumn{1}{c}{$z_\textrm{phot}^{\dagger}$} &
        \multicolumn{1}{c}{$\chi^2$} &
        \multicolumn{1}{c}{$\log_{10}\left(\lfir{}\right)$} \\
        \multicolumn{1}{c}{} &
        \multicolumn{1}{c}{} &
        \multicolumn{1}{c}{} &
        \multicolumn{1}{c}{[$\lsol{}$]} &
        \multicolumn{1}{c}{} &
        \multicolumn{1}{c}{} &
        \multicolumn{1}{c}{} &
        \multicolumn{1}{c}{[$\lsol{}$]}\\
        \hline
        \multicolumn{8}{c}{\dotfill\,\textbf{SGP-28124}\,\dotfill}\\
        \textbf{URG\,J000124.9$-$354212} & $\mathbf{3.4^{+0.1}_{-0.1}}$ & $\mathbf{5.99}$ & $\mathbf{13.50^{+0.02}_{-0.02}}$ & URG\,J000145.0$-$353822 & $2.5^{+0.2}_{-0.2}$ & 0.19 & $13.05^{+0.05}_{-0.06}$\\
        URG\,J00014.2$-$354123 & $3.6^{+2.0}_{-0.8}$ & 0.36 & $12.59^{+0.30}_{-0.19}$ & URG\,J000122.9$-$354211 & $2.5^{+0.2}_{-0.2}$ & 32.37 & $12.95^{+0.05}_{-0.05}$\\
        URG\,J000138.5$-$35442 & $3.7^{+6.3}_{-1.4}$ & 1.02 & $12.38^{+0.64}_{-0.35}$ & URG\,J000115.9$-$35411 & $1.6^{+0.4}_{-0.4}$ & 0.69 & $12.36^{+0.17}_{-0.26}$\\
        URG\,J000129.4$-$354416 & $1.6^{+0.4}_{-0.5}$ & 2.20 & $12.35^{+0.18}_{-0.28}$\\
        \multicolumn{8}{c}{\dotfill\,\textbf{HeLMS-42}\,\dotfill}\\
        \textbf{URG\,J00034.2$+$024114} & $\mathbf{3.2^{+0.2}_{-0.2}}$ & $\mathbf{3.30}$ & $\mathbf{13.26^{+0.04}_{-0.05}}$ & URG\,J000319.2$+$02371$^{\ddagger}$ & \textemdash & \textemdash & \textemdash\\
        \multicolumn{8}{c}{\dotfill\,\textbf{SGP-93302}\,\dotfill}\\
        \textbf{URG\,J000624.4$-$323018} & $\mathbf{3.7^{+0.2}_{-0.2}}$ & $\mathbf{0.14}$ & $\mathbf{13.41^{+0.03}_{-0.03}}$ & URG\,J00067.7$-$322638 & $4.4^{+0.2}_{-0.2}$ & 0.02 & $13.45^{+0.04}_{-0.03}$\\
        URG\,J000621.3$-$32328 & $3.6^{+0.4}_{-0.3}$ & 0.26 & $13.02^{+0.08}_{-0.06}$ & URG\,J000619.9$-$323126 & $2.2^{+0.4}_{-0.4}$ & 0.64 & $12.50^{+0.12}_{-0.15}$\\
        URG\,J00066.1$-$323016 & $1.8^{+0.4}_{-0.5}$ & 1.05 & $12.58^{+0.17}_{-0.29}$ & URG\,J000619.9$-$322847 & $1.9^{+0.4}_{-0.4}$ & 0.42 & $12.43^{+0.14}_{-0.18}$\\
        URG\,J000634.0$-$323138 & $2.3^{+0.8}_{-0.7}$ & 0.13 & $12.33^{+0.22}_{-0.28}$ & URG\,J00068.5$-$323338$^{\ddagger}$ & \textemdash & \textemdash & \textemdash\\
        \multicolumn{8}{c}{\dotfill\,\textbf{ELAISS1-18}\,\dotfill}\\
        \textbf{URG\,J002851.3$-$431353} & $\mathbf{2.9^{+0.2}_{-0.2}}$ & $\mathbf{0.87}$ & $\mathbf{13.03^{+0.05}_{-0.06}}$ & URG\,J00297.7$-$431036 & $2.8^{+0.2}_{-0.2}$ & 0.81 & $13.05^{+0.06}_{-0.07}$\\
        URG\,J002913.4$-$43077 & $6.3^{+3.7}_{-4.1}$ & 1.38 & $12.87^{+0.28}_{-0.71}$ & URG\,J00294.0$-$430737 & $1.4^{+1.1}_{-1.4}$ & 0.37 & $12.08^{+0.44}_{-0.44}$\\
        URG\,J002919.0$-$430817 & $6.3^{+3.7}_{-4.1}$ & 0.75 & $12.52^{+0.29}_{-0.76}$\\
        \multicolumn{8}{c}{\dotfill\,\textbf{ELAISS1-26}\,\dotfill}\\
        \textbf{URG\,J003352.4$-$452015} & $\mathbf{2.8^{+0.3}_{-0.3}}$ & $\mathbf{2.47}$ & $\mathbf{12.88^{+0.07}_{-0.08}}$ & URG\,J003410.4$-$452230 & $2.2^{+0.4}_{-0.5}$ & 1.44 & $12.83^{+0.13}_{-0.18}$\\
        URG\,J003347.9$-$451441 & $2.9^{+0.7}_{-0.7}$ & 0.16 & $12.60^{+0.15}_{-0.20}$\\
        \multicolumn{8}{c}{\dotfill\,\textbf{SGP-208073}\,\dotfill}\\
        \textbf{URG\,J003533.9$-$280260} & $\mathbf{3.6^{+0.3}_{-0.2}}$ & $\mathbf{0.96}$ & $\mathbf{13.19^{+0.05}_{-0.05}}$ & URG\,J003540.1$-$280459 & $2.7^{+0.3}_{-0.3}$ & 0.64 & $12.92^{+0.08}_{-0.09}$\\
        URG\,J003536.4$-$280143 & $2.5^{+0.6}_{-0.6}$ & 1.25 & $12.50^{+0.16}_{-0.20}$\\
        \multicolumn{8}{c}{\dotfill\,\textbf{ELAISS1-29}\,\dotfill}\\
        \textbf{URG\,J003756.6$-$421519} & $\mathbf{2.8^{+0.2}_{-0.3}}$ & $\mathbf{3.89}$ & $\mathbf{12.87^{+0.06}_{-0.07}}$ & URG\,J003831.5$-$421418$^{\ddagger}$ & \textemdash & \textemdash & \textemdash\\
        URG\,J003744.9$-$421240 & $2.0^{+0.3}_{-0.3}$ & 1.15 & $12.70^{+0.10}_{-0.12}$ & URG\,J003811.7$-$42198$^{\ddagger}$ & \textemdash & \textemdash & \textemdash\\
        URG\,J003825.5$-$42128 & $0.9^{+0.5}_{-0.7}$ & 0.34 & $12.34^{+0.35}_{-1.29}$ & URG\,J00388.4$-$421742 & $2.3^{+0.3}_{-0.3}$ & 1.66 & $12.64^{+0.10}_{-0.13}$\\
        \multicolumn{8}{c}{\dotfill\,\textbf{SGP-354388}\,\dotfill}\\
        \textbf{URG\,J004223.7$-$334325} & $\mathbf{4.2^{+0.2}_{-0.2}}$ & $\mathbf{0.19}$ & $\mathbf{13.37^{+0.04}_{-0.03}}$ & URG\,J004223.5$-$334350 & $3.5^{+0.3}_{-0.3}$ & 0.18 & $13.15^{+0.06}_{-0.06}$\\
        URG\,J004233.2$-$33444 & $3.7^{+0.9}_{-0.5}$ & 0.36 & $12.85^{+0.15}_{-0.11}$ & URG\,J004223.2$-$334117 & $3.2^{+0.6}_{-0.5}$ & 1.09 & $12.81^{+0.12}_{-0.11}$\\
        URG\,J004216.1$-$334138 & $1.8^{+0.2}_{-0.2}$ & 0.06 & $12.77^{+0.07}_{-0.09}$ & URG\,J004219.8$-$334435 & $2.6^{+0.3}_{-0.3}$ & 2.39 & $12.72^{+0.08}_{-0.09}$\\
        URG\,J004212.9$-$334544$^{\ddagger}$ & \textemdash & \textemdash & \textemdash & URG\,J004210.1$-$334040$^{\ddagger}$ & \textemdash & \textemdash & \textemdash\\
        URG\,J004228.5$-$334925$^{\ddagger}$ & \textemdash & \textemdash & \textemdash\\
        \multicolumn{8}{c}{\dotfill\,\textbf{SGP-380990}\,\dotfill}\\
        \textbf{URG\,J004614.6$-$321828} & $\mathbf{2.8^{+0.2}_{-0.2}}$ & $\mathbf{4.55}$ & $\mathbf{12.88^{+0.06}_{-0.06}}$ & URG\,J004620.2$-$32209 & $2.7^{+0.3}_{-0.3}$ & 1.34 & $12.77^{+0.09}_{-0.10}$\\
        URG\,J00464.4$-$321844 & $2.0^{+0.7}_{-1.0}$ & 0.23 & $12.43^{+0.24}_{-0.55}$\\
        \multicolumn{8}{c}{\dotfill\,\textbf{HeLMS-10}\,\dotfill}\\
        \textbf{URG\,J005258.6$+$061318} & $\mathbf{3.2^{+0.1}_{-0.2}}$ & $\mathbf{3.56}$ & $\mathbf{13.48^{+0.03}_{-0.04}}$ & URG\,J00532.4$+$061113$^{\ddagger}$ & \textemdash & \textemdash & \textemdash\\
        URG\,J005310.4$+$061510 & $2.5^{+0.5}_{-0.5}$ & 0.12 & $12.97^{+0.13}_{-0.18}$\\
        \multicolumn{8}{c}{\dotfill\,\textbf{SGP-221606}\,\dotfill}\\
        \textbf{URG\,J011918.9$-$294516} & $\mathbf{2.8^{+0.2}_{-0.2}}$ & $\mathbf{1.59}$ & $\mathbf{13.04^{+0.06}_{-0.07}}$ & URG\,J011915.9$-$294748 & $4.4^{+1.7}_{-1.2}$ & 2.72 & $12.65^{+0.22}_{-0.22}$\\
        URG\,J01191.8$-$294342 & $1.3^{+3.7}_{-1.3}$ & 0.56 & $11.71^{+0.99}_{-0.99}$ & URG\,J01199.6$-$294241$^{\ddagger}$ & \textemdash & \textemdash & \textemdash\\
        \multicolumn{8}{c}{\dotfill\,\textbf{SGP-146631}\,\dotfill}\\
        URG\,J013155.8$-$311147 & $2.9^{+0.3}_{-0.3}$ & 2.26 & $12.89^{+0.08}_{-0.09}$ & \textbf{URG\,J01324.5$-$311239} & $\mathbf{2.4^{+0.2}_{-0.2}}$ & $\mathbf{20.97}$ & $\mathbf{13.03^{+0.05}_{-0.06}}$\\
        URG\,J013215.5$-$310837$^{\ddagger}$ & \textemdash & \textemdash & \textemdash\\
        \multicolumn{8}{c}{\dotfill\,\textbf{SGP-278539}\,\dotfill}\\
        \textbf{URG\,J01428.2$-$323426} & $\mathbf{2.9^{+0.3}_{-0.3}}$ & $\mathbf{4.62}$ & $\mathbf{12.94^{+0.07}_{-0.08}}$ & URG\,J014226.2$-$323324$^{\ddagger}$ & \textemdash & \textemdash & \textemdash\\
        URG\,J01421.6$-$323624 & $5.2^{+4.1}_{-1.4}$ & 0.23 & $12.91^{+0.37}_{-0.21}$ & URG\,J014214.4$-$32290 & $3.8^{+2.5}_{-1.6}$ & 0.06 & $12.63^{+0.34}_{-0.41}$\\
        URG\,J014218.2$-$32352$^{\ddagger}$ & \textemdash & \textemdash & \textemdash\\
        \multicolumn{8}{c}{\dotfill\,\textbf{SGP-142679}\,\dotfill}\\
        \textbf{URG\,J014456.9$-$284146} & $\mathbf{2.7^{+0.2}_{-0.2}}$ & $\mathbf{15.33}$ & $\mathbf{13.03^{+0.05}_{-0.06}}$ & URG\,J014448.8$-$283535 & $7.3^{+2.7}_{-2.2}$ & 2.69 & $12.96^{+0.19}_{-0.23}$\\
        URG\,J01456.7$-$284457 & $2.1^{+0.1}_{-0.1}$ & 8.07 & $13.12^{+0.05}_{-0.06}$\\
        \multicolumn{8}{c}{\dotfill\,\textbf{XMM-15}\,\dotfill}\\
        \textbf{URG\,J021745.3$-$030912} & $\mathbf{3.7^{+0.5}_{-0.5}}$ & $\mathbf{0.01}$ & $\mathbf{13.00^{+0.09}_{-0.09}}$ & URG\,J021757.1$-$030753 & $1.2^{+0.4}_{-0.5}$ & 0.09 & $12.51^{+0.23}_{-0.43}$\\
        URG\,J021737.3$-$03128$^{\ddagger}$ & \textemdash & \textemdash & \textemdash\\
        \multicolumn{8}{c}{\dotfill\,\textbf{XMM-30}\,\dotfill}\\
        \textbf{URG\,J022656.6$-$032711} & $\mathbf{3.5^{+0.2}_{-0.2}}$ & $\mathbf{3.23}$ & $\mathbf{13.19^{+0.03}_{-0.03}}$ & URG\,J022644.9$-$032510 & $2.8^{+0.2}_{-0.1}$ & 3.05 & $13.13^{+0.04}_{-0.04}$\\
        URG\,J022630.2$-$032530 & $2.9^{+0.7}_{-0.6}$ & 1.45 & $12.84^{+0.15}_{-0.18}$ & URG\,J02270.8$-$032541 & $2.5^{+0.8}_{-0.9}$ & 0.53 & $12.32^{+0.20}_{-0.32}$\\
        URG\,J022650.0$-$032542 & $1.8^{+0.4}_{-0.4}$ & 0.70 & $12.47^{+0.14}_{-0.21}$\\
        \multicolumn{8}{c}{\dotfill\,\textbf{CDFS-13}\,\dotfill}\\
        \textbf{URG\,J03370.7$-$292148} & $\mathbf{3.0^{+0.2}_{-0.2}}$ & $\mathbf{1.51}$ & $\mathbf{13.21^{+0.05}_{-0.05}}$ & URG\,J03370.3$-$291746 & $3.0^{+2.3}_{-0.8}$ & 13.23 & $12.83^{+0.40}_{-0.25}$\\
        URG\,J033655.2$-$292627 & $2.6^{+1.1}_{-1.1}$ & 0.15 & $12.44^{+0.26}_{-0.47}$\\
        \multicolumn{8}{c}{\dotfill\,\textbf{ADFS-27}\,\dotfill}\\
        \textbf{URG\,J043657.0$-$543813} & $\mathbf{4.4^{+0.4}_{-0.3}}$ & $\mathbf{0.92}$ & $\mathbf{13.23^{+0.06}_{-0.06}}$ & URG\,J043729.9$-$54365 & $4.0^{+0.7}_{-0.6}$ & 0.80 & $13.02^{+0.12}_{-0.12}$\\
        URG\,J04374.7$-$543914$^{\ddagger}$ & \textemdash & \textemdash & \textemdash & URG\,J043717.4$-$54356 & $2.7^{+0.5}_{-0.5}$ & 1.90 & $12.63^{+0.11}_{-0.14}$\\
        URG\,J043717.5$-$543528 & $2.0^{+0.2}_{-0.2}$ & 11.45 & $12.80^{+0.07}_{-0.09}$ & URG\,J04377.5$-$54341 & $1.9^{+0.4}_{-0.5}$ & 0.05 & $12.57^{+0.17}_{-0.28}$\\
        \hline
        \multicolumn{8}{r}{Continued on next page}\\
        \end{tabular}
    \end{scriptsize}
    \end{center}
\end{table*}

\addtocounter{table}{-1} 

\begin{table*}
    \begin{center}
    \caption{Cont...}
    \begin{scriptsize}
        \begin{tabular}{l c c c l c c c}
        \hline
        \hline
        \multicolumn{1}{l}{ID} &
        \multicolumn{1}{c}{$z_\textrm{phot}^{\dagger}$} &
        \multicolumn{1}{c}{$\chi^2$} &
        \multicolumn{1}{c}{$\log_{10}\left(\lfir{}\right)$} &
        \multicolumn{1}{l}{ID} &
        \multicolumn{1}{c}{$z_\textrm{phot}^{\dagger}$} &
        \multicolumn{1}{c}{$\chi^2$} &
        \multicolumn{1}{c}{$\log_{10}\left(\lfir{}\right)$} \\
        \multicolumn{1}{c}{} &
        \multicolumn{1}{c}{} &
        \multicolumn{1}{c}{} &
        \multicolumn{1}{c}{[$\lsol{}$]} &
        \multicolumn{1}{c}{} &
        \multicolumn{1}{c}{} &
        \multicolumn{1}{c}{} &
        \multicolumn{1}{c}{[$\lsol{}$]}\\
        \hline
        URG\,J043649.4$-$54408 & $3.1^{+0.9}_{-0.8}$ & 0.54 & $12.54^{+0.18}_{-0.23}$\\
        \multicolumn{8}{c}{\dotfill\,\textbf{ADFS-32}\,\dotfill}\\
        \textbf{URG\,J044410.1$-$534949} & $\mathbf{3.0^{+0.6}_{-0.6}}$ & $\mathbf{0.45}$ & $\mathbf{12.65^{+0.12}_{-0.15}}$ & URG\,J04450.4$-$53496$^{\ddagger}$ & \textemdash & \textemdash &
        \textemdash \\
        \multicolumn{8}{c}{\dotfill\,\textbf{G09-83808}\,\dotfill}\\
        \textbf{URG\,J090045.7$+$004124} & $\mathbf{4.5^{+0.4}_{-0.3}}$ & $\mathbf{0.23}$ & $\mathbf{13.25^{+0.05}_{-0.05}}$ & URG\,J090032.8$+$004313 & $2.3^{+0.1}_{-0.1}$ & 6.55 & $13.15^{+0.04}_{-0.05}$\\
        URG\,J090019.4$+$004016$^{\ddagger}$ & \textemdash & \textemdash & \textemdash & URG\,J090057.3$+$00415 & $2.1^{+0.3}_{-0.3}$ & 1.20 & $12.60^{+0.09}_{-0.11}$\\
        URG\,J090054.2$+$004343 & $1.9^{+0.5}_{-0.5}$ & 1.16 & $12.34^{+0.16}_{-0.23}$ & URG\,J090057.1$+$004039 & $1.8^{+0.3}_{-0.3}$ & 6.49 & $12.45^{+0.12}_{-0.17}$\\
        URG\,J090037.1$+$003624 & $1.8^{+0.2}_{-0.2}$ & 2.45 & $12.83^{+0.07}_{-0.09}$\\
        \multicolumn{8}{c}{\dotfill\,\textbf{G15-82684}\,\dotfill}\\
        URG\,J14506.3$+$015038 & $3.2^{+0.2}_{-0.2}$ & 1.24 & $13.14^{+0.04}_{-0.05}$ & \textbf{URG\,J145013.1$+$014810} & $\mathbf{3.5^{+0.3}_{-0.2}}$ & $\mathbf{0.05}$ & $\mathbf{13.07^{+0.05}_{-0.05}}$\\
        URG\,J145012.1$+$015158 & $2.7^{+0.3}_{-0.3}$ & 0.58 & $12.93^{+0.07}_{-0.08}$ & URG\,J145015.4$+$015237 & $3.2^{+0.3}_{-0.3}$ & 0.83 & $12.94^{+0.07}_{-0.07}$\\
        URG\,J145025.7$+$015115 & $2.3^{+0.4}_{-0.4}$ & 0.83 & $12.62^{+0.11}_{-0.14}$ & URG\,J145023.8$+$01514 & $2.5^{+0.7}_{-0.7}$ & 2.67 & $12.48^{+0.18}_{-0.25}$\\
        \multicolumn{8}{c}{\dotfill\,\textbf{SGP-433089}\,\dotfill}\\
        \textbf{URG\,J222737.4$-$333835} & $\mathbf{2.5^{+0.3}_{-0.2}}$ & $\mathbf{0.87}$ & $\mathbf{12.77^{+0.08}_{-0.08}}$ & URG\,J222725.2$-$333920 & $2.4^{+0.3}_{-0.3}$ & 0.14 & $12.83^{+0.09}_{-0.10}$\\
        URG\,J222747.9$-$333533 & $2.5^{+0.4}_{-0.3}$ & 0.21 & $12.71^{+0.10}_{-0.10}$ & URG\,J222731.1$-$33404$^{\ddagger}$ & \textemdash & \textemdash & \textemdash\\
        URG\,J222733.7$-$333440 & $1.9^{+0.3}_{-0.3}$ & 0.66 & $12.66^{+0.10}_{-0.12}$ & URG\,J222737.7$-$333727 & $1.5^{+0.2}_{-0.3}$ & 0.81 & $12.57^{+0.12}_{-0.15}$\\
        URG\,J222730.4$-$333534 & $2.3^{+0.6}_{-0.5}$ & 0.16 & $12.57^{+0.16}_{-0.19}$ & URG\,J222750.1$-$334153 & $3.1^{+1.0}_{-0.7}$ & 0.51 & $12.56^{+0.19}_{-0.18}$\\
        URG\,J222753.8$-$333529 & $2.6^{+0.5}_{-0.4}$ & 4.51 & $12.58^{+0.13}_{-0.14}$ & URG\,J222727.8$-$334056 & $2.2^{+0.4}_{-0.4}$ & 1.43 & $12.52^{+0.13}_{-0.15}$\\
        URG\,J222744.7$-$333741 & $2.3^{+0.4}_{-0.4}$ & 6.69 & $12.49^{+0.13}_{-0.13}$\\
        \hline
        \end{tabular}
    \end{scriptsize}
    \end{center}
\vspace{-\baselineskip}

\noindent
$^{\dagger}$ We quote errors based on the $\chi^2 + 1$ values, without the adding the
intrinsic template scatter in quadrature.

\noindent
$^{\ddagger}$ SPIRE non-detections, for which we do not provide any photometric
redshifts; we do not include these in our analysis.

\end{table*}

\end{document}